\documentclass[hyper,a4paper]{JHEP3} % 10pt is ignored!
\usepackage{epsfig}
\usepackage{latexsym}
\usepackage{graphicx}
\usepackage{amsmath}
\usepackage{amsfonts}   % if you want the fonts
\usepackage{amssymb}    % if you want extra symbols
\usepackage{subfigure}

\def\eq#1{Eq.~(\ref{#1})}
\def\eqs#1#2{Eqs.~(\ref{#1}) and (\ref{#2})}

\def\e3{$\epsilon_3$}

\def\ch2{$\chi^2$}

\def\co#1{{\ifmmode{\cal O}_{#1}\else${\cal O}_{#1}$\fi}}

\newdimen\unit
\def\point#1 #2 #3{\vbox to0pt{\kern-#2\unit
  \hbox{\kern#1\unit#3}\vss}
 \nointerlineskip}

\newcommand{\be}{\begin{equation}}
\newcommand{\ee}{\end{equation}}
\newcommand{\bea}{\begin{eqnarray}}
\newcommand{\eea}{\end{eqnarray}}

\newcommand\ps{\mbox{ ps}} 
\newcommand{\mev}{\mbox{ MeV}}
\newcommand{\gev}{\mbox{ GeV}}
\newcommand{\tev}{\mbox{ TeV}}

\newcommand{\cl}{\text{CL}}

\newcommand{\alphaemmz}{\alpha_{\text{em}}(M_Z)^{\overline{MS}}}
\newcommand{\alphas}{\alpha_s(M_Z)^{\overline{MS}}}

\newcount\hour
\newcount\minute
\newtoks\amorpm
\hour=\time\divide\hour by60 \minute=\time{\multiply\hour by60
\global\advance\minute by- \hour}
\edef\standardtime{{\ifnum\hour<12 \global\amorpm={am}%
    \else\global\amorpm={pm}\advance\hour by-12 \fi
    \ifnum\hour=0 \hour=12 \fi
    \number\hour:\ifnum\minute<100\fi\number\minute\the\amorpm}}
\edef\militarytime{\number\hour:\ifnum\minute<100\fi\number\minute}
\def\bold#1{\setbox0=\hbox{$#1$}%
     \kern-.025em\copy0\kern-\wd0
     \kern.05em\copy0\kern-\wd0
     \kern-.025em\raise.0433em\box0 }

\newcommand{\newc}{\newcommand}
\newc\eg{{\rm {e.g.}}}  \newc\etal{{\rm {et al.}}} \newc\ie{{\rm i.e.}}
\newc\etc{{\rm {etc}}}
\newcommand\lsim{\mathrel{\rlap{\lower4pt\hbox{\hskip1pt$\sim$}}
    \raise1pt\hbox{$<$}}}
\newcommand\gsim{\mathrel{\rlap{\lower4pt\hbox{\hskip1pt$\sim$}}
    \raise1pt\hbox{$>$}}}

\newc{\mhalf}{m_{1/2}}      \newc{\mzero}{m_0}
\newc{\tanb}{\tan\beta}
\newc{\azero}{A_0}
\newc{\at}{A_t} \newc{\ab}{A_b} \newc{\atau}{A_\tau}
\newc{\bmu}{B\mu}           \newc{\sgn}{{\rm sgn}}
\newc{\mone}{M_1}           \newc{\mtwo}{M_2}

 \newc{\hu}{H_u}       \newc{\hd}{H_d}
 \newc{\mhu}{m_{H_u}}       \newc{\mhd}{m_{H_d}}
 \newc{\mhuew}{m^{\ast}_{H_u}}       \newc{\mhdew}{m^{\ast}_{H_d}}
 \newc{\mhuewsq}{m^{\ast\, 2}_{H_u}}       \newc{\mhdewsq}{m^{\ast\, 2}_{H_d}}
 \newc{\mhuast}{m^{\ast}_{H_u}}       \newc{\mhdast}{m^{\ast}_{H_d}}
\newc{\charone}{\chi_1^\pm} \newc{\mcharone}{m_{\chi_1^\pm}}

\newc{\hl}{h}               \newc{\mhl}{m_{\hl}}   \newc{\gammahl}{\Gamma_{\hl}}
\newc{\hh}{H}               \newc{\mhh}{m_{\hh}}   \newc{\gammahh}{\Gamma_{\hh}}
\newc{\ha}{A}               \newc{\mha}{m_{\ha}}   \newc{\gammaha}{\Gamma_{\ha}}
\newc{\hpm}{H^{\pm}}        \newc{\mhpm}{m_{\hpm}} \newc{\gammahpm}{\Gamma_{\hpm}}
\newc{\hp}{H^{+}} \newc{\mhp}{m_{\hp}} \newc{\hm}{H^{-}}
\newc{\mhm}{m_{\hm}}
\newc{\xt}{X_{t}}           \newc{\xb}{X_{b}}

\newc{\qzero}{Q_0}          \newc{\qstop}{Q_{\widetilde t}}
\newc{\amu}{a_{\mu}}        \newc{\amususy}{a_{\mu}^{\text{SUSY}}}
\newc{\amuexpt}{a_{\mu}^{\text{expt}}}        \newc{\amusm}{a_{\mu}^{\text{SM}}}
\newc{\deltaamususy}{\delta a_{\mu}^{\text{SUSY}}}
\newc\gmtwo{(g-2)_{\mu}} %lr \newc\deltaamu{\Delta a_{\mu}}
\newc\deltagmtwo{\delta (g-2)_{\mu}} 
\newc\deltaamu{\Delta a_{\mu}}
\newc{\msbar}{\overline{MS}} \newc{\drbar}{\overline{DR}}
\newc{\yt}{h_t} \newc{\yb}{h_b} \newc{\ytau}{h_{\tau}}

\newc{\mtop}{m_t}               \newc{\mtpole}{M_t}
\newc{\mtaupole}{m_{\tau}^{\text{pole}}}
\newc{\mtmtsmmsbar}{m_t(m_t)^{\msbar}_{{\text{SM}}}}
\newc{\mtmtsmdrbar}{m_t(m_t)^{\drbar}_{{\text{SM}}}}
\newc{\mtmtmssmdrbar}{m_t(m_t)^{\drbar}_{{\text{SUSY}}}}

\newc{\mbmbmsbar}{m_b(m_b)^{\msbar} }

\newc{\mbmbsmmsbar}{m_b(m_b)^{\msbar}_{{\text{SM}}}}
\newc{\mbmzsmmsbar}{m_b(\mz)^{\msbar}_{{\text{SM}}}}
\newc{\mbmzsmdrbar}{m_b(\mz)^{\drbar}_{{\text{SM}}}}
\newc{\mbmzmssmdrbar}{m_b(\mz)^{\drbar}_{{\text{SUSY}}}}

\newc{\mtaumzsmmsbar}{m_{\tau}(\mz)^{\msbar}_{{\text{SM}}}}
\newc{\mtaumzsmdrbar}{m_{\tau}(\mz)^{\drbar}_{{\text{SM}}}}
\newc{\mtaumzmssmdrbar}{m_{\tau}(\mz)^{\drbar}_{{\text{SUSY}}}}

\newc{\mgut}{M_{\rm GUT}}
\newc{\mplanck}{M_{\rm P}}      \newc{\mpl}{M_{\text{Pl}}}
\newc{\msusy}{M_{\rm SUSY}}      \newc{\ms}{M_{\text{S}}}
\newc{\jxf}{J({\xf})}
\newc{\jxfexact}{J_{\rm exact}({\xf})}  \newc{\jxfexp}{J_{\rm exp}({\xf})}
\newc{\VEV}[1]{\langle #1 \rangle}
\newc{\xf}{x_f}
\newc\vrel{v_{\rm rel}}
\newcommand\mchi{m_{\chi}}              
\newc\sell{{\widetilde e}_L}      \newc\msell{m_{\sell}}
\newc\selr{{\widetilde e}_R}      \newc\mselr{m_{\selr}}
\newc\snue{{\widetilde \nu}_e}      \newc\msnue{m_{\snue}}
\newc\snutau{{\widetilde \nu}_\tau}      \newc\msnutau{m_{\snutau}}
\newc\supl{{\widetilde u}_L}      \newc\msupl{m_{\supl}}
\newc\supr{{\widetilde u}_R}      \newc\msupr{m_{\supr}}
\newc\sdl{{\widetilde d}_L}      \newc\msdl{m_{\sdl}}
\newc\sdr{{\widetilde d}_R}      \newc\msdr{m_{\sdr}}

\newcommand\stopone{{\widetilde t}_1}   \newcommand\mstopone{m_{\stopone}}
\newcommand\stoptwo{{\widetilde t}_2}   \newcommand\mstoptwo{m_{\stoptwo}}

\newcommand\mgluino{m_{\widetilde g}}

\newc\sfermion{\tilde f}  \newc\msfermion{m_{\sfermion}}
\newc\cmeter{{\rm cm}} \newc\meter{{\rm m}} \newc\kmeter{{\rm km}}
\newc\second{{\rm sec}}

\newc\sr{{\rm sr}}

\newc{\gstar}{g_\ast}           \newc{\gsstar}{g_{s\ast}}
\newc{\geff}{g_{\rm eff}}
\newcommand\mz{m_{Z}}

\newc{\sthw}{\sin\theta_W}              \newc{\cthw}{\cos\theta_W}
\newc{\bino}{\widetilde B}              \newc{\wino}{\widetilde W_30}
\newc{\higgsinob}{{\widetilde H}^0_b}   \newc{\higgsinot}{{\widetilde H}^0_t}
\newc{\abund}{\Omega h^2}
\newc{\abundchi}{\Omega_\chi h^2}
\newc{\abundcdm}{\Omega_{\text{CDM}} h^2}
\newc{\omegam}{\Omega_{M}}       \newc{\abundm}{\Omega_{M} h^2}
\newc{\omegab}{\Omega_{b}}       \newc{\abundb}{\Omega_{b} h^2}
\newc{\omegacdm}{\Omega_{CDM}}
\newc{\omegatot}{\Omega_{TOT}}
\newc{\rhocrit}{\rho_{crit}}
\newc{\rhochi}{\rho_{\chi}}
\newcommand\pb{\,\mbox{pb}} 

\newc\pc{\,\mbox{pc}} \newc\kpc{\,\mbox{kpc}}
\newc\mpc{\,\mbox{Mpc}} \newc\gpc{\,\mbox{Gpc}}

\newc\BR{BR}
\newc\bsgamma{b\rightarrow s \gamma }
\newc\bxsgamma{\overline{B}\rightarrow X_{s}\gamma}
\newc\brbsgamma{\BR(\overline{B}\rightarrow X_s\gamma)}

\newcommand\brbsmumu{\BR(\overline{B}_s\to\mu^+\mu^-)}

\newcommand\delmbs{\Delta M_{B_s}}

\newc{\beq}{\begin{equation}}
\newc{\eeq}{\end{equation}}

\renewcommand\[{\left[}
\renewcommand\]{\right]}
\newcommand\vs{{\it {vs.}}}

\newc\stoponetwo{{\widetilde t}_{1,2}}
\newc\sbotonetwo{{\widetilde b}_{1,2}}
\newc\stauonetwo{{\widetilde \tau}_{1,2}}

\newc{\sigsip}{\sigma^{SI}_{p}} \newc{\sigsin}{\sigma^{SI}_{n}}
\newc{\sigsiN}{\sigma^{SI}_{N}}
\newc{\sigsdp}{\sigma^{SD}_{p}} \newc{\sigsdn}{\sigma^{SD}_{n}}
\newc{\sigsiA}{\sigma^{SI}_{A}}

\newc{\pbar}{\bar{p}}

\newc{\egamma}{E_{\gamma}}
\newc{\flux}[1]{\Phi_{#1}}
\newc{\dfluxde}[1]{\frac{d\Phi_{#1}}{d E_{#1}}}

\newc{\fluxg}{\Phi_{\gamma}}
\newc{\dfluxgde}{\frac{d\Phi_{\gamma}}{d\egamma}}
\newc{\dfluxgdetext}{ d\Phi_{\gamma} / d\egamma}

\newc{\eplus}{e^+}
\newc{\epos}{E_{\eplus}}
\newc{\eps}{\varepsilon}

\newc{\npos}{n_{\eplus}} \newc{\Npos}{N_{\eplus}}
\newc{\dnposde}{\frac{d n_{\eplus}}{d\epos}}
\newc{\dnposdeps}{\frac{d n_{\eplus}}{d\eps\phantom{_{\eplus}}}}
\newc{\dnposdepstext}{ d n_{\eplus} / d\eps}
\newc{\fluxpos}{\Phi_{\eplus}}  \newc{\fluxelec}{\Phi_{e^{-}}}
\newc{\dfluxposde}{\frac{d\Phi_{\eplus}}{d\epos}}
\newc{\dfluxposdetext}{ d\Phi_{\eplus} / d\epos}

\newc{\nfwc}{{\text{NFW+ac}}} \newc{\moorec}{{\text{Moore+ac}}}

\newc{\chisq}{\chi^2}  \newc{\chisqred}{\chi^2_{\text{red}}}

\newc\xilim{\xi_{\rm lim}} %\newc\xilim{\xi_{\star}}
\newc\tlim{t_{\rm lim}} %\newc\tlim{t_{\star}}
\newc\zetalim{\zeta_{\rm lim}} %\newc\zetalim{\zeta_{\star}}

\newc\zetah{\zeta_h}
\newc{\relprobone}[1]{p({#1} \vert d)}
\newc{\relprobtwo}[2]{p({#1},{#2} \vert d)}
\renewcommand\[{\left[}
\renewcommand\]{\right]}

\long\def\begincomment#1\endcomment{%
        \begingroup\sf\baselineskip12pt#1\endgroup}
\newcommand{\squishlist}{
   \begin{list}{$\bullet$}
    { \setlength{\itemsep}{0pt}      \setlength{\parsep}{3pt}
      \setlength{\topsep}{3pt}       \setlength{\partopsep}{0pt}
      \setlength{\leftmargin}{1.em} \setlength{\labelwidth}{1em}
      \setlength{\labelsep}{0.5em} } }
\newcommand{\squishend}{
    \end{list}  }
\newcommand\mZ{m_{Z}}        \newcommand\mW{m_{W}}
\newcommand{\data}{d}

\newcommand{\basis}{m}
\newcommand{\derived}{\xi}

\newcommand{\sineff}{\sin^2 \theta_{\text{eff}}}

\newcommand\jcap[3]   { %lr \@spires{00124%2C#1%2C#3}
		{{\it JCAP\ }{\bf #1} (#2) #3}}

\title{Global fits of the Non-Universal Higgs Model}
\author{Leszek Roszkowski\\
        Department of Physics and Astronomy, University of Sheffield,\\
        Sheffield S3 7RH, UK,\\ and The Andrzej
Soltan Institute for Nuclear Studies, Warsaw, Poland\\
        E-mail: \email{L.Roszkowski@sheffield.ac.uk}}
\author{Roberto Ruiz de Austri\\
        Instituto de F\'isica Corpuscular, IFIC-UV/CSIC,\\ Valencia, Spain\\
        E-mail: \email{rruiz@ific.uv.es}}
\author{Roberto Trotta\\
        Astrophysics Group, Imperial College London \\
	  Blackett Laboratory, Prince Consort Road, London SW7 2AZ, UK
        E-mail: \email{r.trotta@imperial.ac.uk}}
\author{Yue-Lin Sming Tsai\\
        Department of Physics and Astronomy, University of Sheffield,\\
        Sheffield S3 7RH, UK
        E-mail: \email{php06yt@sheffield.ac.uk}}
\author{Tom A. Varley\\
        Department of Physics and Astronomy, University of Sheffield,\\
        Sheffield S3 7RH, UK
        E-mail: \email{php06tav@sheffield.ac.uk}}

      \abstract{We carry out global fits to the Non-Universal Higgs
        Model (NUHM), applying all relevant present-day
        constraints. We present global probability maps for the NUHM
        parameters and observables (including collider signatures,
        direct and indirect detection quantities), both in terms of
        posterior probabilities and in terms of profile likelihood
        maps. We identify regions of the parameter space where the
        neutralino dark matter in the model is either bino-like, or
        else higgsino-like with mass close to $1\tev$ and
        spin-independent scattering cross section $\sim
        10^{-9}-10^{-8}$ pb. We trace the occurrence of the
        higgsino-like region to be a consequence of a mild focusing
        effect in the running of one of the Higgs masses, the
        existence of which in the NUHM we identify in our
        analysis. Although the usual bino-like neutralino is more
        prominent, higgsino-like dark matter cannot be excluded,
        however its significance strongly depends on the prior and
        statistics used to assess it.  We note that, despite
        experimental constraints often favoring different regions of
        parameter space to the Constrained MSSM, most observational
        consequences appear fairly similar, which will make it
        challenging to distinguish the two models experimentally.}

\begin{document}
%%%%%%%%%%%%%%%%%%%%%%%%%%%%%%%%%%%%%%%%%%%%%%%%%%%%%%%%%%%%%%%%

%%%%%%%%%%%%%%%%%%%%%%%%%%%%%%%%%%%%%%%%%%%%%%%%%%%%%%%%%%%%%%%%%%%%%%%%%%%%%%
\section{Introduction}\label{sec:intro}
%%%%%%%%%%%%%%%%%%%%%%%%%%%%%%%%%%%%%%%%%%%%%%%%%%%%%%%%%%%%%%%%%%%%%%%%%%%%%%

Softly broken low-energy supersymmetry (SUSY) has many attractive
features \cite{susy-reviews}. For example, unlike the Standard Model
(SM), it provides an 
elegant solution to the gauge hierarchy problem  and a natural
weakly-interacting dark matter (DM) candidate, in addition to accounting for
gauge coupling unification. On the other hand, SUSY itself has to be
(softly) broken in order to make contact with reality, which in the
general Minimal Supersymmetric Standard Model (MSSM) introduces a
large number of new free parameters, namely the soft masses. Because of
SUSY's natural link with grand unification theories (GUTs), one
often explores SUSY models by imposing various boundary conditions at
the GUT scale.  The most popular model of this class is the Constrained MSSM
(CMSSM)~\cite{kkrw94}, in which not only do gaugino soft masses unify to
$\mhalf$ but also soft masses of all the sfermions and Higgs doublets
unify to $\mzero$. These parameters, along with a common tri-linear
mass parameter $\azero$ and the ratio of Higgs vacuum expectation
values $\tanb$, form the four continuous parameters of the CMSSM. The relative
simplicity of the model makes a very attractive playground for many
studies.

On the other hand, precisely because of its economy, the CMSSM may be
missing some features of unified models with less restrictive boundary
conditions at the unification scale.  In particular, the assumption of
Higgs (soft) mass unification with those of the sfermions does not
seem strongly motivated since the Higgs and matter fields belong to
different supermultiplets.  One explicit example where this is the
case is a minimal $SO(10)$ supersymmetric model
(MSO$_{10}$SM)~\cite{raby-s010}, which is well motivated and opens up
a qualitatively new region of parameter space~\cite{drrr1+2}.  Models
like this provide a good motivation for exploring a wider class of
phenomenological models in which the soft masses $\mhu$ and $\mhd$
(as defined at the GUT scale) of the two Higgs doublets are treated as
independent parameters and which come under the name of the
Non-Universal Higgs Model
(NUHM)~\cite{nuhmbasics}. \footnote{A
reduced version of NUHM with $\mhu=\mhd$ has also been examined in
several papers, eg~\cite{Baer:2004fu, Baer:2005bu}.}

We assume, as ever, a Minimal Flavor Violating (MFV) scenario with no additional flavor violating terms appearing beyond the SM ones. In the NUHM, there are therefore six continuous free parameters:
\beq
\mzero, \mhalf, \tanb, \azero, \mhu~{\rm and}~ \mhd.
\label{nuhmpars:eq}
\eeq
The Renormalization Group Equations (RGEs) are then used to evaluate
masses and couplings at the electroweak scale and the Higgs potential
is minimized in the usual way. 
Electroweak Symmetry Breaking (EWSB) conditions for NUHM read:
\bea
\label{ewsb1:eq}
\mu^2&=& \frac{\mhd^2 - \mhu^2\, \tan^2\beta}{\tan^2\beta -1} -\frac{1}{2}\mz^2,\\
\mha^2 &=& \mhd^2 + \mhu^2+2\mu^2,
\label{ewsb2:eq}
\eea
where $\mha$ stands for the mass of the pseudoscalar Higgs $\ha$ and
$\mu$ is the SUSY-preserving Higgs/higgsino mass parameter. In the
above equations all the parameters are evaluated at the usual electroweak
scale 
$\msusy\equiv \sqrt{\mstopone\mstoptwo}$ (where
$m_{\stopone,\stoptwo}$ denote the masses of the scalar partners of
the top quark), chosen so as to minimize higher order loop
corrections. At $\msusy$ the (1-loop corrected) conditions of
EWSB are imposed and the SUSY spectrum
is computed at $\mz$.

Like in the CMSSM the sign of $\mu$ remains undetermined. On the other
hand, in contrast to the CMSSM, because of the larger number of free
parameters~\eqs{ewsb1:eq}{ewsb2:eq} allow one to treat both $\mu$ and
$\mha$ as independent parameters in place of high scale parameters
$\mhu$ and $\mhd$.  This will have an important impact on the
properties of the model. In particular, the ability to effectively
choose the position of the $A$ funnel or to tune $\mu$ for a given
point in the parameter space to give a correct relic abundance of the
neutralino DM will lead to very different phenomenological
predictions, as we will see below.

The moderately increased number of free parameters of the NUHM has
been shown to lead to a rich and distinct phenomenology (see, for
example~\cite{Ellis:2002wv,Ellis:2002iu, Ellis:2007by, Baer:2008ih}
and references therein). In particular, there is a larger variety of
choices for the lightest superpartner (LSP). In the CMSSM the LSP is
either the lightest neutralino or the lighter stau or, in some
relatively rare cases, the lighter stop. In contrast, in the NUHM, the
LSP can in addition be a sneutrino or right handed
selectron~\cite{Ellis:2002iu}.  In this, as in previous analyses, we
do not consider the possibility of a gravitino LSP.  Assuming the LSP
to be the dark matter in the Universe eliminates states that are not
electrically neutral and leads to a non-trivial constraint on the
parameter space. However, the near degeneracy of many states with the
LSP leads to a great variety of co-annihilation channels. Also, given
that $\mha$ can now be treated effectively as a free parameter, the
resonance channel can be important in different ways from the CMSSM.
Also, as we will see, there are sizable regions of the parameter space
where the neutralino is actually higgsino-like while giving the
correct dark matter abundance.  Such regions cannot be excluded
although their statistical significance is presently difficult to
determine, as we will show.  Although in general a higgsino-dominated
LSP in the mass range of several hundred GeV, or less, underproduces
dark matter~\cite{Ellis:lighthiggsino}, as its mass increases the
transition to higgsino dark matter can provide an acceptable relic
density. In general these effects will be important to give the
correct relic density in different parts of the parameter space of the
model.

%lr* The larger number of parameters makes a full exploration of the NUHM
%lr* parameter space even more challenging than that of the
%lr* CMSSM. Additionally, varying relevant SM parameters can lead to
%lr* important consequences, which has been shown in recent studies of the
%lr* CMSSM with the help of Markov Chain Monte Carlo (MCMC) scanning
%lr* techniques coupled to Bayesian statistics~\cite{al05,rtr1}. There are
%lr* significant advantages in doing this, such as being able to
%lr* efficiently explore the parameter space varying all inputs
%lr* simultaneously and being able to incorporate errors, both theoretical
%lr* and experimental, as well as relevant SM parameters correctly. This
%lr* will allow us to efficiently explore regions of the model's parameters
%lr* that were not easily accessible in the usual fixed-grid scans.

The larger number of parameters makes a full exploration of the NUHM
parameter space even more challenging than that of the CMSSM. There
are several important reasons why it might be advantageous to adopt a
Markov Chain Monte Carlo (MCMC) sampling technique to perform a global
scan, as we do in the present paper. First, fixed grid scans of the
likelihood become rapidly inefficient with the increase of the
dimensionality of the model's parameter space, as the computational
effort required scales exponentially with the number of dimensions of
the parameter space. Secondly, typically such scans do not have
sufficient resolution to map out in detail the finer structure of the
likelihood surface. The combination of those two aspects means that
fixed grid scans are typically limited to exploring 2-dimensional
slices of the likelihood at the time, while fixing all other
parameters to arbitrary values. This leads to a large underestimation
of the uncertainty in the parameters one is interested in studying and
to limiting the ability to perform global scans of the model's
parameter space. For example, fixed-grid scans identify in the CMSSM
so-called "WMAP strips" in various 2-dimensional slices, \ie, tightly
constrained regions of parameter space where the relic abundance is in
good agreement with the WMAP value. However, the use of global scans
which fit all model parameters simultaneously allows one to reveal
that such strips only arise because the other parameters of the model
(typically, $\azero$ and $\tanb$) have been fixed at arbitrary
values. Once those quantities are incorporated in a global scan, the
WMAP strips disappear and become unified in much wider error regions.

Recently, such difficulties have been overcome thanks to the
introduction of MCMC scanning techniques coupled to Bayesian
statistics (for recent studies of the CMSSM,
see~\cite{al05,rtr1}).\footnote{An alternative $\chi^2$-based approach
  has been pursued in Ref.~\cite{buchmueller_chisq}.}  This methodology
presents several significant advantages -- as the computational effort
of MCMC scales approximately linearly with the number of dimensions
being scanned, all relevant parameters can be included simultaneously
in the fit. This allows one to incorportate important residual
uncertainties in the SM parameters, as well. The more efficient
exploration of the likelihood allows one to efficiently explore
regions of the model’s parameters that are not easily accessible in
the usual fixed-grid scans. Another important aspect of the technique
being used here is that, once the samples from the posterior have been
accumulated, one can investigate both Bayesian maps (in terms of
posterior distributions) and frequentist ones (the profile
likelihood), in order to assess the robustness of the result with
respect for example to different priors or choices of
statistics. Whenever we consider the profile likelihood, we have of course removed the prior weight and are therefore effectively showing samples from the likelihood alone. 
Finally, it becomes possible to produce probability maps
of any derived quantity that one is interested in, and of its
correlations with any other variable of interest. Altogether, those
advantages make MCMC methods a powerful tool for the global
exploration of the NUHM parameter space.

In this paper we investigate global fits of the NUHM parameter space
with a modified version of the \texttt{SuperBayeS}~package
\cite{superbayes}. Our fits include all the relevant constraints
coming from experiment, in particular, the branching ratio of
$\bsgamma$, the difference $\deltagmtwo$ between the experimental and
SM values of the magnetic moment of the muon, the LEP limits on
sparticle and Higgs masses, the 5 year WMAP limits on the relic
density $\abundcdm$ and several other measured but imprecisely known
quantities that SUSY can contribute to. It should be noted that we
take a wider range of parameters than previous scans in the literature
(up to $4\tev$ in the case of the soft masses), and the MCMC technique
allows us to vary all our parameters such that we also consider
$\azero \ne 0$, while at the same time varying relevant SM parameters.
As our statistical measures we will apply a Bayesian posterior
probability density function and profile likelihood, both to be
introduced below.

The paper is organised as follows. In sec.~\ref{sec:Bayes} we
summarise the statistical formalism that we employ and list the
constraints that we apply in our numerical analysis. Next, in sec.~\ref{sec:nuhmpars} we present
the constraints on NUHM parameters resulting from our global scan and
discuss the main features, including the possibility of higgsino DM
and a mild focusing effect. We also discuss some implications for
phenomenology, including direct and indirect SUSY searches in
colliders, while in sec.~\ref{sec:resultsdm} we discuss prospects of
direct and indirect detection of the neutralino dark matter in the
model.  We conclude and summarize our results in
sec.~\ref{sec:summary}. We also illustrate the prior dependence of
some of our results in Appendix~\ref{sec:priors}.

%%%%%%%%%%%%%%%%%%%%%%%%%%%%%%%%%%%%%%%%%%%%%%%%%%%%%%%%%%%%%%%%%%%%%%%%%%%%%%
%\section{Outline of the Bayesian method}\label{sec:Bayes}
\section{Outline of the statistical treatment}\label{sec:Bayes}
%%%%%%%%%%%%%%%%%%%%%%%%%%%%%%%%%%%%%%%%%%%%%%%%%%%%%%%%%%%%%%%%%%%%%%%%%%%%%%

In comparison to earlier analyses of the
CMSSM~\cite{al05,rtr1,rrt2} here we have a larger base parameter
set, defined by
\beq
\theta = (\mhalf, \mzero,\azero, \tanb,\mhu, \mhd),
\label{eq:cmssm}
\eeq
where we fix $\sgn(\mu)$= +1. As the relevant SM parameters, when
varied over their experimental ranges, have impact on the observable
quantities, fixing them at their central values would lead to
inaccurate results. Instead, here we incorporate them explicitly as
free parameters (which are then constrained using their measured
values), which we call {\em nuisance parameters} $\psi$, where
\beq
\psi = (\mtpole,m_b(m_b)^{\msbar},\alpha_{em}(\mz)^{\msbar}, \alpha_{s}(\mz)^{\msbar}).
\label{eq:nuis}
\eeq
In \eq{eq:nuis} $\mtpole$ denotes the pole top quark mass, while the other three
parameters:  $\mbmbmsbar$ -- the bottom
quark mass evaluated at $m_b$, $\alphaemmz$ and $\alphas$ -- respectively the
electromagnetic and the strong coupling constants evaluated at the $Z$ pole mass
$\mZ$ -  are all computed in the $\msbar$ scheme.
Using notation consistent with previous analyses we define our now ten
{\em basis  parameters}  as 
\beq
\basis=(\theta,\psi)
\label{eq:basis}
\eeq
which we will be scanning simultaneously with the MCMC technique. For
each choice of $\basis$ various colliders or cosmological observables
will be calculated. These derived variables are denoted by
$\derived=(\xi_1,\xi_2,\ldots)$, which are then compared with the
relevant available data $\data$.  The quantity we are interested in is
the {\em posterior probability density function} (pdf), or simply
posterior, $p(\basis|\data)$ which gives the probability of the
parameters after the constraints coming from the data have been
applied. The posterior follows from Bayes' theorem,
\beq 
p(\basis|\data)=\frac{p(\data|\derived) \pi(\basis)}{p(\data)},
\label{eq:bayesthm}
\eeq
where $p(\data|\derived)$, taken as a function of $\derived$ for {\em
  fixed data} $\data$, is called the {\em likelihood} (where the
dependence of $\derived(\basis)$ is understood).  The likelihood is
the quantity that compares the data with the derived
observables. $\pi(m)$ is the {\em prior} which encodes our state of
knowledge of the parameters before comparison with the data. This
state of knowledge is then updated by the likelihood to give us the
posterior. $p(d)$ is called the {\em evidence} or {\em model
  likelihood}, and in our analysis can be treated as a normalisation
and hence is ignored subsequently. (See instead~\cite{BenModelComp}
for an example of how the evidence can be used for model comparison
purposes.)

Our inference problem is fully defined once we specify the prior on
the rhs of Bayes' theorem~\eq{eq:bayesthm}. We present our main results below for the following
choice of priors (which we call our {\em log prior} case): a uniform
prior on $\log\mhalf$, $\log\mzero$, $\log\mhu$ and $\log\mhd$, in the
range $50\gev<\mzero,\mhalf<4\tev$ and $0<\mhu,\mhd<4\tev$, and a
uniform prior on $\azero$ and $\tanb$ in the ranges $|A_0|<7\tev$ and
$2<\tanb<62$. Notice that by this range of $\mhu$ and $\mhd$, and
taking $\sgn(\mu)>0$, we automatically satisfy the GUT stability
constraint of ref.~\cite{Ellis:2002iu}.

Our choice of a prior uniform in the log of the masses is dictated by
both physical and statistical reasons. From the physical point of
view, log priors explore in much greater detail the low-mass region,
which exhibits many fine-tuned points that can easily be missed
otherwise. From the statistical point of view, log priors give the
same {\em a priori} weight to all orders of magnitude in the masses,
and thus appear to be less biased to giving larger statistical {\em a
  priori} weights to the large mass region, which under a flat prior
has a much larger volume in parameter space.

In the ideal case where the likelihood is sharply peaked within the
range of the prior, the posterior is dominated by the likelihood and
the details of the prior choice do not matter. However, given
present-day data even the CMSSM exhibits a certain amount of prior
dependence~\cite{tfhrr1}, which is however expected to be strongly
reduced once future LHC data become available~\cite{rrt4}. We thus
expect that the NUHM will also be affected by a certain amount of
prior dependence. In order to assess whether our results are robust
wrt the prior choice we also investigate a different statistical
measure, namely the profile likelihood. It is obtained from our scan
by maximising over the parameters not shown (rather than integrating
over them, as one does to obtain the Bayesian posterior).\footnote{The
  procedure for evaluating the profile likelihood that we use can be
  found in Ref.~\cite{tfhrr1}.} 
  
  The profile likelihood has the advantage of being a prior-independent quantity. However, it is a difficult quantity to evaluate numerically, for it requires a maximisation along the directions one is profiling over. Therefore, it is very difficult to ensure that sparse-sampling schemes (such as MCMC) have gathered a sufficient number of samples to guarantee that the profile likelihood obtained from them can be deemed to be stable and robust. We stress that the MCMC algorithm we adopt in our study is not designed to achieve a maximisation of the likelihood -- rather, the MCMC exploration is targeted at obtaining samples from modes of large posterior mass. This is of special concern for multi-dimensional parameter spaces with a fragmented, multi-modal likelihood function such as the one being considered here. Indeed, the analysis of Ref.~\cite{GA} showed that a genetic algorithm can be used to find better fitting points than those returned by MCMC (albeit with a computational effort which is about 10 times larger). Related statistical issues in connection with profile likelihood coverage properties have also been recently discussed in~\cite{coverage}. Since profile likelihood confidence intervals depend on the value of the best fit likelihood, this implies that the results we present below for profile likelihood intervals have to be interpreted with great caution. In particular, the best-fit points we report below (and their $\chi^2$ values) should be considered as representative of the the various physical regions in parameter space, rather than the absolute global best fits. The aim of this work is to achieve a first global exploration of the NUHM, and therefore we leave the detailed discussion of the stability of the profile likelihood reconstruction from MCMC methods to a future, dedicated study~\cite{PL}. 
  
In fact, the profile likelihood and the Bayesian posterior ask two
different statistical questions of the data: the latter evaluates
which regions of parameter space are most plausible in the light of
the measure implied by the prior (and is thus in general prior dependent except when the data are sufficiently constraining to overrule any sensible choice of prior); the former singles out regions of
high quality of fit, independently of their extent in parameter space and of the prior choice,
thus disregarding the possibility of them being highly fine-tuned. The
information contained in both is relevant and interesting in
evaluating the viability of the underlying physical
%lr mechanisms, 
properties of the model (possible collider signatures, \etc)
which of course do not depend on our statistical tools
{\em per se}. However, our conclusions about the plausibility of such
physical 
%lr mechanisms, 
properties might be different depending on the exact
statistical question asked of the data, e.g., whether one considers a
posterior pdf or a profile likelihood (in the simple case of a
Gaussian distributed quantity, both the pdf and the profile likelihood
are identical and thus the question of which to choose does not
arise). In Appendix~\ref{sec:priors} we examine the robustness of our
results with respect to a change of priors in the mass variables.

The SM parameters are assigned flat priors over a sufficiently wide range and are then constrained by
applying Gaussian likelihoods representing the experimental
observations (see table~\ref{tab:meas}). The Gaussian likelihoods being much more sharply peaked than the flat prior range, we find that the posterior is completely dominated by the likelihood for the SM parameters. 

The predictions for the observable quantities are obtained by using
SoftSusy 2.0.5 and DarkSusy 4.0 \cite{softsusy, darksusy} as
implemented in the \texttt{SuperBayeS} code. The likelihoods for the
relevant observables are taken as Gaussian (for measured observables)
with mean $\mu$, experimental errors $\sigma$ and theoretical errors
$\tau$ (see the detailed explanation in~\cite{rtr1}). In the case
where there only an experimental limit is available, this is given,
along with the theoretical error. The smearing out of bounds and
combination of experimental and theoretical errors is handled in an
identical manner to \cite{rtr1}.
 
\begin{table} 
 \centering
\begin{tabular}{|l | l l | l|}
\hline
SM (nuisance) parameter  &   Mean value  & \multicolumn{1}{c|}{Uncertainty} & ref. \\
 &   $\mu$      & ${\sigma}$ (exper.)  &  \\ \hline
$\mtpole$           &  172.6 GeV    & 1.4 GeV&  \cite{topmass:mar07} \\
$m_b (m_b)^{\overline{MS}}$ &4.20 GeV  & 0.07 GeV &  \cite{pdg06} \\
$\alpha_{\text{s}}(M_Z)^{\overline{MS}}$       &   0.1176   & 0.002 &  \cite{pdg06}\\
% $1/\alpha_{\text{em}}(M_Z)^{\overline{MS}}$  & 127.958 & 0.048
$1/\alpha_{\text{em}}(M_Z)^{\overline{MS}}$  & 127.955 & 0.018 &  \cite{pdg06} \\ \hline
\end{tabular}
\caption{Experimental mean $\mu$ and standard deviation $\sigma$
 adopted for the likelihood function for SM (nuisance) parameters,
 assumed to be described by a Gaussian distribution.
\label{tab:meas}}
\end{table}
%%%%%%%%%%%%%
\begin{table}
\centering
\begin{tabular}{|l | l l l | l|}
\hline
Observable &   Mean value & \multicolumn{2}{c|}{Uncertainties} & ref. \\
 &   $\mu$      & ${\sigma}$ (exper.)  & $\tau$ (theor.) & \\\hline
 $\mW$     &  $80.392\gev$   & $29\mev$ & $15\mev$ & \cite{lepwwg} \\
$\sineff{}$    &  $0.23153$      & $16\times10^{-5}$
                & $15\times10^{-5}$ &  \cite{lepwwg}  \\
$\deltagmtwo \times 10^{10}$       &  27.5 & 8.4 &  1 & \cite{gm2alt}\\
 $\brbsgamma \times 10^{4}$ &
 3.55 & 0.26 & 0.21 & \cite{bsgexp} \\
$\delmbs$     &  $17.33\ps^{-1}$  & $0.12\ps^{-1}$  & $4.8\ps^{-1}$
& \cite{cdf-deltambs} \\
%$\brbtaunu \times 10^{4}$ &  $1.32$  & $0.49$  & $0.38$
%& \cite{bsgexp} \\
$\abundchi$ &  0.1099 & 0.0062 & $0.1\,\abundchi$& \cite{wmap5yr} \\\hline
   &  Limit (95\%~\cl)  & \multicolumn{2}{r|}{$\tau$ (theor.)} & ref. \\ \hline
$\brbsmumu$ &  $ <5.8\times 10^{-8}$
& \multicolumn{2}{r|}{14\%}  & \cite{cdf-bsmumu}\\
$\mhl$  & $>114.4\gev$\ ($91.0\gev$)  & \multicolumn{2}{r|}{$3 \gev$}
& \cite{lhwg} \\
$\zetah^2$
& $f(m_h)$ & \multicolumn{2}{r|}{negligible}  & \cite{lhwg} \\
sparticle masses  &  \multicolumn{3}{c|}{See table~4 in ref.~\cite{rtr1}.}  & \\ \hline
%  &  & \multicolumn{2}{r|}{}  & \\
\end{tabular}
\caption{Summary of the observables used in the analysis. Upper part:
Observables for which a positive measurement has been
made. $\deltagmtwo$ denotes the discrepancy between
the experimental value and the SM prediction of the anomalous magnetic
moment of the muon $\gmtwo$. For central values of the SM input
parameters used here, the SM value of $\brbsgamma$ is
$3.11\times10^{-4}$, while the theoretical error of
$0.21\times10^{-4}$ includes uncertainties other than the parametric
dependence on the SM nuisance parameters, especially on $\mtpole$ and
$\alphas$.  For each quantity we use a
likelihood function with mean $\mu$ and standard deviation $s =
\sqrt{\sigma^2+ \tau^2}$, where $\sigma$ is the experimental
uncertainty and $\tau$ represents our estimate of the theoretical
uncertainty (see~\cite{rtr1} for details). Lower part: Observables for which only limits currently
exist.  The likelihood function is given in
ref.~\cite{rtr1}, including in particular a smearing out of
experimental errors and limits to include an appropriate theoretical
uncertainty in the observables. $\mhl$ stands for the light Higgs mass
while $\zetah^2= g^2(\hl ZZ)_{\text{MSSM}}/g^2(\hl ZZ)_{\text{SM}}$,
where $g$ stands for the Higgs coupling to the $Z$ and $W$ gauge boson
pairs.
\label{tab:measderived}}
\end{table}
%%%%%%%%%%%%% 
Any points that fail to provide radiative EWSB, give us tachyonic
sleptons or provide the LSP which is not the lightest neutralino are
excluded.  As in previous works~\cite{rtr1, rrt2, rrt3}, we adopt a
Metropolis-Hastings MCMC algorithm to sample the parameter space. We
have also cross-checked our results by employing the more recently
implemented MultiNest algorithm~\cite{Feroz:2007kg,tfhrr1} and the
findings are compatible (up to numerical noise). The results presented
in the rest of this paper are obtained from the 25 chains that were
used, garnering a total of $3\times10^5$ samples each, with an
acceptance rate of around 4\%. Convergence criteria are the same as in
our previous papers~\cite{rtr1, rrt2, rrt3}.

%%%%%%%%%%%%%%%%%%%%%%%%%%%%%%%%%%%%%%%%%%%%%%%%%%%%%%%%%%%%%%%%%%%%%%%%%%%%%%
\section{Global constraints on NUHM parameters}\label{sec:nuhmpars}
%%%%%%%%%%%%%%%%%%%%%%%%%%%%%%%%%%%%%%%%%%%%%%%%%%%%%%%%%%%%%%%%%%%%%%%%%%%%%%

In this section we present some results from our global scans of the NUHM
parameter space. 

\subsection{Probability maps of NUHM parameters}

\begin{figure}[tbh!]
\begin{center}
\begin{tabular}{c c c}
  \includegraphics[width=0.31\textwidth]{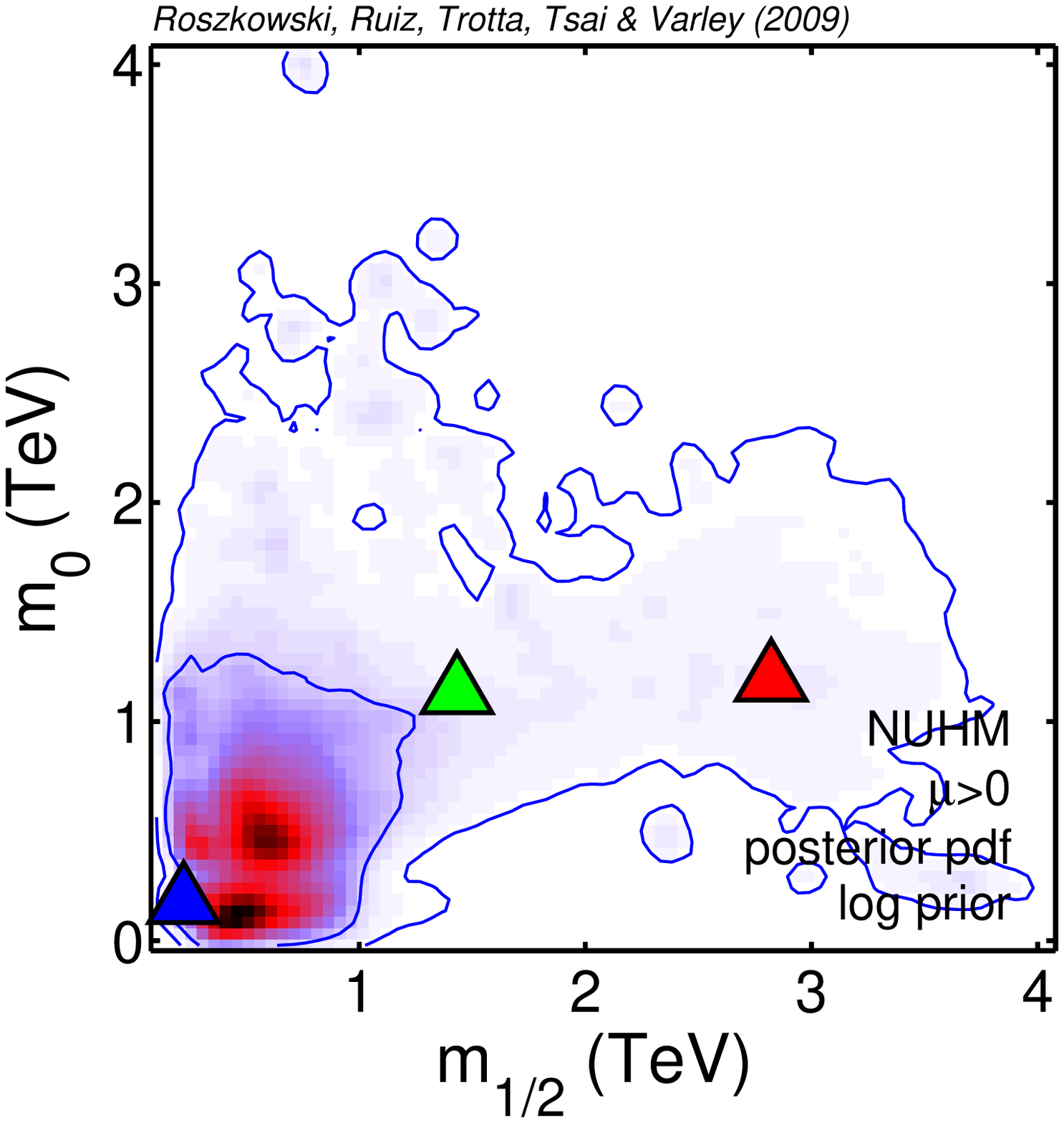}
 & \includegraphics[width=0.31\textwidth]{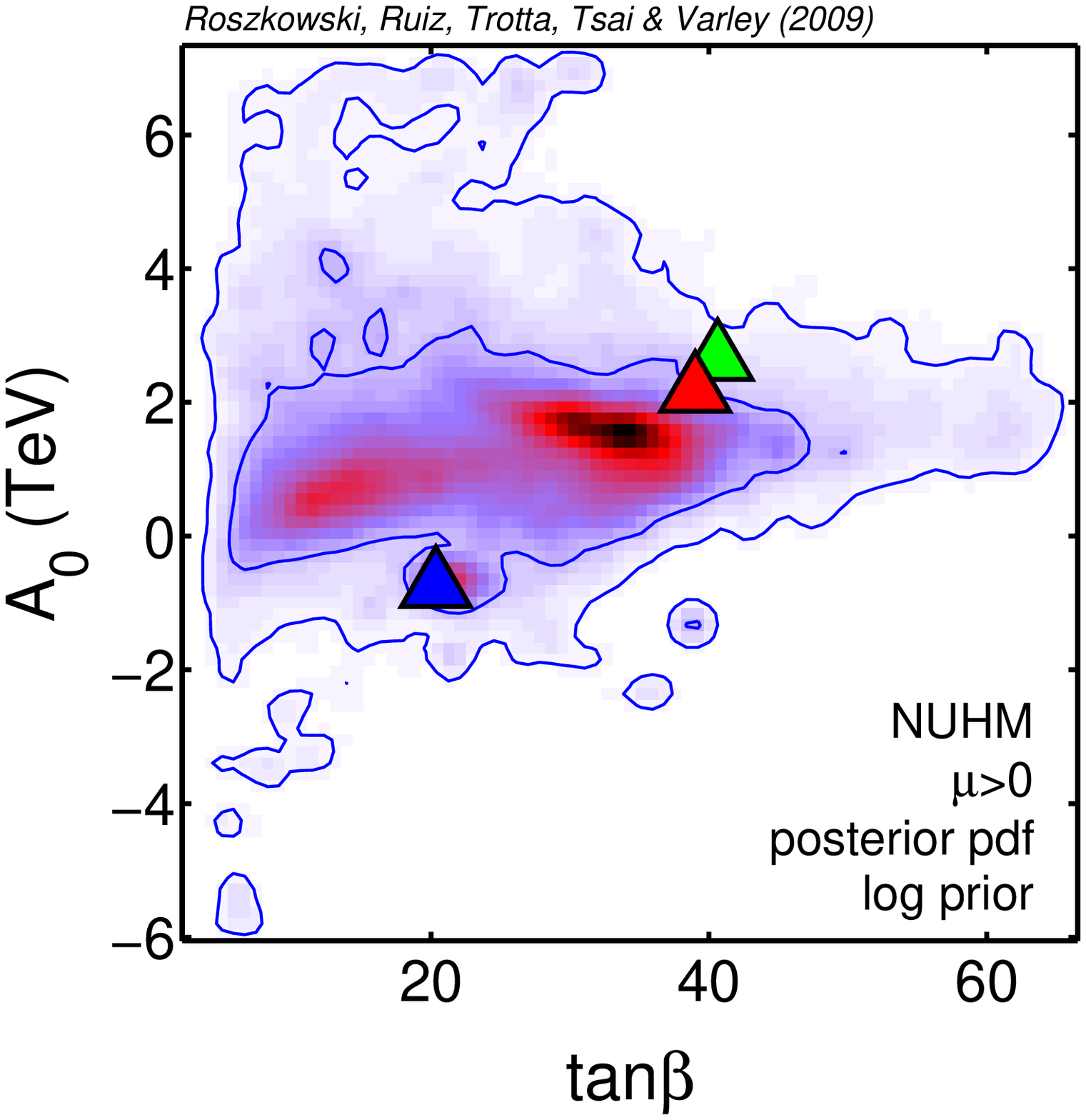}
 & \includegraphics[width=0.31\textwidth]{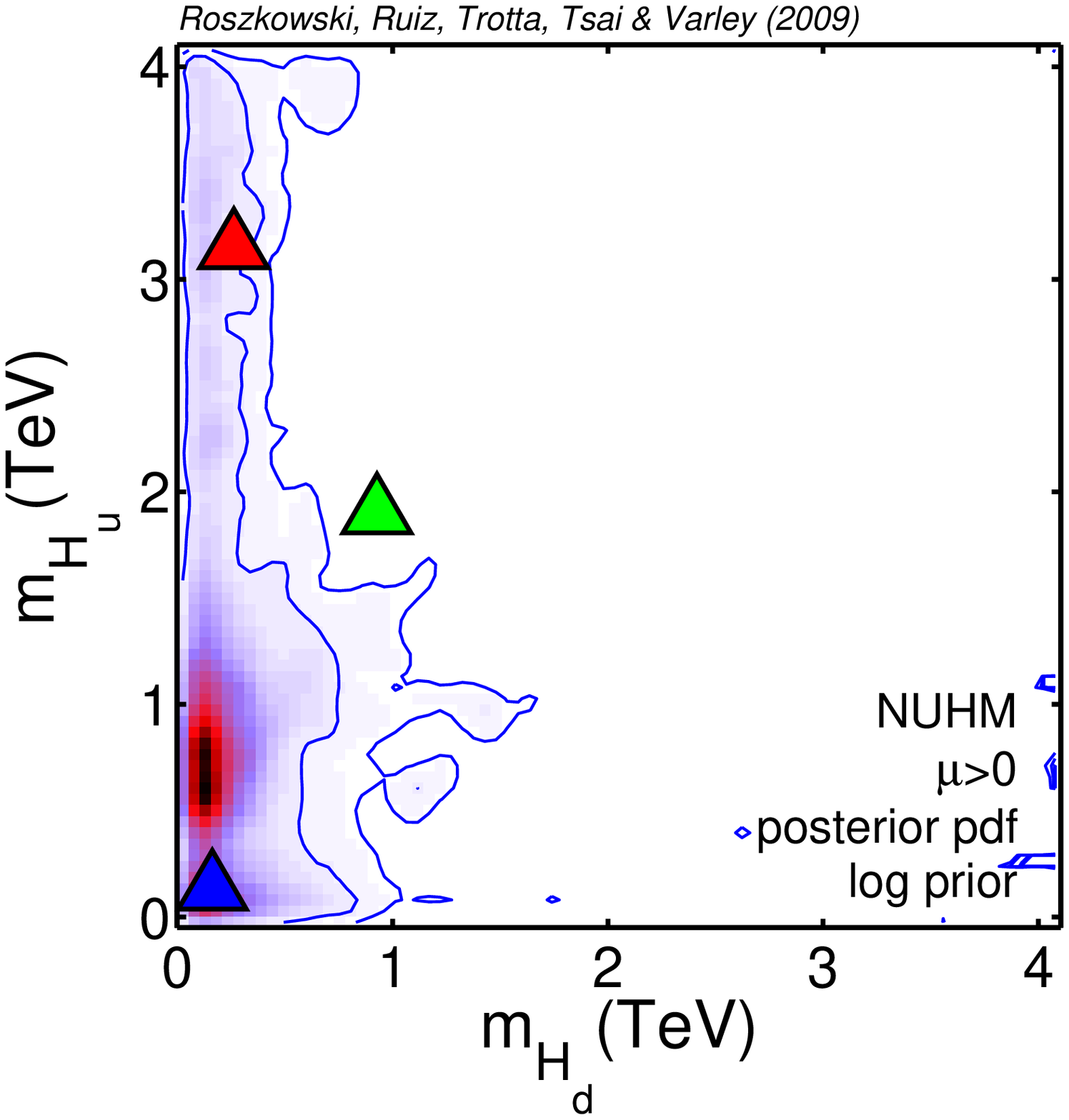}\\
 \multicolumn{3}{c}{\includegraphics[width= 0.3\textwidth]{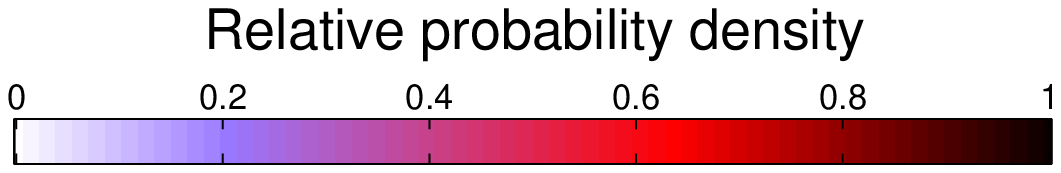}}\\ 
   \includegraphics[width=0.31\textwidth]{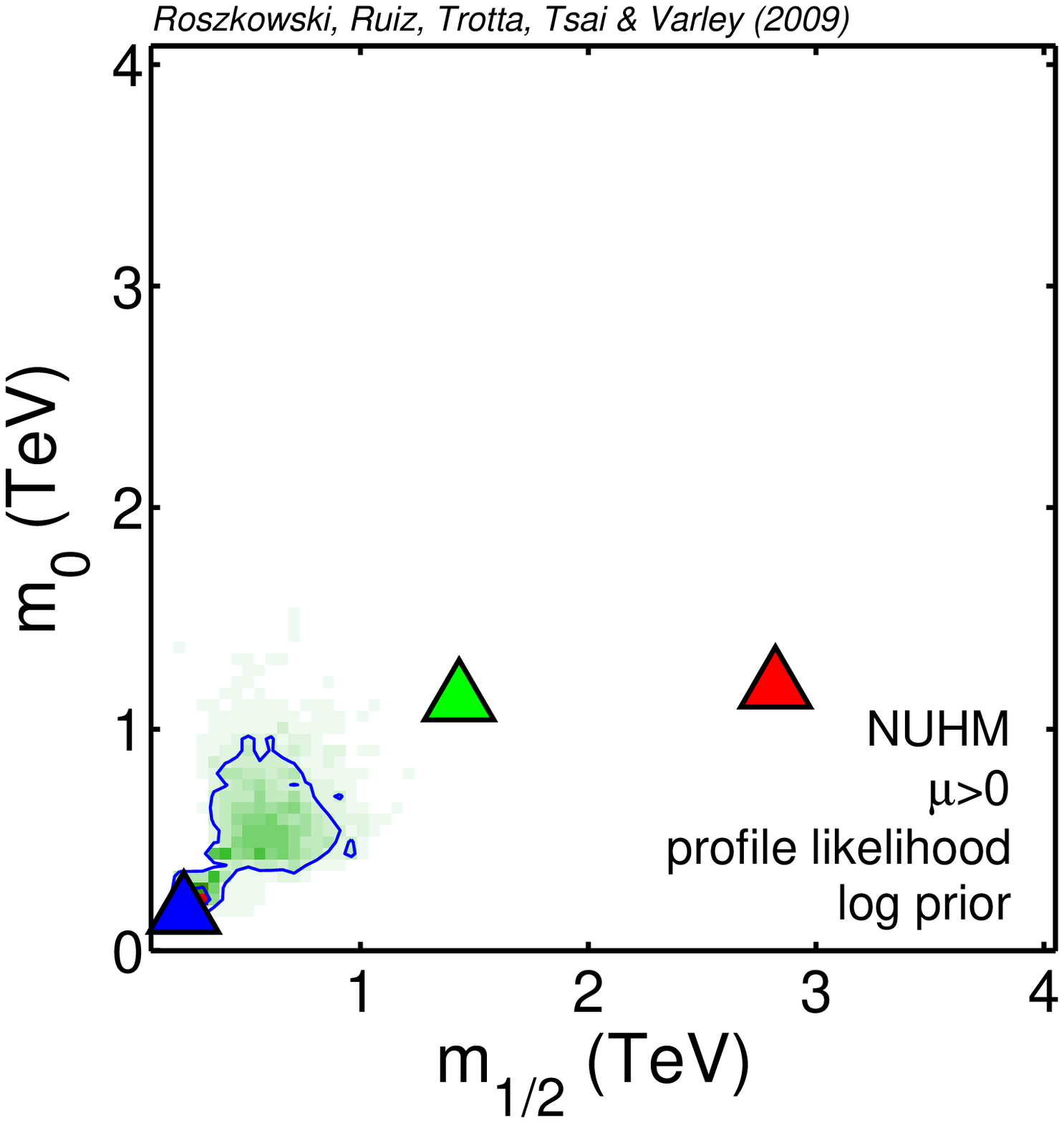}
 & \includegraphics[width=0.31\textwidth]{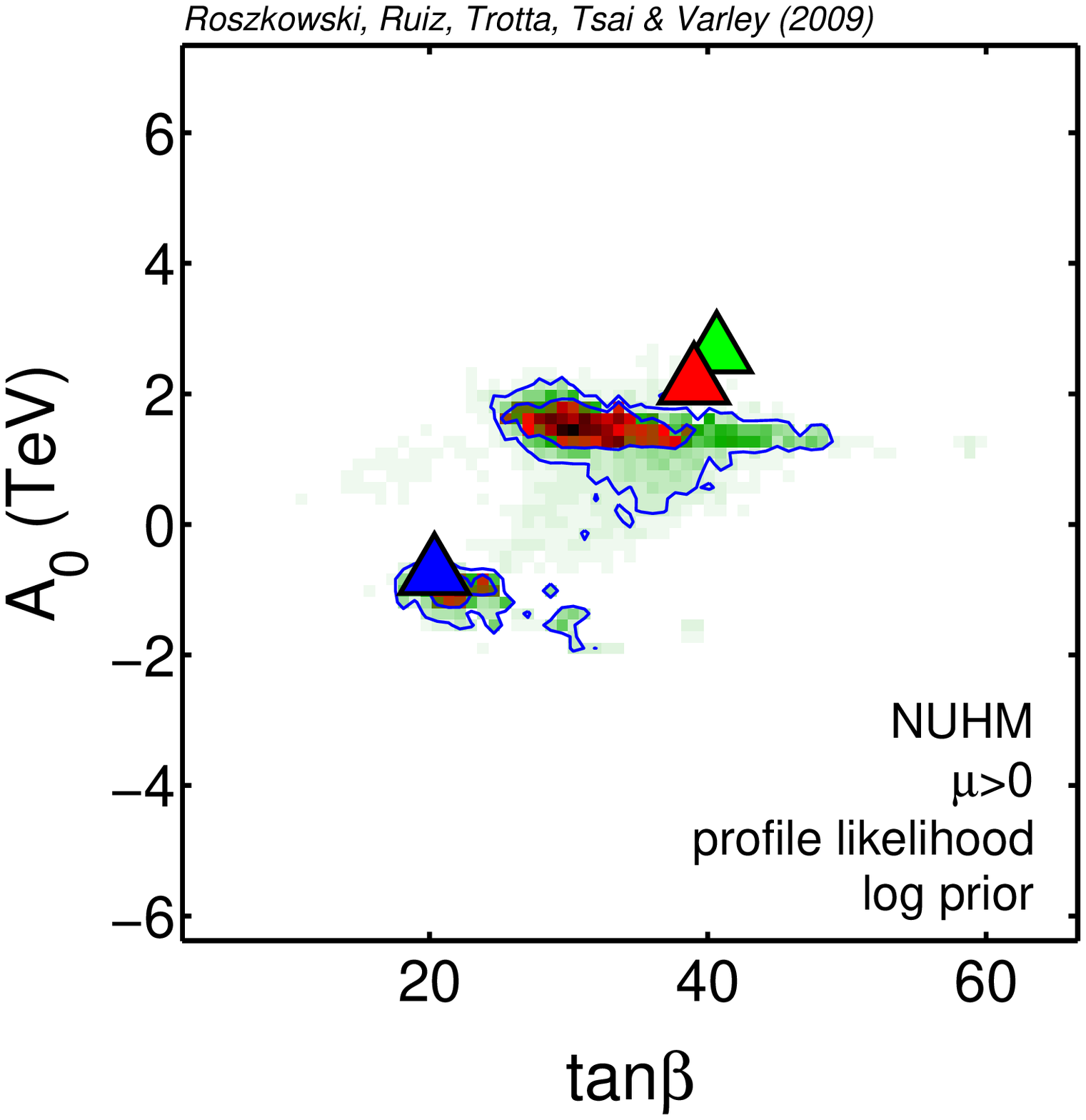}
 & \includegraphics[width=0.31\textwidth]{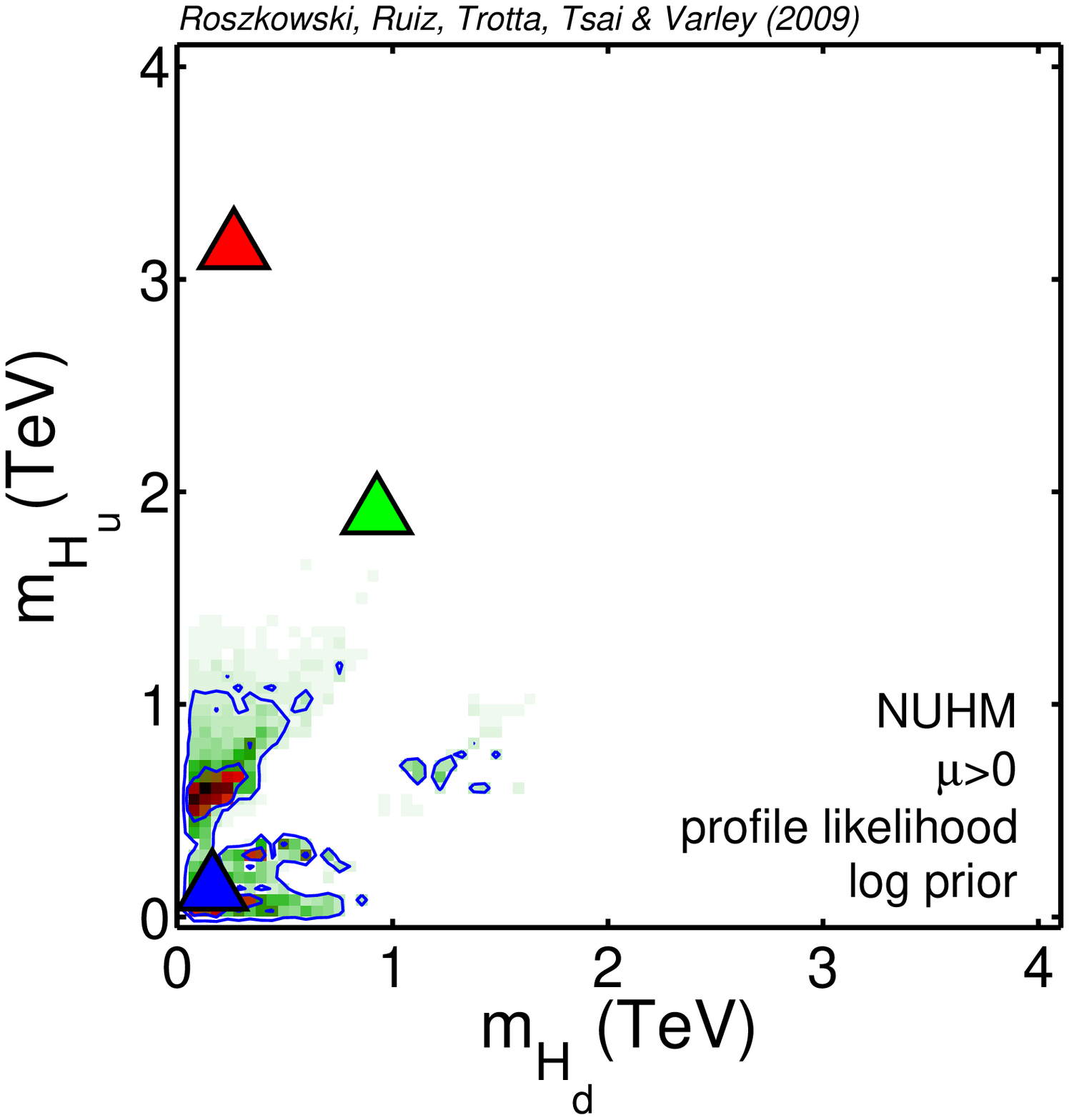}\\
  \multicolumn{3}{c}{\includegraphics[width= 0.3\textwidth]{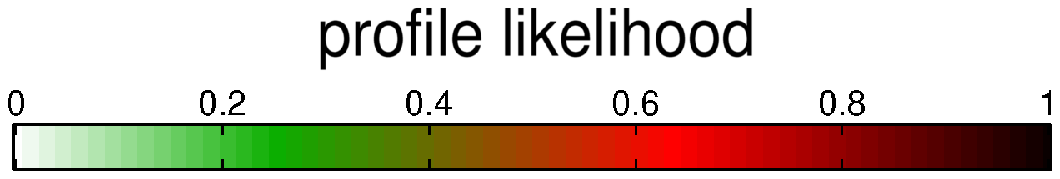}}\\ 
\end{tabular}

\end{center}
\caption{\label{fig:nuhmps2dpost_log_mup} Top panels: the 2D posterior
  for the log prior choice in the planes spanned by three pairs of
  NUHM parameters: $(\mhalf,\mzero)$, $(\tanb,\azero)$ and
  $(\mhd,\mhu)$ for $\mu>0$. The inner (outer) blue solid contours
  delimit regions encompassing 68\% and 95\% of the total posterior
  probability, respectively. Triangles mark the location of the
  best-fit points for each of the three different DM compositions:
  mostly gaugino (blue), mostly higgsino (red) and mixed (green); see
  text for details. The overall best-fit is the blue triangle, see
  Table~\protect\ref{table:best_fit} and discussion below. Bottom panel:
  profile likelihood maps for the same quantities, where all other
  parameters have been maximised over. Contours denote 68\% and 95\%
  confidence regions from the profile likelihood. }
\end{figure}

To start with, in the top row of fig.~\ref{fig:nuhmps2dpost_log_mup}
we plot joint 2D posteriors for some combinations of the NUHM base
parameters, marginalising over the parameters not shown. In the bottom
row, we plot instead the profile likelihood, where the other
parameters have been maximised over.  In terms of the posterior pdf,
we can see that the 68\% probability region (inner contour) for
$\mhalf$ and $\mzero$ is remarkably well confined to mostly a fairly
low mass region of $\mhalf\lsim1\tev$ and $\mzero\lsim1.4\tev$, where
also the overall best fit point, marked by a blue triangle is located
(see below for further discussion). However, the 95\% posterior region
(outer contour) is much wider and extends to much larger ranges of
both parameters. This signals that the constraining power of the data
is not sufficient to strongly confine the posterior.  Turning to the
middle panel, we can see a preference for moderately large
$\tanb\lsim40$, as well as for positive $\azero$, although zero or
negative values of $\azero$ are not excluded.  Finally, regarding the
new parameters beyond the CMSSM, we see that $\mhu$ is fairly poorly
constrained while $\mhd$ favors rather low values, $\sim 1\tev$ at
95\% probability.

It is worth comparing the posterior regions of the NUHM with those for
the analogous parameters in the CMSSM.  The two top left panels in
fig.~\ref{fig:nuhmps2dpost_log_mup} are directly comparable with the
corresponding bottom panels in Fig.~13 of~\cite{tfhrr1} for the
CMSSM. It is clear that, while in both the CMSSM
and the NUHM high probability regions are given by rather low $\mhalf$
and $\mzero$ (assuming the log prior, see Appendix~\ref{sec:priors}
for further comments on prior dependence), but actually they are
different and favored by different physical mechanisms. As is well
known, in the CMSSM this is mostly the neutralino-stau coannihilation
region of $\mzero\ll\mhalf\lsim1\tev$~\cite{tfhrr1} (plus a tiny
vertical region of $Z$ and $\hl$ pole annihilation), while in the NUHM
the analogous (mentioned above) ``low-mass'' ranges $\mhalf\lsim1\tev$
and $\mzero\lsim1.4\tev$ corresponds to the ``bulk region'' where
$\ha$ funnel annihilation play a dominant r\^{o}le. In both models,
the favored ranges (and also best fit points) of $\tanb$ are rather
moderate (although in the CMSSM very large values of around 55 are
also allowed at 68\%, in contrast to the NUHM). Finally, $\azero$
in both models shows a mild preference for positive values but
otherwise is not well constrained.  Also shown in
fig.~\ref{fig:nuhmps2dpost_log_mup} (as triangles) are the best-fit
points corresponding to the different regions divided by three
different neutralino DM compositions: mostly gaugino ($Z_g>0.7$),
mixed ($0.7<Z_g<0.3$) and mostly higgsino ($Z_g<0.3$), where
$Z_g=Z^2_{11}+Z^2_{12}$ and $Z^2_{11}$ and $Z^2_{12}$ are the
respective bino and wino fractions. We will come back to this point
below when we discuss dark matter aspects of the model.

\begin{figure}[tbh!]
\begin{center}
\begin{tabular}{c c}
  \includegraphics[width=0.45\textwidth]{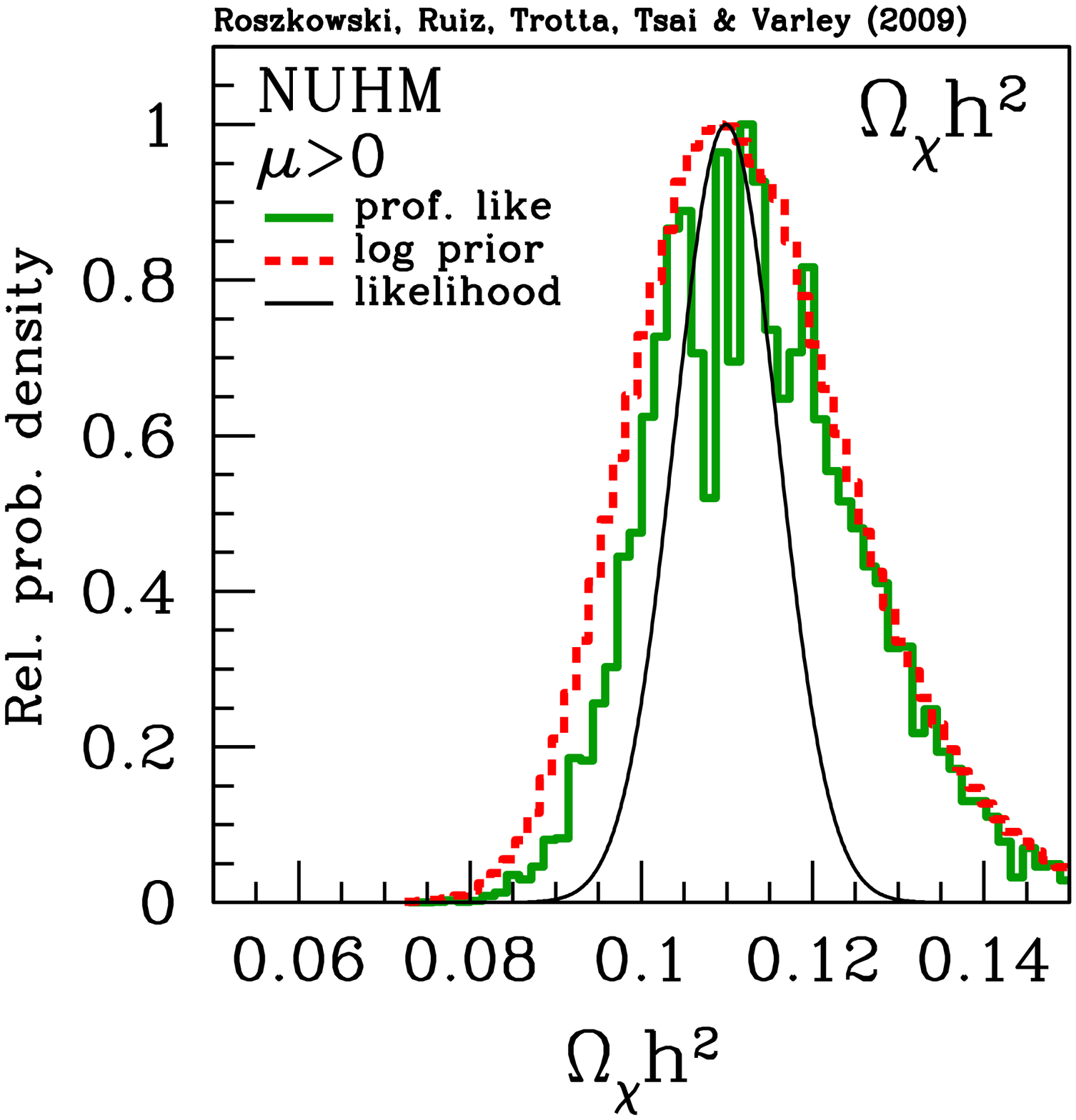}
  & \includegraphics[width=0.45\textwidth]{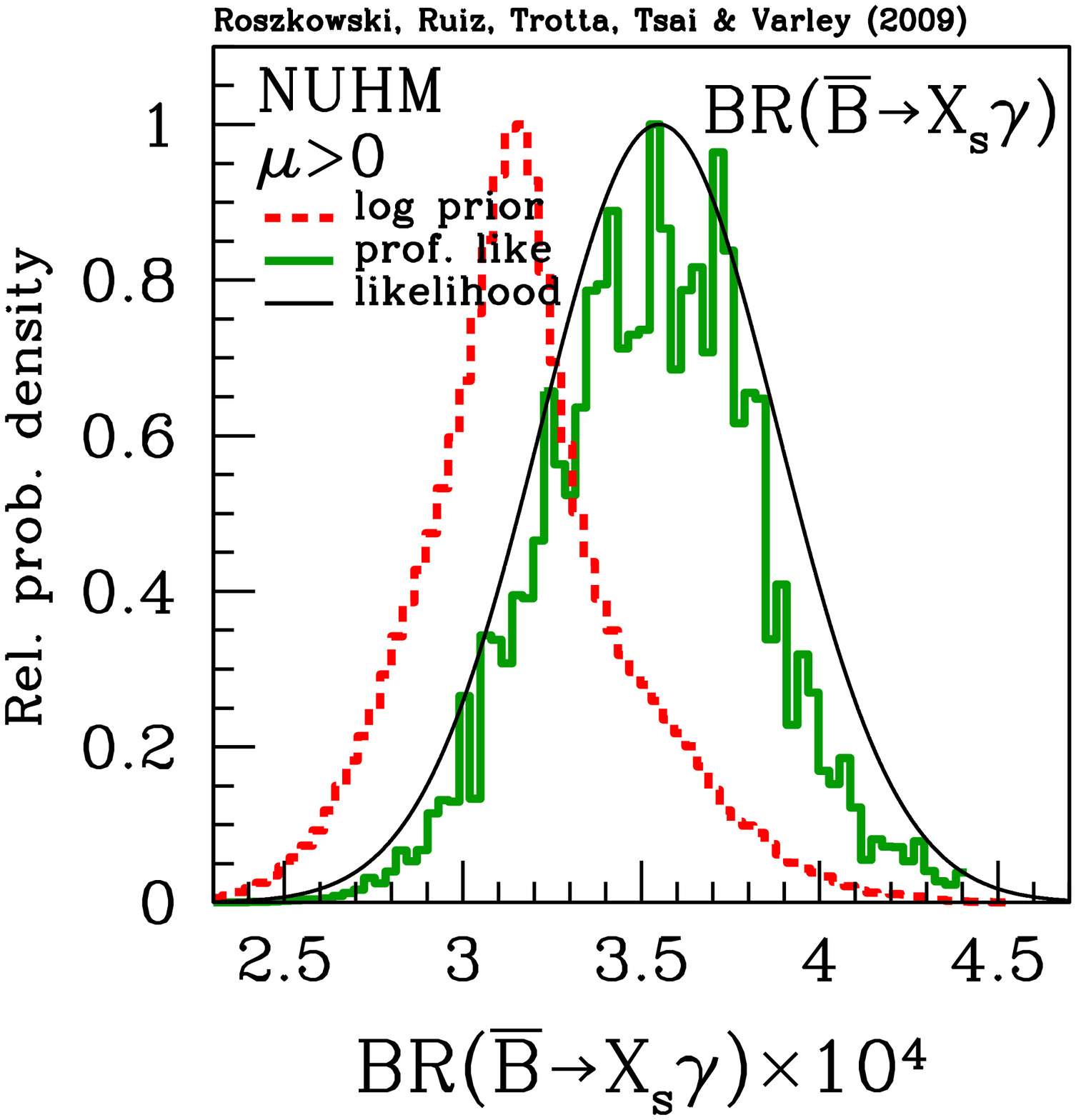}\\
  \includegraphics[width=0.45\textwidth]{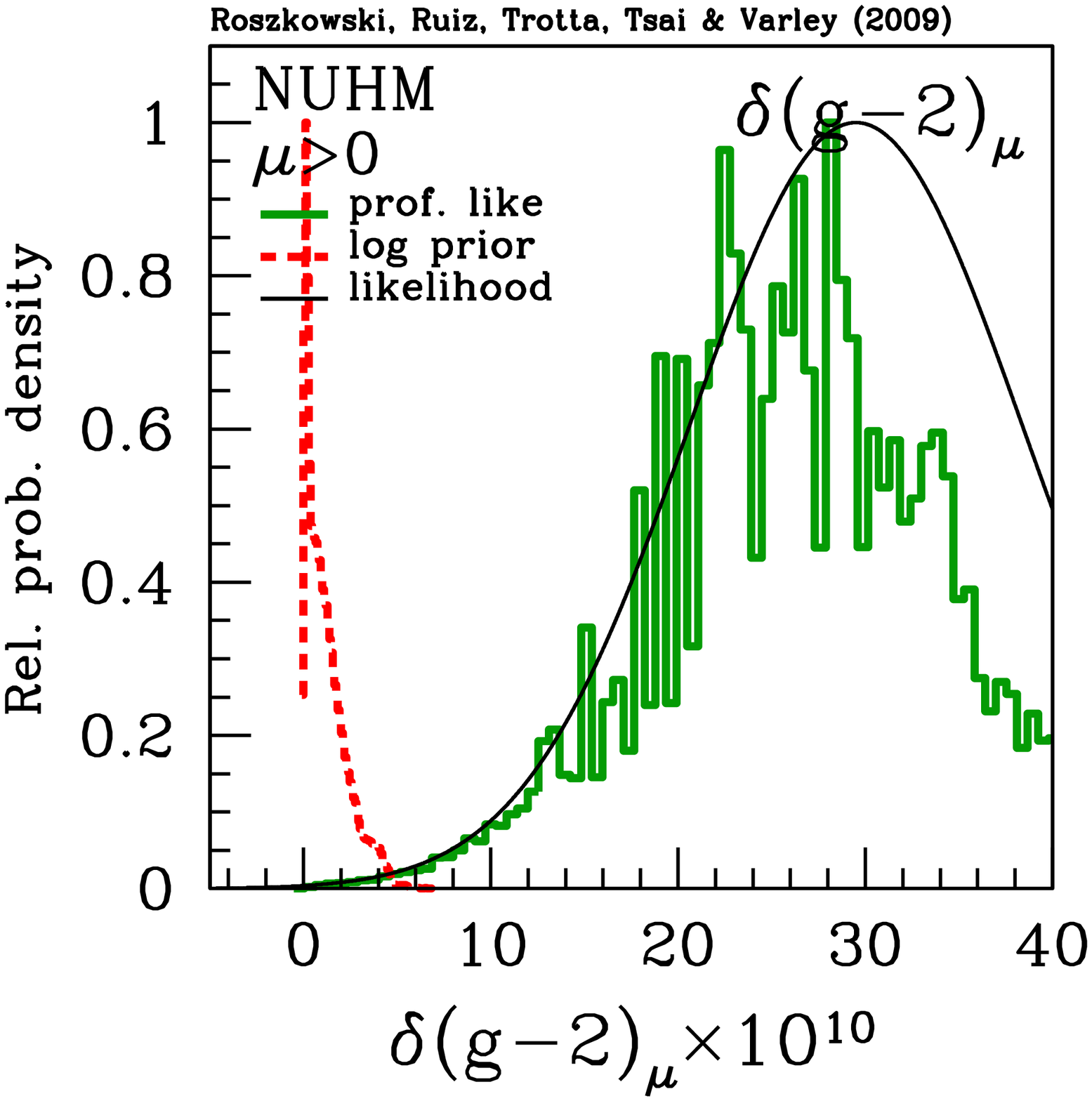}
  & \includegraphics[width=0.45\textwidth]{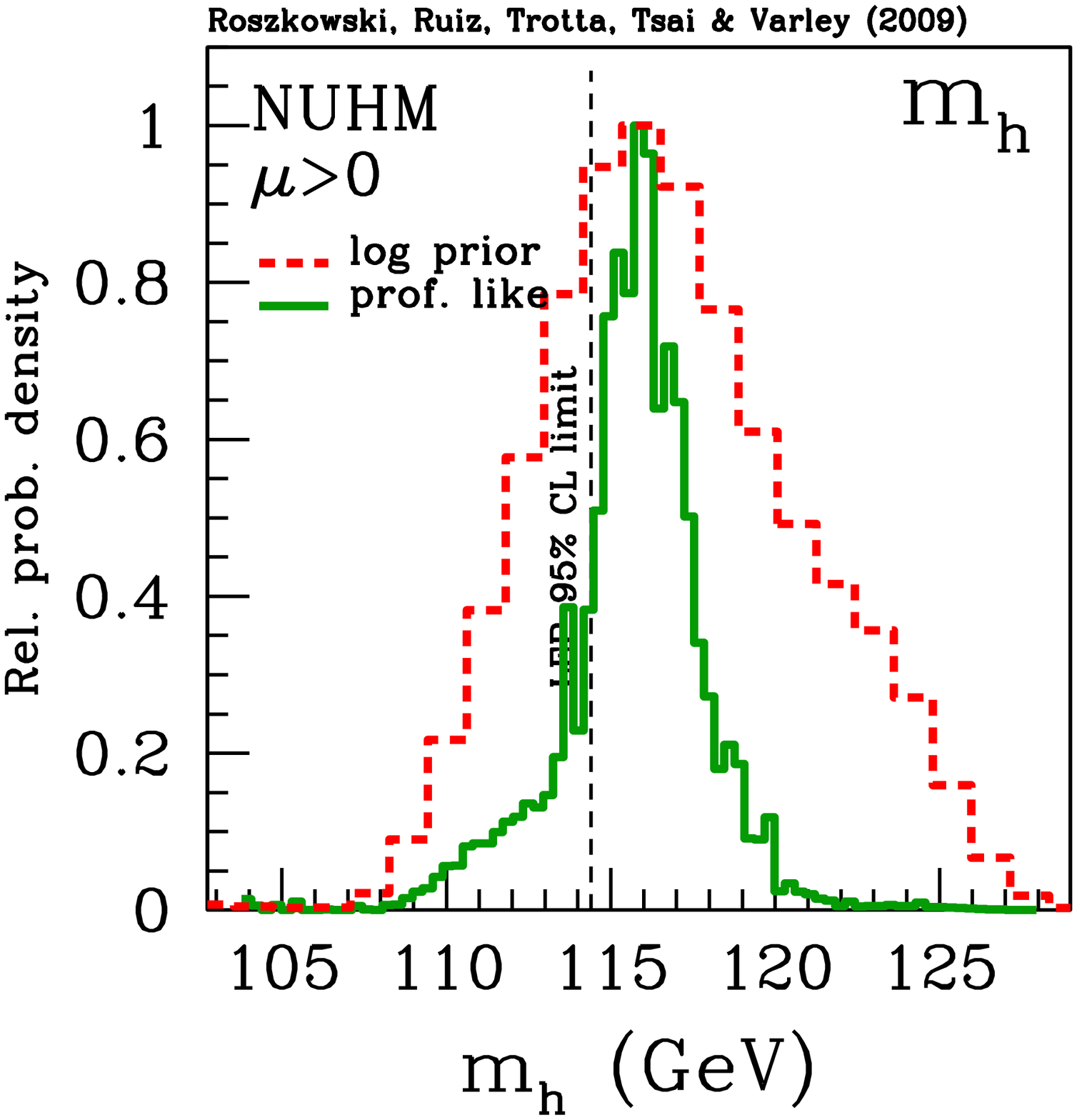}\\
\end{tabular}
\end{center}
\vspace{2\baselineskip}
\caption{\label{fig:nuhm_oh2+bsg+gm2+mhl_1d} The 1D posterior
  probability density (red, from our log prior scan) and the profile
  likelihood (green) for $\abundchi$ (upper left panel), $\brbsgamma$
  (upper right panel), $\deltagmtwo$ (lower left panel) and the light
  Higgs mass $\mhl$ (lower right panel).  In each panel we also plot
  the the likelihood function used to constrain the observable (solid
  black). }
\end{figure}

\begin{figure}[tbh!]
\begin{center}
\begin{tabular}{c c c}
   \includegraphics[width=0.31\textwidth]{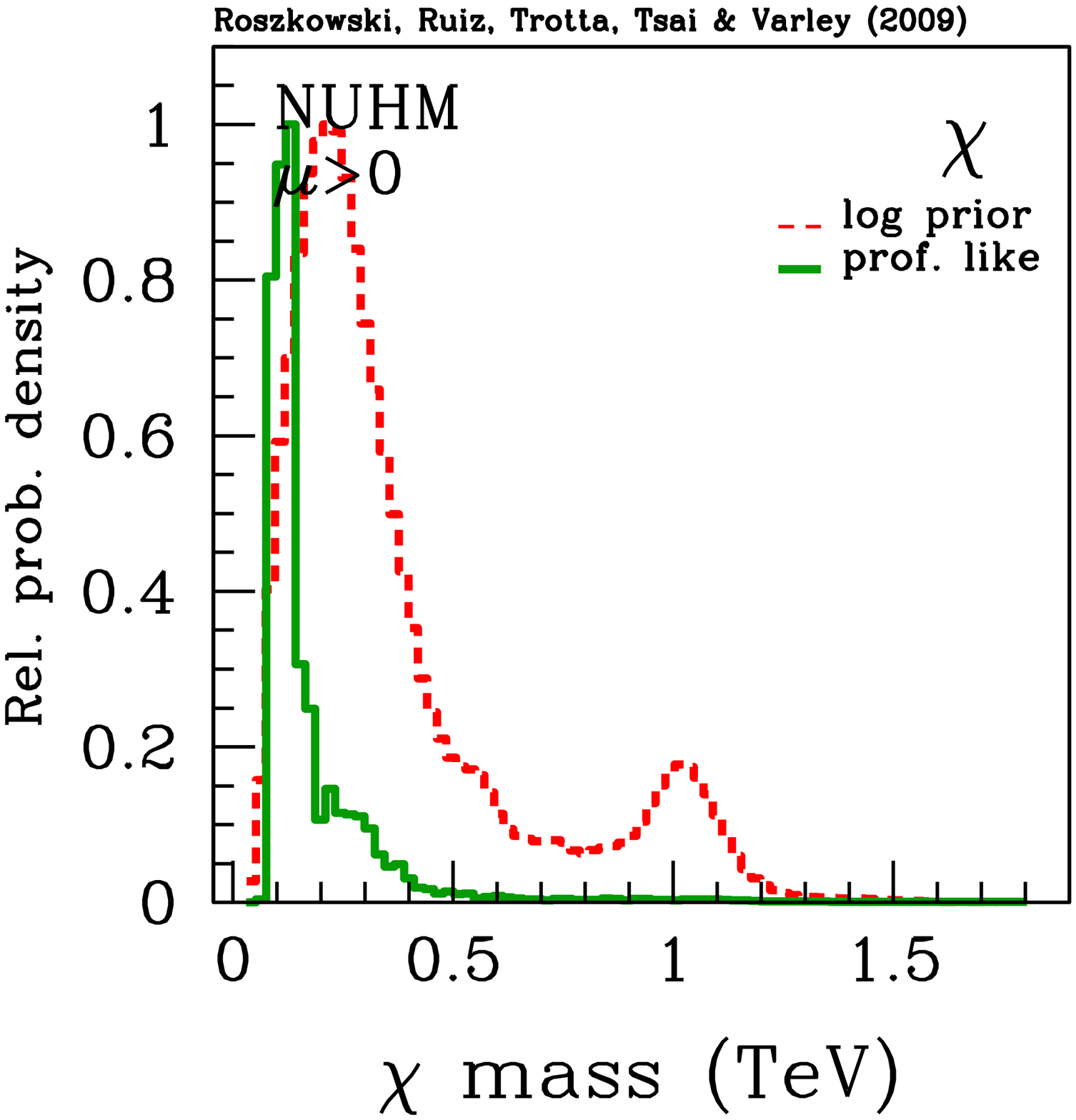}
 & \includegraphics[width=0.31\textwidth]{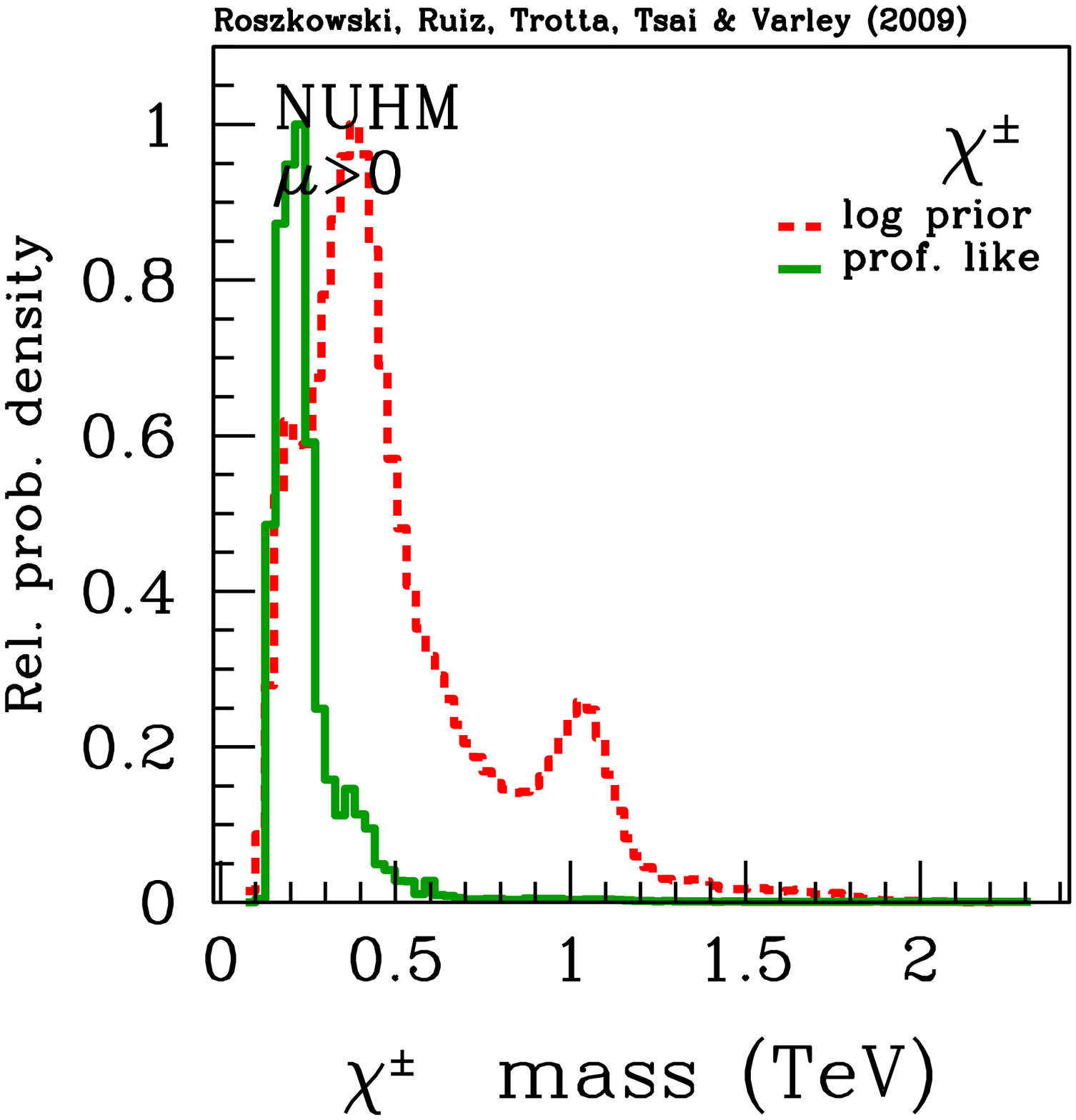}
 & \includegraphics[width=0.31\textwidth]{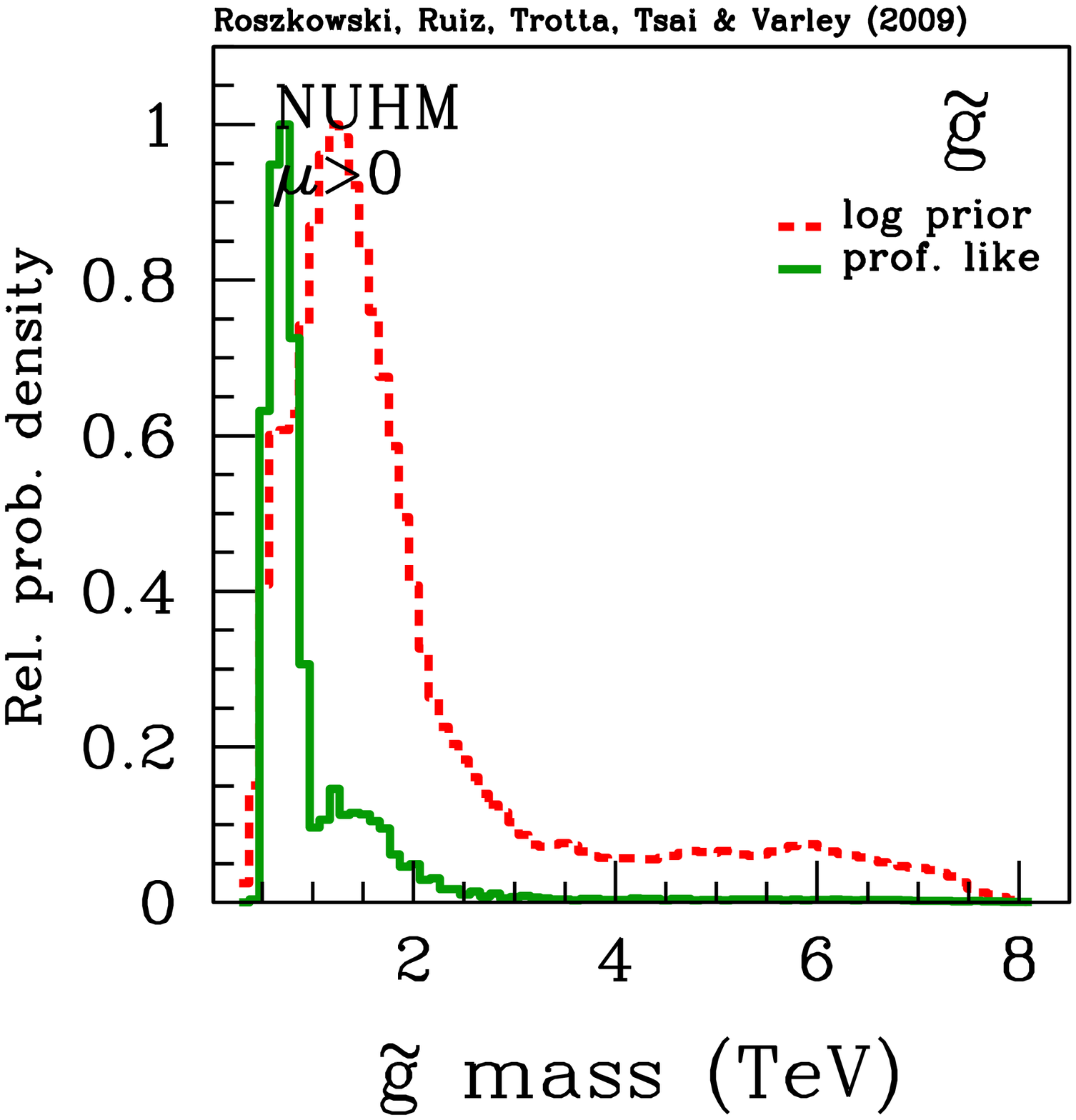}\\
    \includegraphics[width=0.31\textwidth]{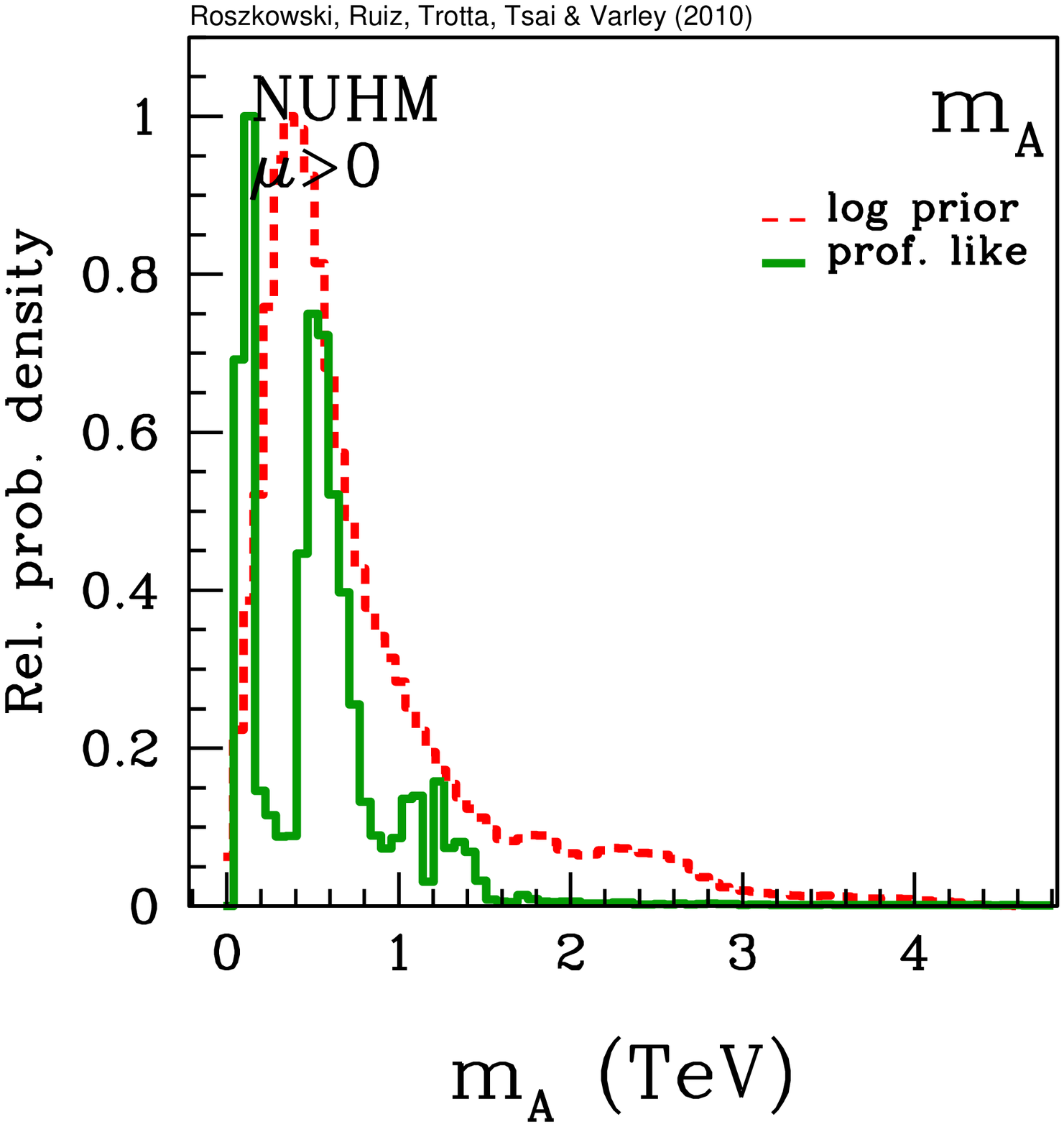}
 & \includegraphics[width=0.31\textwidth]{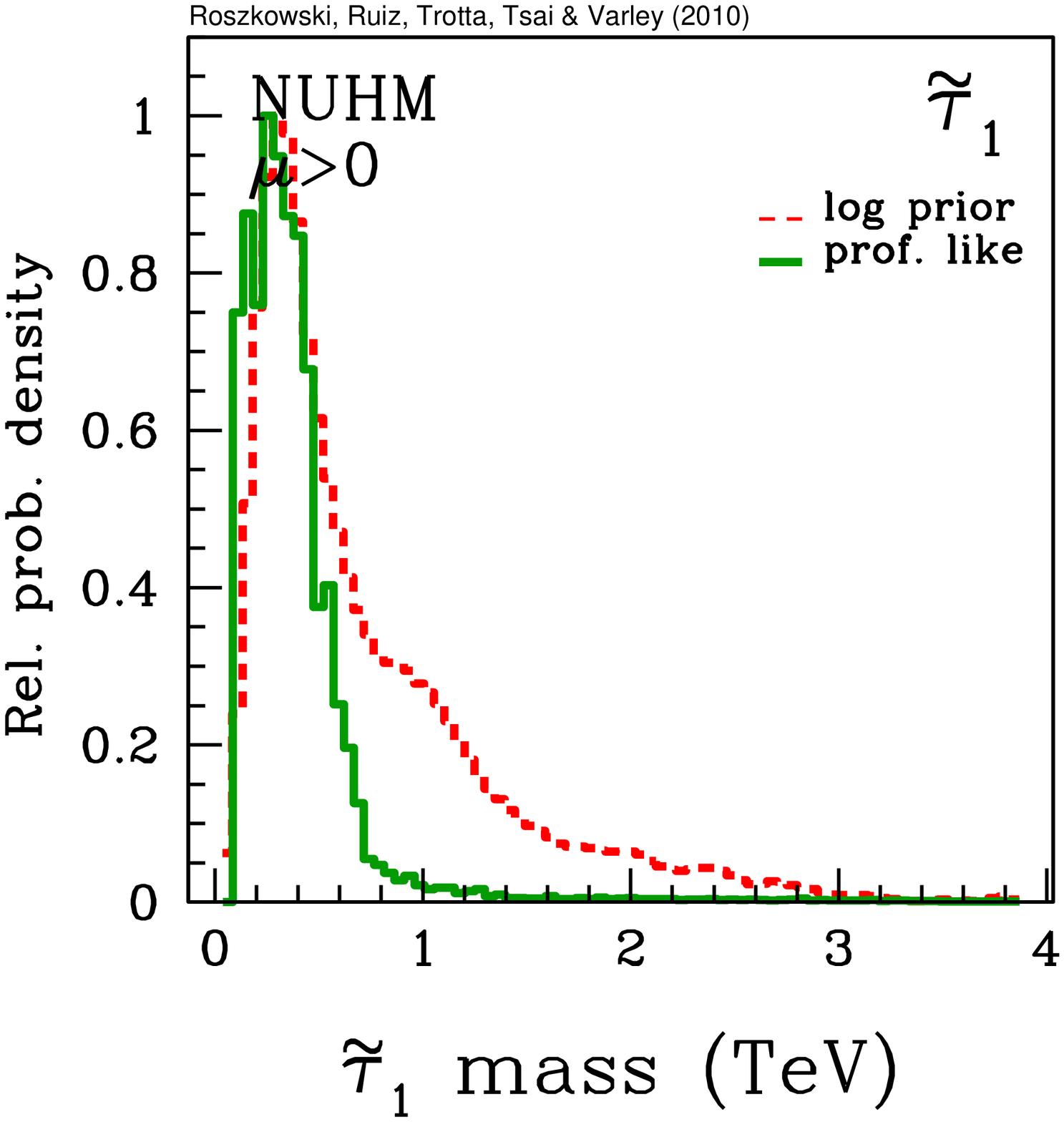}
 & \includegraphics[width=0.31\textwidth]{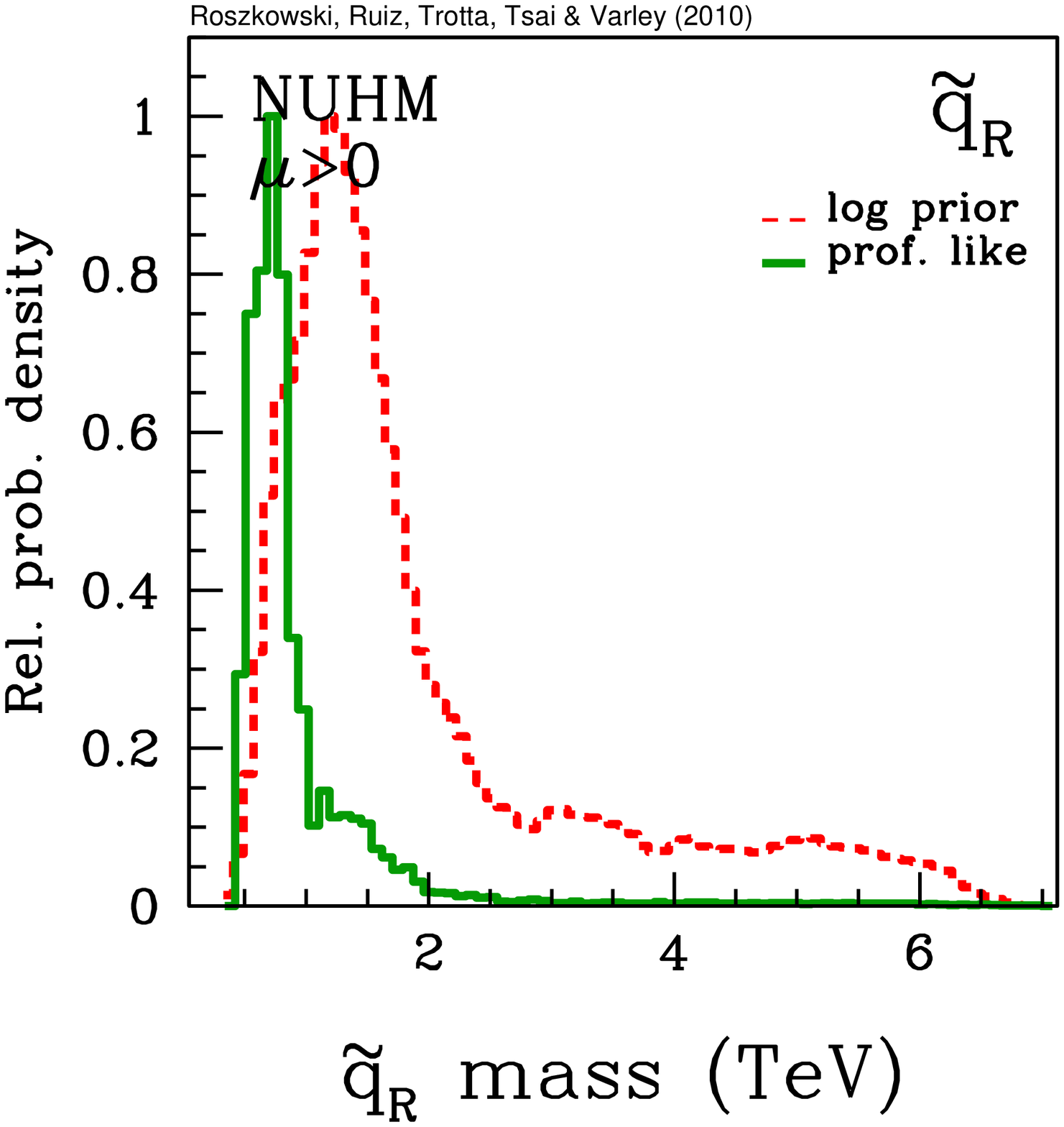}
\end{tabular}
\end{center}
\caption{\label{fig:nuhm_mchi_1d} Top row: The 1D posterior
  probability and the profile likelihood for the lightest neutralino
  mass $\mchi$ (left panel), the lightest chargino $\mcharone$ (middle
  panel) and the gluino $\mgluino$ (right panel). Bottom row: the same
  quantities for the mass of the lightest neutralino $m_A$ (left
  panel), the first stau $m_{\tilde{\tau}_1}$ (middle panel) and a
  squark $m_{\tilde{q}_r}$ (right panel).}
\end{figure}

Turning now to the profile likelihood maps (bottom panels of
Fig.~\ref{fig:nuhmps2dpost_log_mup}, we see that the confidence
contours from this statistical measure are much more strongly
confined than the posterior. The favoured region at 95\% from the
profile likelihood is always inside the 68\% posterior region obtained
from the posterior. This constraints all of the NUHM soft masses
to be below $\sim 1\tev$, with $\azero$ confined within $\pm 2\tev$
around zero and moderate values of $\tanb$ between some 20 and 40
being preferred. This striking difference between the posterior and
the profile likelihood can be understood with the help of
Fig.~\ref{fig:nuhm_oh2+bsg+gm2+mhl_1d} which investigates some of the
observables that play a key r\^{o}le in determining the favored
regions of NUHM parameter space using either statistics, namely the
neutralino relic abundance $\abundchi$, $\brbsgamma$, the SUSY
contribution to $\gmtwo$ and the light Higgs mass $\mhl$. The tight
WMAP constraint on $\abundchi$ in the likelihood is well matched by
the shape of the 1D pdf as well as by the profile likelihood, showing
that this particular observable can be well fitted in the model. On
the contrary, we observe a certain discrepancy between the 1D pdf of
$\brbsgamma$ and the likelihood function. The posterior shows a rather
strong preference for what is basically the SM value of the
observable. The same effect can be seen in the posterior for $\gmtwo$,
which is strongly peaked at very small values. Both those effects are
a consequence of the fact that, under the prior measure chosen, the
vast majority of points in the NUHM parameter space leads to a
prediction for $\brbsgamma$ and $\gmtwo$ which is very close to their
SM value. The likelihood function is not strong enough to completely
override this preference, and hence the posterior remains influenced
by the prior. This statistical effect has already been noticed in
the case of the CMSSM~\cite{aclw07,rrt3,tfhrr1}.  The
profile likelihood for $\brbsgamma$ and $\gmtwo$ instead follows
closely the values of the likelihood function. This shows that there
are indeed a small number of samples in our chains which achieve both
the correct $\brbsgamma$ and $\gmtwo$ values. As we shall show below,
it is in fact the $\gmtwo$ constraint that mostly drives the
fit. However, since the number of the samples which provide a good fit
to both observables is so small, their posterior probability is
suppressed, as in Bayesian statistics points that are finely tuned wrt
the prior measure are penalized. We therefore conclude that the tight
constraints on the NUHM parameters obtained from the profile
likelihood are largely driven by the need to fit both $\brbsgamma$ and
$\gmtwo$, two observables that appear to be in some
tension~\cite{fhrrt1}.

Finally, the posterior for $\mhl$ still allows fairly low values of
the Higgs mass. Interesting, unlike in the CMSSM, the lightest Higgs
is not necessarily SM-like and therefore the 95\% LEP lower bound on
SM Higgs mass should only be considered as indicative. In fact, our
analysis fully accounts for the possibility of non SM-like Higgs in
the likelihood. The same conclusion remains qualitatively valid even
in terms of the profile likelihood statistics.

Moving on to the sparticle spectrum of the best-fit point, we present
in the top row of fig.~\ref{fig:nuhm_mchi_1d} the 1D posterior and
profile likelihood for the lightest neutralino (left panel), the
lighter chargino (middle panel) and the gluino (right panel).  In each
case, the secondary bump observed in the posterior at $\mchi \sim 1
\tev$, $\mcharone \sim 1 \tev$ and $\mgluino \sim 6 \tev$ is a
reflection of the parameter space region leading to higgsino DM, as we
will discuss in detail below. In the bottom row, we show the posterior
and profile likelihood for the pseudoscalar Higgs and sleptons. The
non-universality of $\mhu$ and $\mhd$ in the NUHM can lead to a large
positive value for the $S$ parameter, defined in the RGEs as: 
%lr\beq 
$S = m_{H_{u}}^2 - m_{H_{d}}^2 + \text{Tr}\left[
  \mathbf{m_Q^2-m_L^2-2m^2_{\bar{u}}+m^2_{\bar{d}}+m^2_{\bar{e}}}
\right] $, where the parameters in boldface denote $3\times3$ soft
mass parameters.
%lr \eeq 
In general the $S$ parameter is a fixed point in the RGEs of the
CMSSM, but in the NUHM it can be nonzero and make large contributions
to the running of many of the scalars, leading to, for example several
light sleptons. However, we do not find this to be the case.
\begin{table} 
\centering
\begin{tabular}{| l | l |  l | l  |}
\hline
$Z_g$ composition& $Z_g>0.7$ & $0.7<Z_g<0.3$ & $Z_g<0.3$ \\
\hline
\multicolumn{4}{|c|}{Base parameters}\\
\hline
$\mhalf$ &  0.224\tev   &	 1.43\tev &	2.83\tev  \\
$\mzero$ &  0.174\tev     & 1.13\tev &	1.19\tev	\\
$\mhu$ &  0.129\tev & 1.90\tev &  	3.15\tev\\
$\mhd$ &  0.162\tev &0.927\tev	& 0.263\tev	\\
$\azero$ &  1.56\tev &  2.64\tev &  2.17\tev	\\
$\tanb$ &  20.4  &  40.6	& 39.0 \\
\hline
\multicolumn{4}{|c|}{Observables}\\
\hline
$\abundchi$ & 0.111    & 0.112  &   0.108 \\
$Z_g$      & 0.989   &  0.655  &   0.00256  \\
$\mchi$    & 88.6\gev  & 593\gev& 1.04\tev\\
$m_{\chi}^{\pm}$   & 488\gev  & 640\gev& 1.05\tev\\
$\sigsip$ (pb) & $2\times10^{-9}$ & $1.44\times10^{-7}$ &   $1.81\times10^{-8}$ \\
\hline
\multicolumn{4}{|c|}{Annihilation channels}\\
\hline
$\langle \sigma_{ann} v \rangle$ (Dom.) &  $\chi^{0}_{1} \chi^{0}_{1}
\to \tau \tau$   (57\%)    & $\chi^{0}_{1} \chi^{0}_{1} \to t \bar{t}$   (15\%) & $\chi^{0}_{1} \chi^{+}_{1} \to u\bar{d} (8\%)$ \\
$\langle \sigma_{ann} v \rangle$ (Sub.) &  $\chi^{0}_{1} \tilde{\tau}_1 \to A \tau$   (11\%)    & $\chi^{0}_{1} \chi^{0}_{1} \to  b \bar{b} $
(13\%) & $\chi^{0}_{1} \chi^{+}_{1} \to c\bar{s} (8\%)$ \\
\hline
\multicolumn{4}{|c|}{Pulls for observables}\\
\hline
$\chi^2_{\abundchi}$ &$<0.01$ & 0.04 &  0.03 \\
$\chi^2_\text{Higgs}$ & 0.84 & 0.08 &  $<0.01$ \\
$\chi^2_\text{sparticles}$ & $<0.01$ & 0.0 & 0.0 \\
$\chi^2_\text{nuisance}$ &$<0.01$  & 0.68 &  0.77 \\
$\chi^2_{\gmtwo}$ & $<0.01$ & 8.10 &  9.80 \\
$\chi^2_{\bsgamma}$ & $<0.01$ & 0.08 & 0.06 \\
$\chi^2_{m_W}$ & 0.56 & 1.42 &  1.22 \\
$\chi^2_{\sineff}$ & 0.36 & 0.06 &  0.05 \\
$\chi^2_{\delmbs}$ & $<0.01$ & 0.16 & 0.23\\
\hline\multicolumn{4}{|c|}{Quality of fit and parameter space fraction}\\
\hline
$\chi^2_\text{tot}$ & 1.76 & 10.62 &  12.16 \\
Parameter space & 80.5\%  & 7.4\% & 12.1\% \\
\hline
\end{tabular}
\caption{\label{table:best_fit} Best-fit points in each of the three
  different regions regarding the neutralino dominant composition:
  mostly gaugino ($Z_g>0.7$), mixed ($0.7<Z_g<0.3$) and mostly
  higgsino ($Z_g<0.3$) regions. The overall best-fit is in the
  gaugino region, where the neutralino is mostly bino (left column).  
  The bottom section shows the corresponding $\chi^2$ value and the
  parameter space fraction covered by each region.}  
\end{table}

The first column of Table~\ref{table:best_fit} gives the best fit
values for the NUHM base parameters and for a number of quantities of
particular interest, as well as the overall $\chi^2$ value and the
pull of each observable. The dominant role of the $\gmtwo$ constraint in
driving the fit towards the small mass region will be discussed in
more detail at the end of the next subsection where we examine the
higgsino-dominated DM and address the question of its statistical viability.  

\subsection{Higgsino dark matter in the NUHM}

\begin{figure}[tbh!]
\begin{center}
\begin{tabular}{c c}
\includegraphics[width=0.45\textwidth]{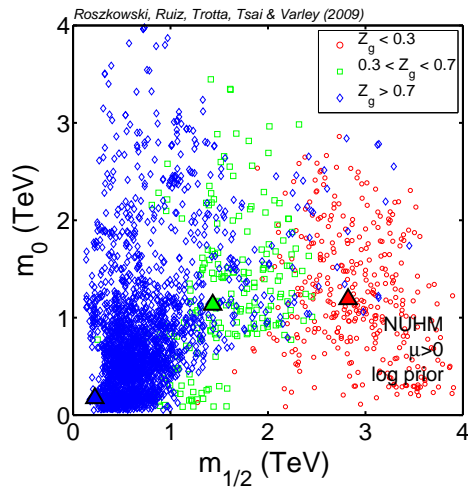}
\end{tabular}
\end{center}
\caption{\label{fig:nuhm3D_mup} A distribution of the gaugino fraction
  $Z_g$ in the plane of $(\mhalf,\mzero)$ for samples uniformly
  selected from our MC chains.  The color coding is as follows: red
  dots correspond to $Z_g<0.3$ (mostly higgsino), green squares to
  $0.3<Z_g<0.7$ and blue diamonds to $Z_g > 0.7$ (mostly gaugino).
  The triangles denote the best fit point for each cloud of samples of
  a given respective gaugino fraction (of corresponding color) taken
  separately. The overall best-fit is in the gaugino-like DM region.}
\end{figure}

An interesting feature of the NUHM is the possibility of higgsino-like
neutralino DM, as we have already mentioned above.\footnote{The
  possibility of higgsino-like LSP in the NUHM has also been noticed
  in~\cite{baer95} but not explored in more detail.}  That this
possibility exist in the NUHM should come as no surprise, since the
$\mu$ parameter can now be chosen as a free parameter, and thus
adjusted such as to give the correct amount of the relic density.  On
general ground, in order to satisfy this constraint in the MSSM-type
models, the LSP neutralino must either be mostly bino-like (like in
the CMSSM) if the bino soft mass $\mone<|\mu|$, a sufficiently heavy
higgsino-like state with $|\mu|<\mone$, or a mixed region in between
the two.

\begin{figure}[tbh!]
\begin{center}
\begin{tabular}{c c}
   \includegraphics[width=0.45\textwidth]{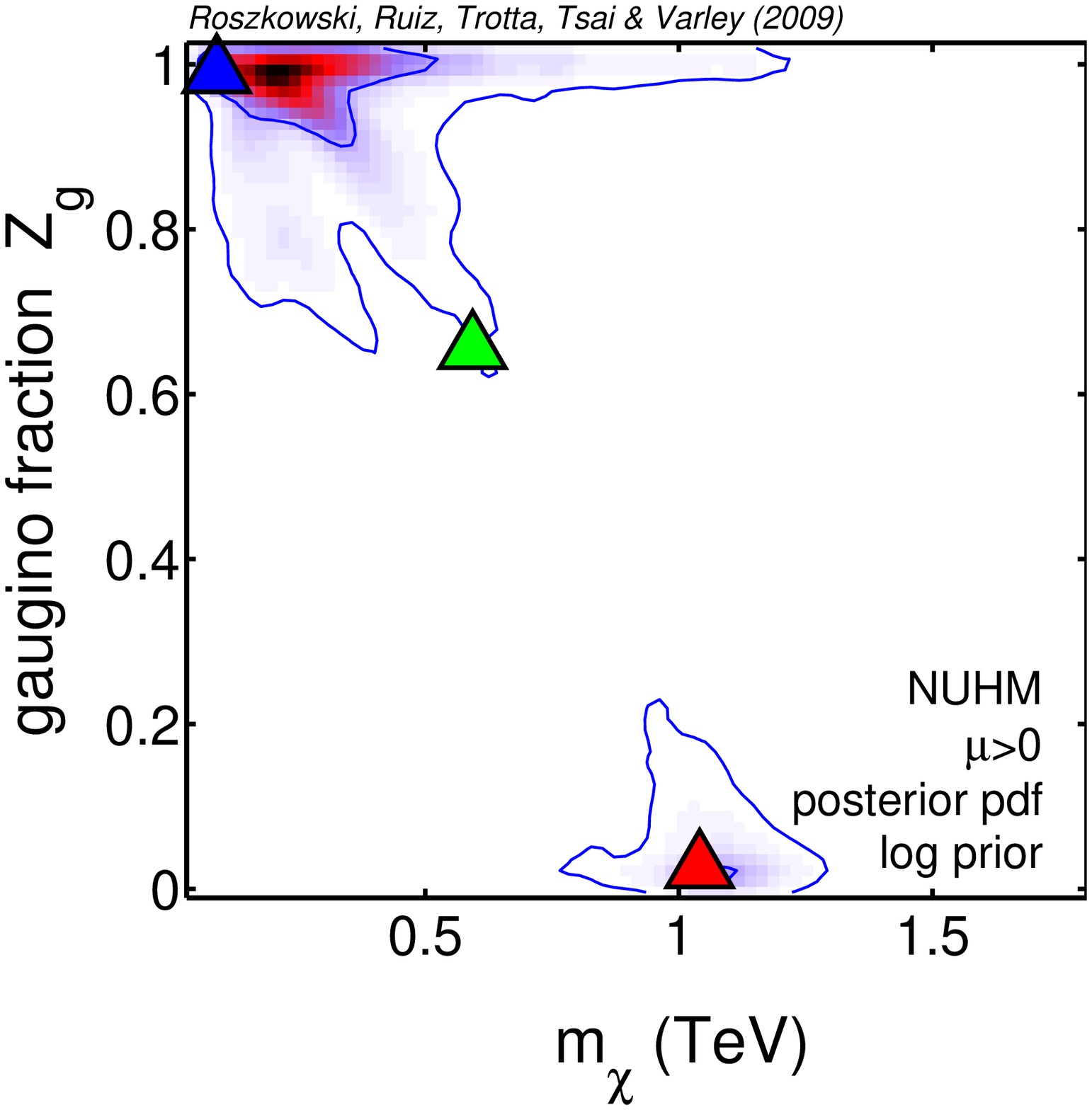} 
   \includegraphics[width=0.45\textwidth]{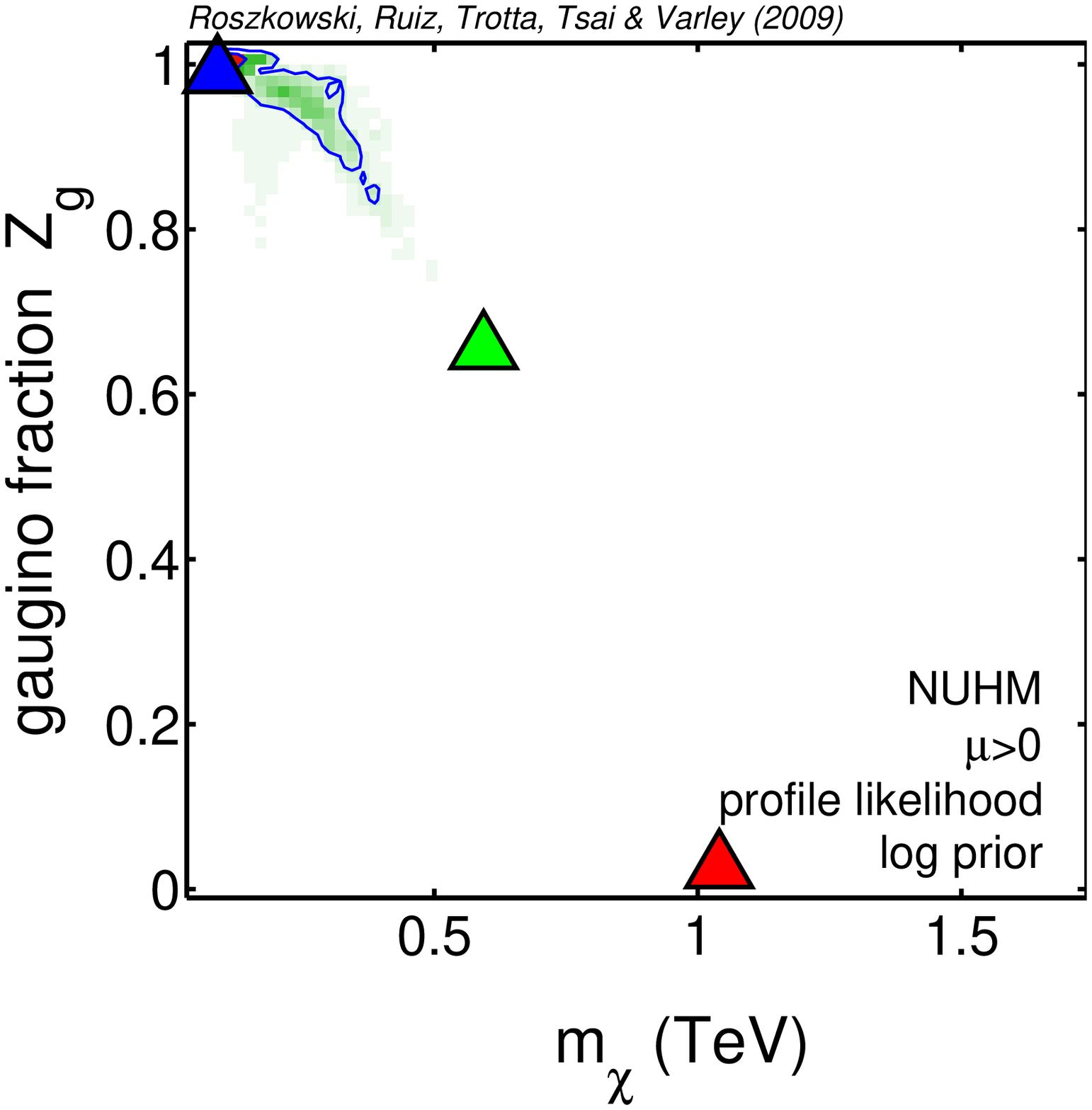} \\
 \hfill  \includegraphics[width= 0.3\textwidth]{figs/colorbar.ps}
 \hfill \includegraphics[width= 0.3\textwidth]{figs/colorbar-proflike.ps}
 \hfill 
  \end{tabular}
\end{center}
\caption{\label{fig:mchi_vs_gaug_mup} Left panel: Posterior
  probability distribution for the neutralino mass $\mchi$ and its
  gaugino fraction $Z_g$. Right panel: corresponding profile
  likelihood. As above, triangles mark the location of the best-fit
  points for each of the three different DM compositions: mostly
  gaugino (blue), mostly higgsino (red) and mixed (green). The overall
  best-fit is given by the blue triangle. }
\end{figure}

On the other hand, it is not at all clear to what extent satisfying
the relic abundance condition in a specific unified model like the
NUHM is allowed by the other constraints that are currently
available. This is an interesting issue, since the viability of the
higgsino region in the NUHM could potentially lead to a
phenomenological differences with the CMSSM, where the neutralino is
mostly a bino.

To start with, in Fig.~\ref{fig:nuhm3D_mup} we show in the plane
$(\mhalf,\mzero)$ a distribution of samples uniformly selected from
our MC chains, which are color-coded according to the gaugino fraction
$Z_g$ of the lightest neutralino.
Red circles correspond to a mostly higgsino
state, $Z_g<0.3$, green squares to a mixed state ($0.3<Z_g<0.7$) and
blue diamonds to mostly gaugino neutralino, $Z_g>0.7$. Notice that,
differently from usual ``random scans'' of the parameter space, in the
case of Fig.~\ref{fig:nuhm3D_mup} the density of samples reflects
their relative posterior probability (as a consequence of them having
been drawn using MCMC), hence we can make quantitative probabilistic
statements about the relative viability of the different regions given
our choice of prior.

The higgsino DM region corresponds to large values of $\mhalf$ (within
the $2\sigma$ posterior contour in the left panel of
Fig.~\ref{fig:nuhmps2dpost_log_mup}). As $\mhalf$ becomes smaller, the
bino-dominated fraction takes over, since in this region the
neutralino mass is approximated by $\mone$, which scales with
$\mhalf$. In between the two, we find a relatively smaller sample of
mixed-type neutralino cases. The triangles denote the best fit point for
each cloud of samples of a given respective gaugino fraction (of
corresponding color) taken separately.

Fig.~\ref{fig:mchi_vs_gaug_mup} shows the posterior pdf (left panel)
and the profile likelihood (right panel) for the gaugino fraction of
the neutralino \vs its mass. In the posterior distribution, the upper
left island of probability corresponds to bino-like LSP, while the
bottom right region around $1\tev$ corresponds to the higgsino
case. However, as mentioned above, the higgsino-like region is strongly
disfavoured by the profile likelihood, as can be seen in the
right panel. In this particular projection, the mixed region has too
little statistical weight to be visible.

While the presence in the NUHM of a sizable region of parameter space
where the LSP has a large higgsino fraction, as shown in
Fig.~\ref{fig:nuhm3D_mup}, can easily be understood at the electroweak
scale if one treats $\mu$ as a free parameter, it is interesting to
investigate the underlying mechanism at the unification scale. Below
we show that in the NUHM the appearance of the higgsino-like LSP is a
consequence of an interesting feature of the model which is the
existence of a mild focusing effect in the RG running of $\mhu$, akin,
but identical, to that in the CMSSM~\cite{focuspoint-orig}.

\begin{figure}[tbh!]
\begin{center} 
\begin{tabular}{c c}
 \includegraphics[width=0.45\textwidth]{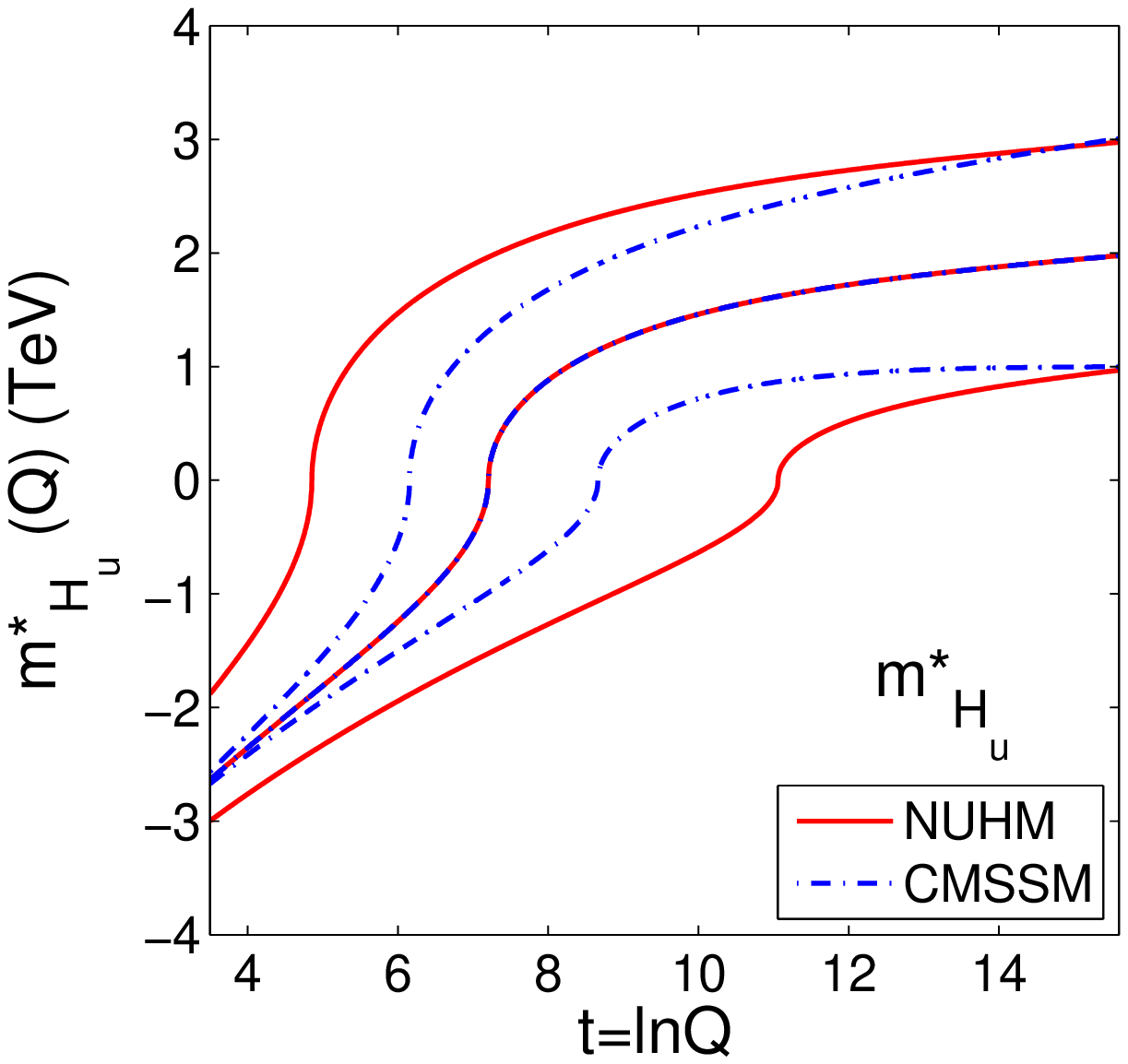}
\includegraphics[width=0.45\textwidth]{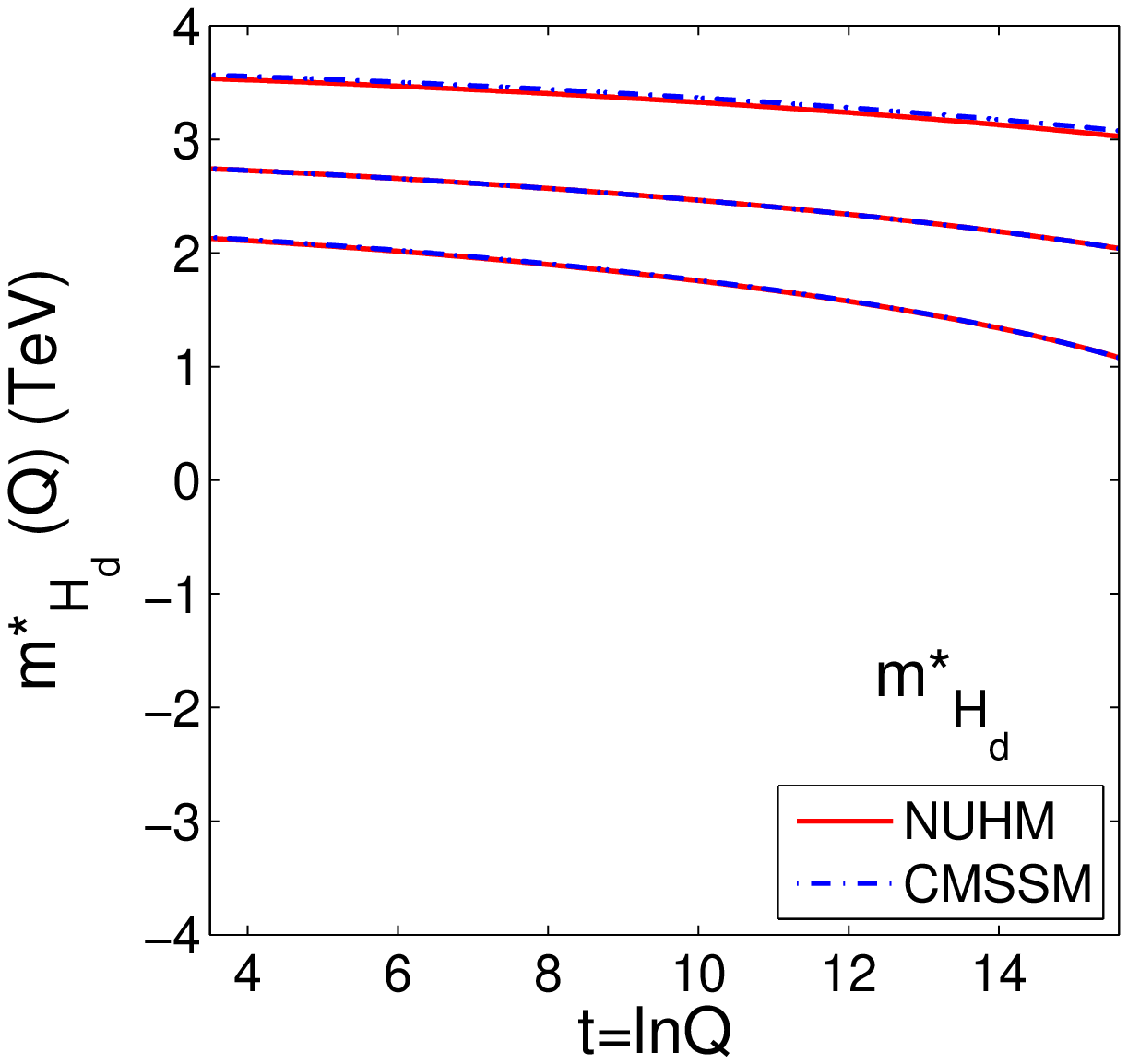}\\
\end{tabular}
\end{center}
\caption{\label{fig:running-fp} The running of $\mhuew(Q)$ (left
  panel) and $\mhdew(Q)$ (right panel) with the log of energy scale
  $t=\ln\, Q$. We take $\azero=4\tev$, $\mhalf=3\tev$, $\tanb=5$ and,
  in the NUHM case (red solid lines), $\mzero=2\tev$.  In order to
  facilitate comparison with the CMSSM (blue dash-dotted lines), for
  each curve we initially set $\mhd=\mhu$ and then evolve 
  differently for each model.  In the CMSSM, for each curve we take
  $\mzero=\mhd=\mhu$, which is why the middle curves overlap.  }
\end{figure}

For further discussion it
will be convenient to introduce the running parameters $\mhuew(Q)$ and
$\mhdew(Q)$, where $Q$ is the energy scale, defined as 
\beq
m^{\ast}_{H_{u,d}}(Q)=
\sgn\left(m^2_{H_{u,d}}(Q)\right)\, |m^2_{H_{u,d}}(Q)|^{1/2}.
\label{eq:mhudastdef}
\eeq
Since in the RGEs the running Higgs soft mass parameters $\mhu(Q)$ and
$\mhd(Q)$ appear only in squares, which can become negative, the
parameters $m^{\ast}_{H_{u,d}}(Q)$ are convenient to deal with in the
sense that they adequately reflect both the magnitude
and the sign of the respective parameters $m^2_{H_{u,d}}(Q)$. In
particular, the quantities $m^{\ast}_{H_{u,d}}$, without any
arguments, will denote the respective running quantities evaluated at
$Q=\msusy$,
\beq
m^{\ast}_{H_{u,d}}\equiv m^{\ast}_{H_{u,d}}(Q=\msusy).
\label{eq:mhudewdef}
\eeq

The focusing effect that we have identified is illustrated in
fig.~\ref{fig:running-fp} where red solid lines show the running of
the NUHM parameters $\mhuew(Q)$ (left panel) and $\mhdew(Q)$ (right
panel) with the log of energy scale $t=\ln\, Q$. For each case we take
$\azero=4\tev$, $\mhalf=3\tev$ and $\tanb=5$. In order to facilitate
comparison with the CMSSM (blue dash-dotted lines), for each curve we
initially set $\mhd=\mhu$ and then evolve differently in each
model. In the NUHM we fix also $\mzero=2\tev$ while in the CMSSM, for
each curve we take $\mzero=\mhd=\mhu$, which is why the curves in the
middle case overlap. One can see that, in the NUHM the running of
$\mhu$ is stronger for larger GUT values of the parameter, but it is
not as strong as in the CMSSM.

\begin{figure}[tbh!]
\begin{center}
\begin{tabular}{c c}
 \includegraphics[width=0.45\textwidth]{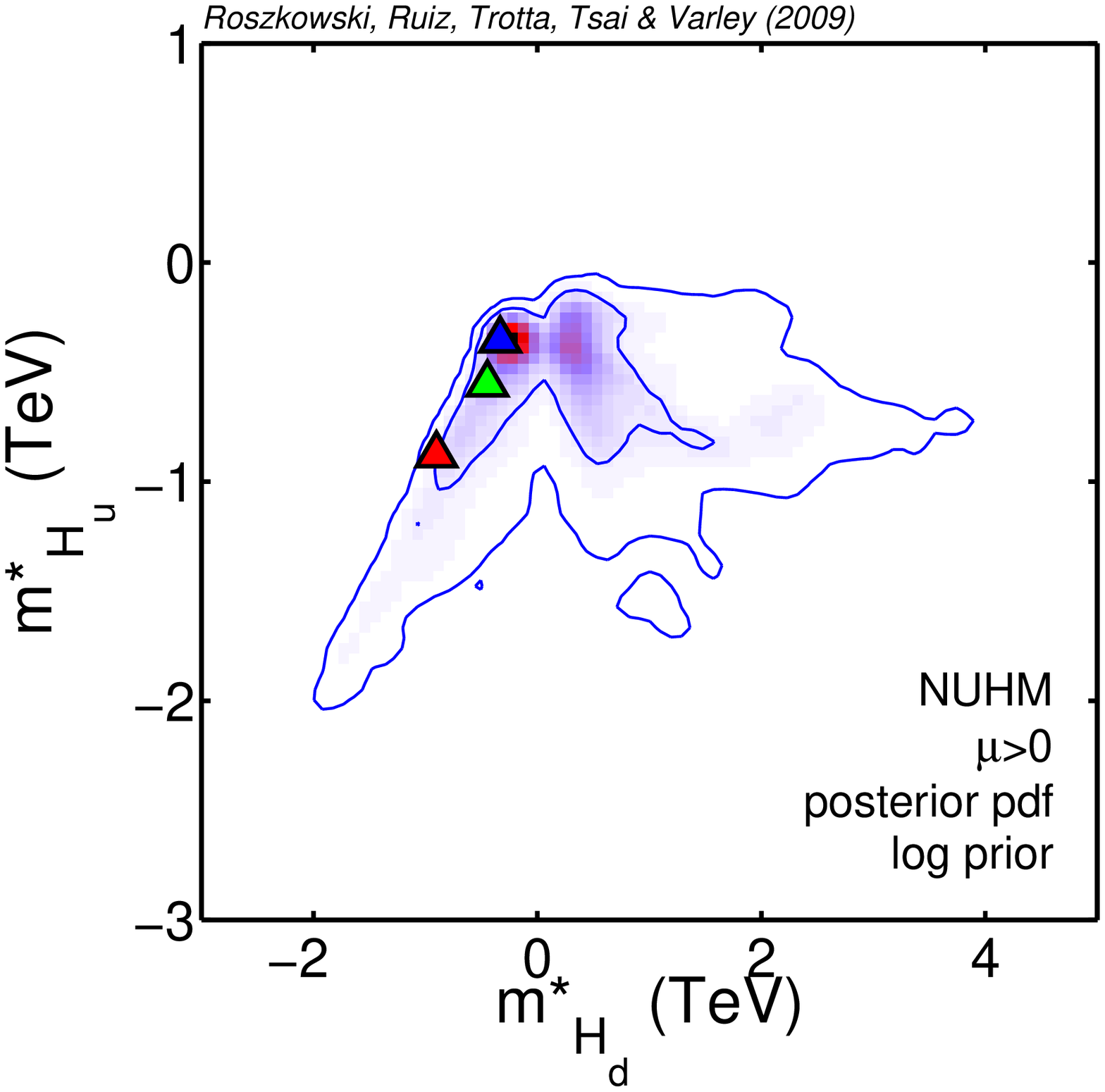}
 \includegraphics[width=0.45\textwidth]{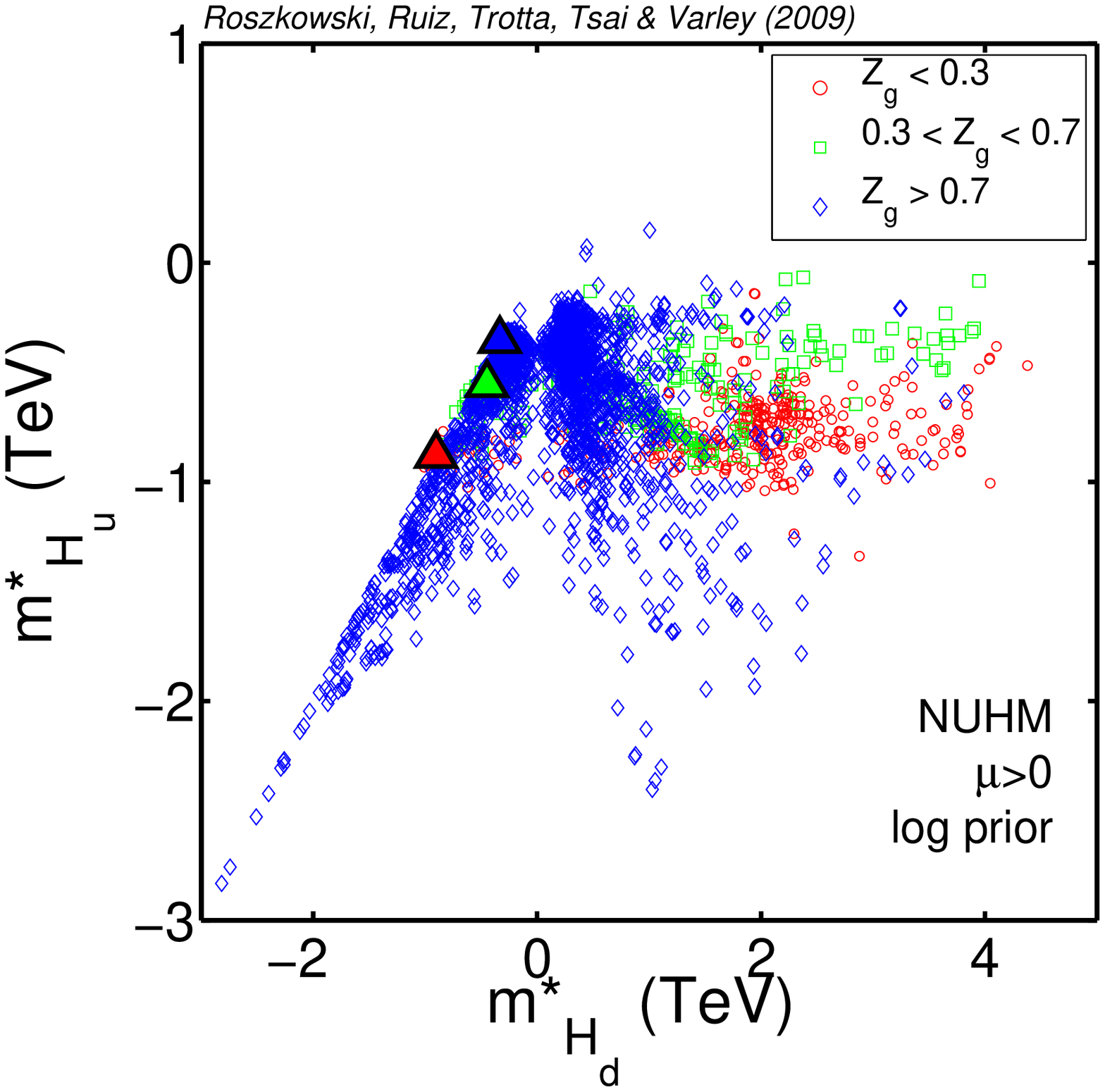}
\end{tabular}
\end{center}
\caption{\label{fig:nuhm_mhuew_vs_mhdew_mup} Left panel: posterior
  probability for $\mhuew$ and $\mhdew$, with the best-fit points for
  each of the three DM composition types marked by triangles as
  above. Right panel: A distribution of the
  gaugino fraction $Z_g$ in the same plane of samples uniformly
  selected from our MC chains.  The color coding is as before.  The
  triangles denote the best fit point for each cloud of samples of a
  given respective gaugino fraction (of corresponding color) taken
  separately. The overall best-fit is in the gaugino-like DM region.
}
\end{figure}

As a result, we can see some ``squeezing'' of $\mhuew$ compared to the
GUT values $\mhu$, while this is not the case with the $\hd$ soft mass
parameter. This is shown in the left panel of Fig.~\ref{fig:nuhm_mhuew_vs_mhdew_mup}
where islands of 68\% posterior probability region of $\mhuew$ between some
$-1\tev$ and $-0.5\tev$ (close to the location of the best fit points)
correspond to large values of $\mhu$ (compare
fig.~\ref{fig:nuhmps2dpost_log_mup}), close to the assumed upper limit
of the prior, nearly independently of the $\hd$ soft mass parameter.
It is clear  that $\mhuew$ is to a large extent constrained
by the focusing effect in the RG running. On the other hand, regions
where $\mhu$ tends to be small correspond to a bino-like neutralino
with the correct relic density, as we will show below. In this case,
however, $\mhd$ is confined to preferably fairly small values
($\lsim2\tev$), as otherwise one ends up with tachyonic sleptons.  The
corollary to this is that, in general we have more freedom in
obtaining a phenomenologically desired range of values at the
electroweak scale by appropriately choosing $\mhu$ at the GUT
scale. Clearly that is not so in the CMSSM as one can never attain
smaller $\mu$ here due to the stronger FP behaviour essentially
focusing to nearly one point at the EW scale.  The two distinct
branches visible in the left panel of fig.~\ref{fig:nuhm_mhuew_vs_mhdew_mup} correspond
to the two distinct neutralino regimes, as shown in the right panel of
of the Figure. In the horizontal branch the
neutralino is mostly a higgsino (red dots), turning into mixed
(green squares) while the other, an inverted V-shaped region around
$\mhdew = 0$ gives us a mostly bino (blue diamonds).

Above the reach of current collider limits, the high probability
regions of $\mhu$, $\mhd$ and the other soft parameters are primarily
determined by requiring the correct dark matter
abundance.\footnote{Exploratory runs with the constraint switched off
yield much wider ranges of parameters.}  
The
higgsino case is obtained in the NUHM, because, as
explained above, by starting with large enough $\mhu$ at the GUT scale
one arrives, via the mild focusing effect (see 
fig.~\ref{fig:running-fp}) at less negative values of $\mhuew$ at the
EW scale. In the limit $|\mhdew| \ll |\mhuew|$ and large enough
$\tanb$, \eq{ewsb1:eq} would imply $\mu\simeq |\mhuew|$. In reality,
in the NUHM this limit is often violated and as a result $\mu$ comes
out somewhat larger.  On the
other hand, $\mchi\simeq|\mu|$ has to be large enough to give an
acceptable relic density, as mentioned before. Numerically, this leads
to $\mu\gsim 0.8\tev$, which translates (via $0.4\,\mhalf\simeq
\mone>|\mu| $) to $\mhalf\gsim2\tev$. This explains the position
of the broad higgsino region at large $\mhalf$ in
Figs.~\ref{fig:nuhmps2dpost_log_mup} and~\ref{fig:nuhm3D_mup}.
The second EWSB
condition,~\eq{ewsb2:eq}, then implies the approximate relation $\mha^2
\simeq \mhdewsq + \mu^2$ which we have checked numerically to hold.

\begin{figure}[tbh!]
\begin{center}
\begin{tabular}{c c}
\includegraphics[width=0.31\textwidth]{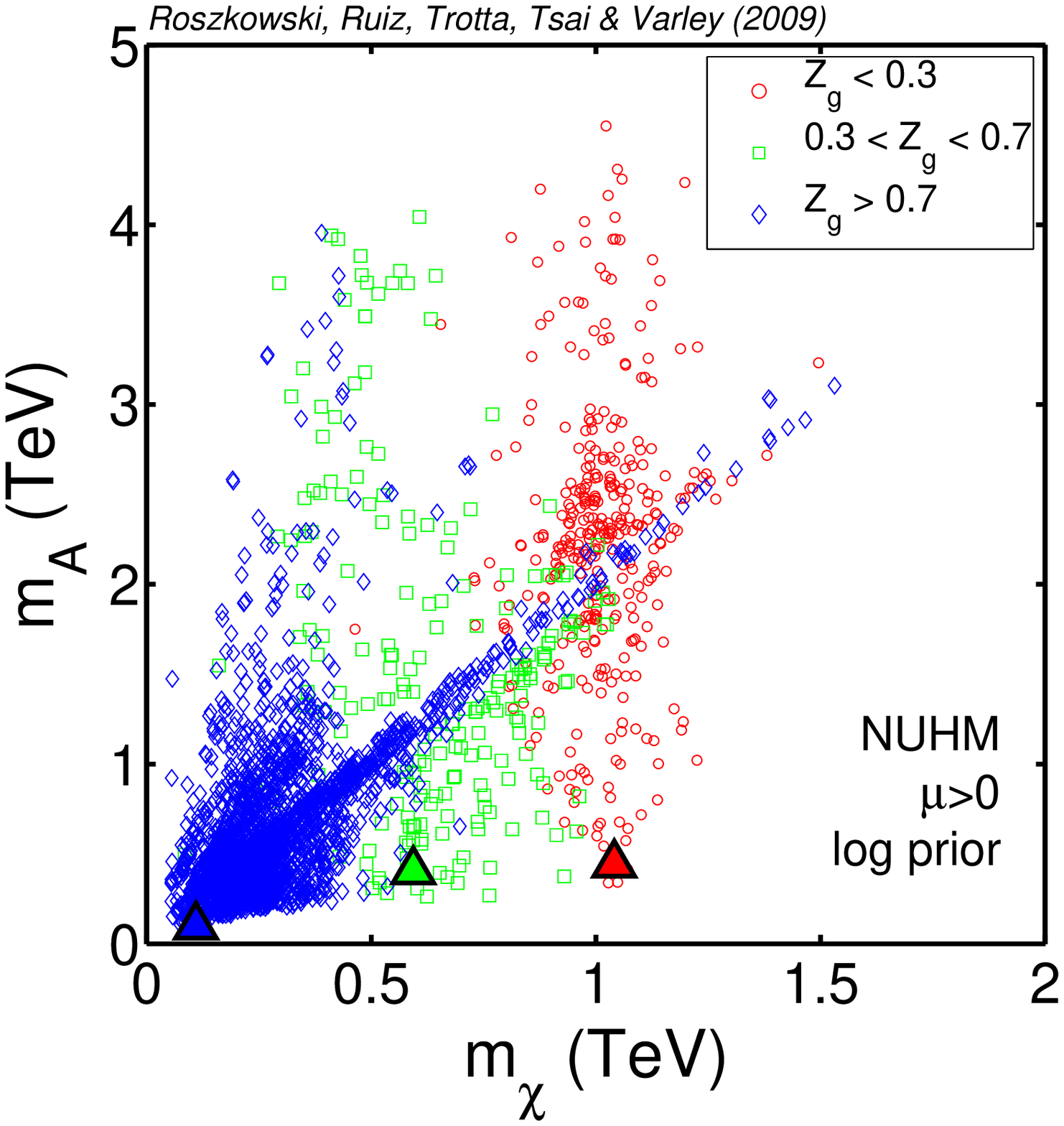}
\includegraphics[width=0.31\textwidth]{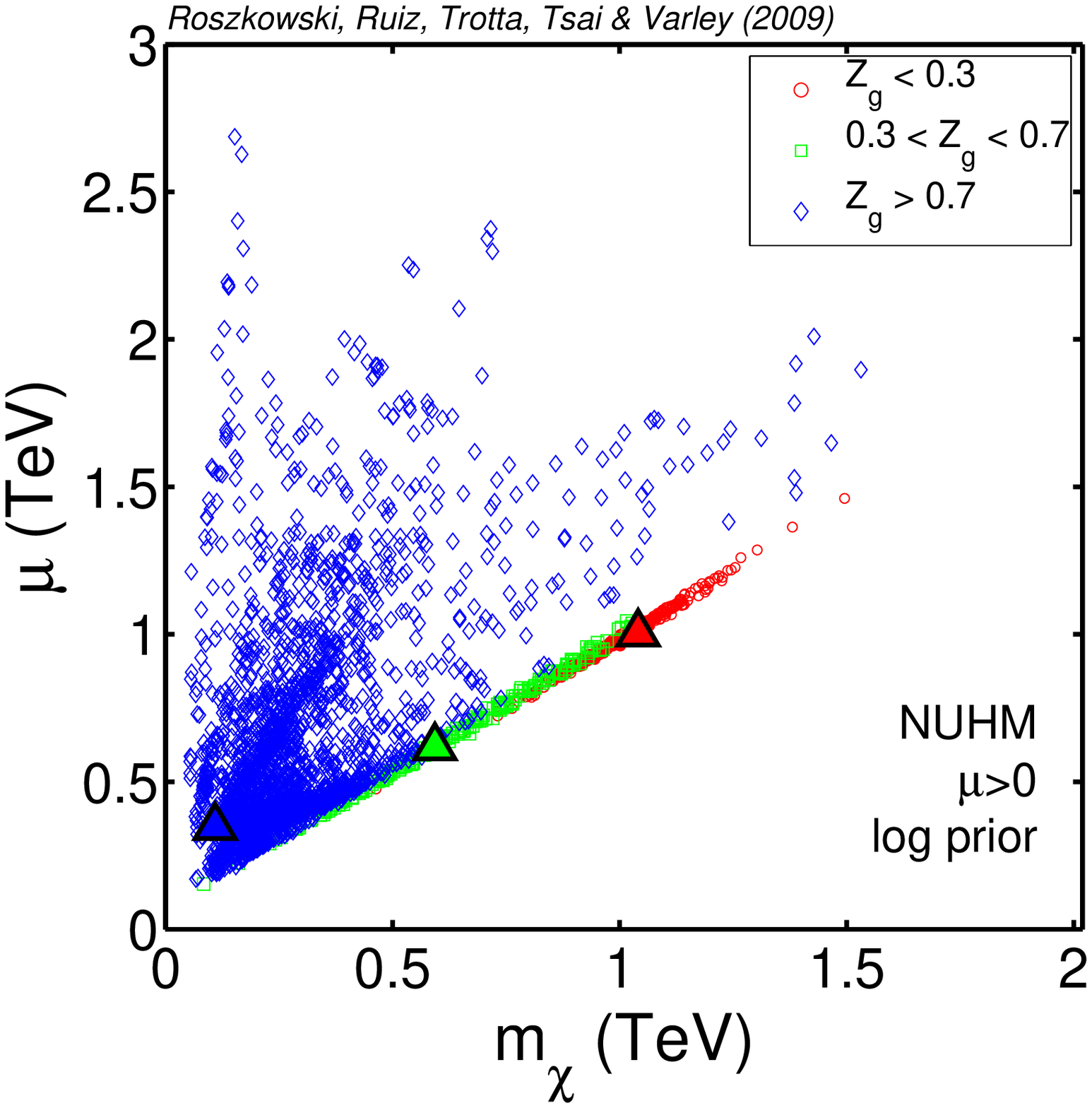}
\includegraphics[width=0.31\textwidth]{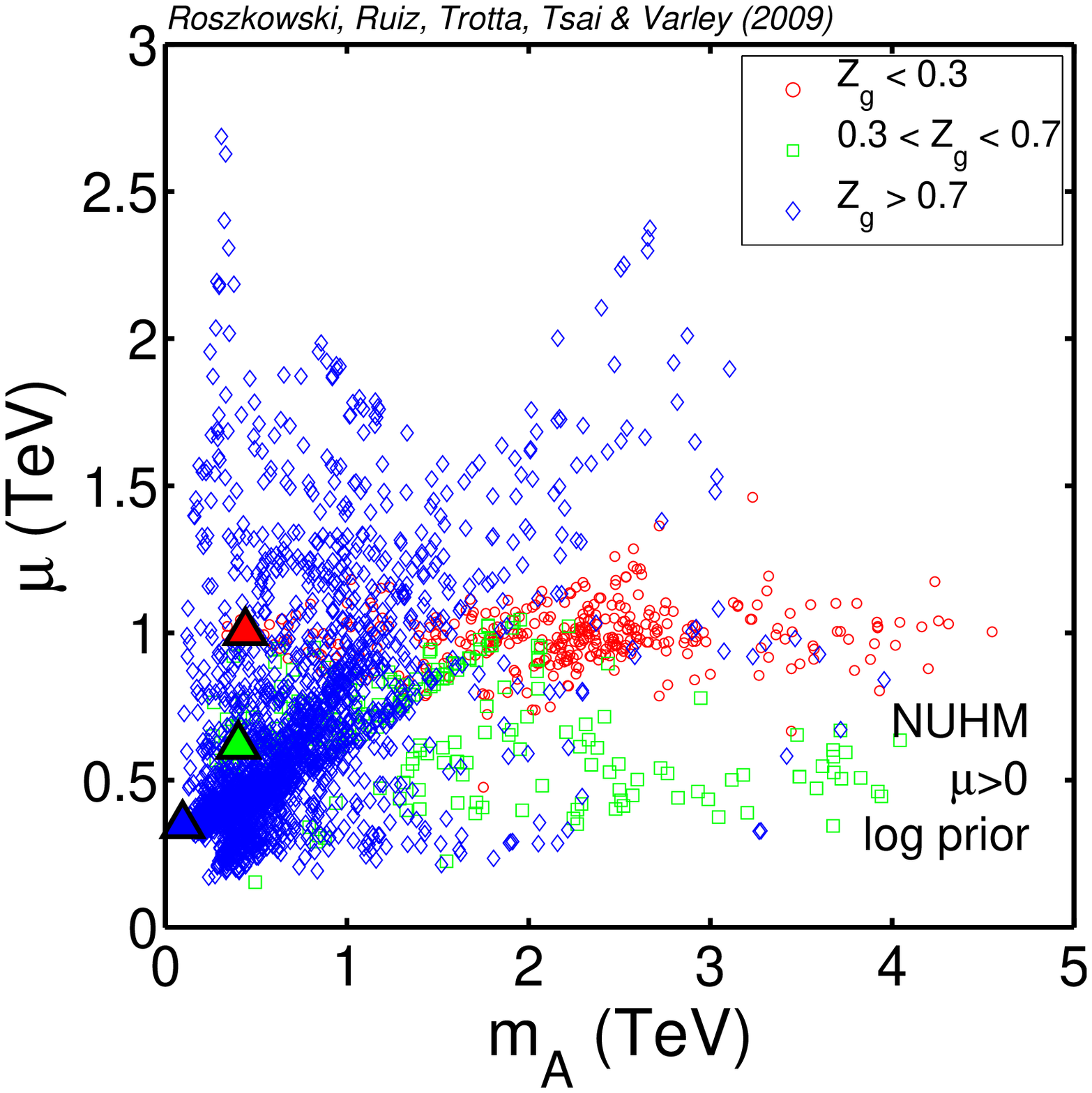}
\end{tabular}
\end{center}
\caption{\label{fig:nuhm_ma_mchi_gaug_3D_mup}
From left to right, values of the gaugino   fraction $Z_g$ in the planes of: $(\mchi,\mha)$,
  $(\mchi,\mu)$ and $(\mha,\mu)$, for samples uniformly
  selected from our MC chains obtained. The color coding is as before.
%lr   follows: red dots correspond to $Z_g<0.3$ (mostly higgsino), green
%lr   squares to $0.3<Z_g<0.7$ and blue diamonds to $Z_g > 0.7$
%(mostly   % gaugino).  
}
\end{figure}

The condition $\mu\gsim 0.8\tev$ is reflected in the vertical red band
(higgsino-like DM) in the left panel of Fig.~\ref{fig:nuhm_ma_mchi_gaug_3D_mup} where we
show the distribution of samples (color-coded according to the gaugino
fraction) in the plane $(\mchi,\mha)$. On the other hand, as $\mu$
increases, so does $\abundchi$ which quickly becomes unacceptably
large. As a result one finds a strong concentration at
$\mu\simeq1\tev$, corresponding to higgsino LSP (compare middle panel). On the other hand,
the diagonal branch in fig.~\ref{fig:nuhm_ma_mchi_gaug_3D_mup}
corresponds to the second way of arriving at the correct relic
density, namely via an $\ha$
resonance annihilation of bino-like neutralino which is clearly also
realized in the NUHM. In this case the LSP mass $\mchi$ extends over a
range of values and the correct relic density is achieved when
$\mha\simeq 2\mchi$, as one can clearly see in the left panel of
fig.~\ref{fig:nuhm_ma_mchi_gaug_3D_mup}. On the other hand, there
appear to be little correlation between $\mha$ and $\mu$ (right panel
of the Figure).

%%
%lr* \begin{figure}[tbh!]
%lr* \begin{center}
%lr* \begin{tabular}{c c}
%lr* \includegraphics[width=0.45\textwidth]{figs/nuhm1_ma_vs_mchi_log_gaug_3d.ps}
%lr* \end{tabular}
%lr* \end{center}
%lr* \caption{\label{fig:nuhm_ma_mchi_gaug_3D_mup} Values of the gaugino
%lr*   fraction $Z_g$ in the plane of $(\mchi,\mha)$ for samples uniformly
%lr*   selected from our MC chains obtained. The color coding is as
%lr*   follows: red dots correspond to $Z_g<0.3$ (mostly higgsino), green
%lr*   squares to $0.3<Z_g<0.7$ and blue diamonds to $Z_g > 0.7$ (mostly
%lr*   gaugino).  }
%lr* \end{figure}
%%

%lr \subsection{The statistical viability of higgsino dark matter}\label{sec:viability}

As we have seen, a great majority of viable points in our scans of the
NUHM parameter space give bino-like DM, while it is the higgsino-like
cases that could lead to a potentially striking distinction from the
CMSSM. It is therefore worth assessing the statistical viability of
higgsino DM in the NUHM.

In terms of posterior probability, the relative probability of each DM
composition (mostly higgsino, mostly bino or mixed) can be determined
by counting the number of independent samples of each type. We find
that, under the assumption of the log prior, the probability of the
higgsino region is $\sim 12\%$, while the gaugino region has a
probability of $\sim 74\%$, as reported in
Table~\ref{table:best_fit}. However, those results are expected to be
prior-dependent, as we have already seen that the constraints do not
appear to be sufficiently strong to completely override the prior
choice. Therefore, we also investigate the values of the best-fit
$\chisq$ values in each of the three regions separately, see
Table~\ref{table:best_fit}. The overall best-fit is in the region
where the LSP is mostly gaugino, where the $\chisq =1.6$. The value
increases to $\chisq = 10.52$ for the best-fit in the mixed
region, and to $\chisq = 12.12$ for the higgsino
region. Therefore, from the point of view of a goodness-of-fit test,
the higgsino region would appear to be completely excluded. However,
the detailed breakdown of the total $\chisq$ into the contribution
from each observable reveals that most of the penalty for the best-fit
points in the mixed and higgsino region comes from $\gmtwo$,
see bottom section of Table~\ref{table:best_fit}. In fact, it is only
in the bino region that an excellent fit to the observed anomalous
magnetic moment can be achieved. This has a very simple physical
explanation, namely the fact that the region where the neutralino is
mostly bino is in the bulk region, hence the value of the SUSY masses
is low. 

If one removes $\deltagmtwo$ from the analysis, we find that the
best-fit $\chisq$ values for the three different regions become very
close to each other: $\chisq = 1.60, 2.42, 2.32$ for the bino, mixed
and higgsino regions, respectively. With a $\Delta\chisq \lsim 0.8$
wrt the overall best-fit, the best-fit higgsino DM point can no longer
be ruled out even at the 1$\sigma$ level if one removes the $\gmtwo$
constraint. This demonstrates that the overall best-fit value and the
profile likelihood maps are being driven to a large extent by the
(somewhat controversial) $\deltagmtwo$ constraint. We stress once more at this point, however, that the best-fit values found above in each region have been derived from our MCMC samples, and therefore they cannot be considered as necessarily being the global best fits that could be obtained with a dedicated algorithm, optimized for likelihood maximisation. We can however consider those points as representative of the quality of fit in each region.

At the same time, all three regions achieve very satisfactory fits of
the WMAP relic abundance, albeit via quite different physical
mechanisms.  In the mixed case, annihilation to gauge bosons is
required to give the right relic density, while for higgsino
domination there are coannihilation processes with the
next-to-lightest neutralino and the lightest chargino, along with
$t\bar t$ and gauge boson pair final states contributing to
the required reduction of the relic DM density. This is a consequence
of the masses of the lightest chargino and of the two lightest
neutralinos becoming very close to each other. Indeed, the two most
important annihilation channels reported in Table~\ref{tab:meas} for the higgsino
best-fit point are only just ahead of about a dozen
coannihilation-based processes.

In summary, the parameters of the NUHM exhibit a rather complex
structure, with different regions of parameter space leading to very
different DM compositions. We have identified two dominant regimes:
one corresponds mostly to large
values of the soft mass parameters and gives a $\sim1\tev$
higgsino-like neutralino, while the other is more similar
to the situation seen in the CMSSM, with bino dark matter
and a lighter spectrum. In between the two there is a fairly
statistically insignificant number of cases corresponding to a rather quick
transition between the higgsino and bino dominated DM compositions.
Our detailed statistical analysis finds that
the latter scenario is favoured, although it is not possible to rule
out robustly the former possibility. In particular, we have found that
the difference in the best-fit $\chisq$ for the two cases is largely
due to the $\gmtwo$ constraint, and as such it should be
evaluated with some care.

%%%%%%%%%%%%%%%%%%%%%%%%%%%%%%%%%%%%%%%%%%%%%%%%%%%%%%%%%%%%%%%%%%%%%%%%%%%%%%
\section{Direct and indirect dark matter detection signatures}\label{sec:resultsdm}
%%%%%%%%%%%%%%%%%%%%%%%%%%%%%%%%%%%%%%%%%%%%%%%%%%%%%%%%%%%%%%%%%%%%%%%%%%%%%%

We now proceed to examine implications from our global scans for the
detection of the lightest neutralino as dark matter, considering both
direct detection via its elastic scatterings with targets in
underground detectors, as well as indirect signatures of neutralino
pair annihilation resulting in an additional component of diffuse
gamma radiation from the Galactic center and of positron flux from the
Galactic halo.  The underlying formalism for both direct and indirect
search modes can be found in several sources. (See,
\eg,~\cite{dn93scatt:ref,susy-dm-reviews,efo00,knrr1}.) In our
analysis we have followed the procedure as well as hadronic matrix
elements inputs as presented in our earlier
work~\cite{rtr1,rrt3,rrst1}. Some investigations into this area in the
case of non-universality have also been done in the literature, see
for example refs.~\cite{Baer:2005bu, Mambrini:2004ke}.

\subsection{Direct detection prospects} 

\begin{figure}[tbh!]
\begin{center}
\begin{tabular}{c c}
     \includegraphics[width=0.45\textwidth]{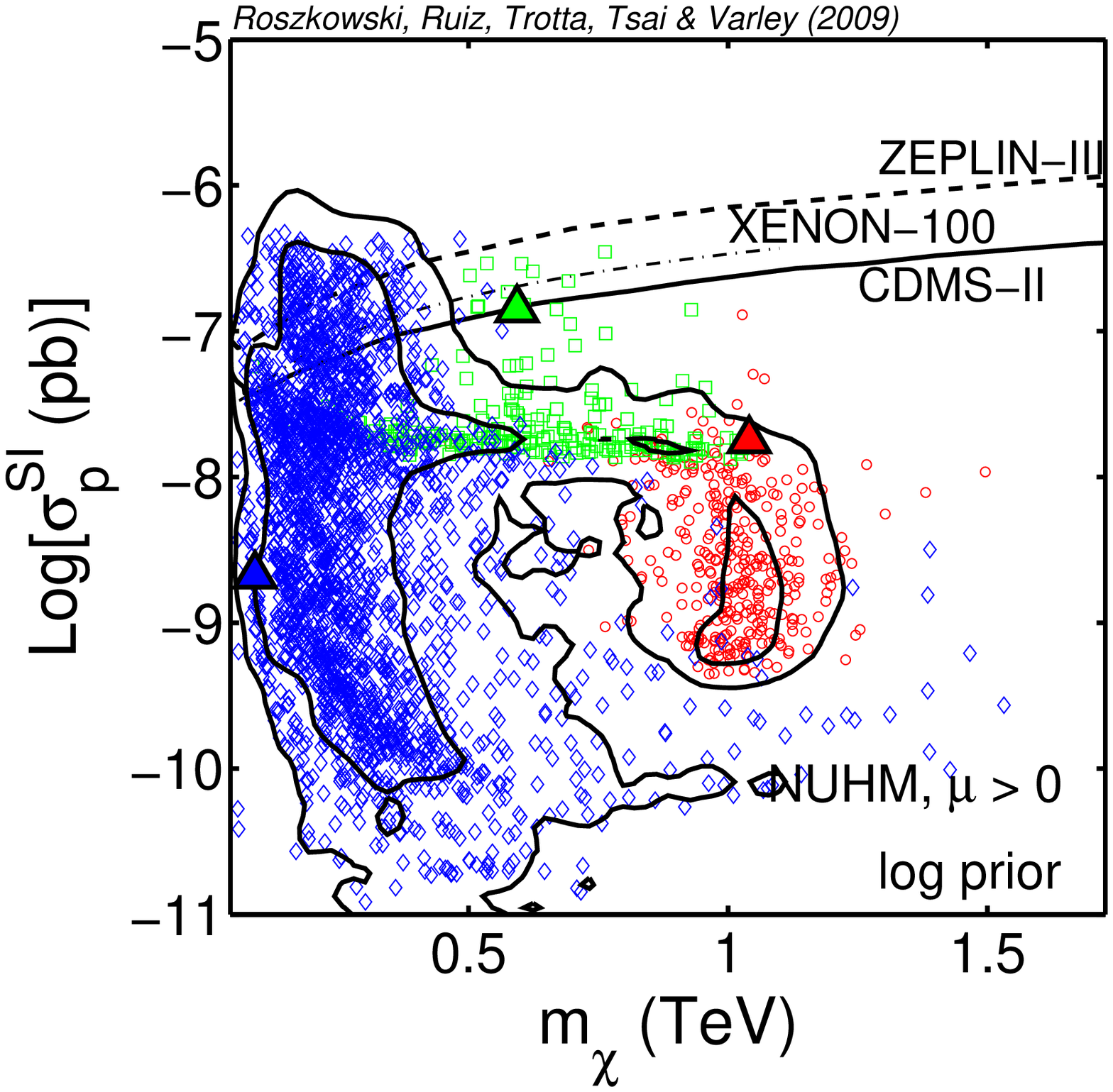} 
   \includegraphics[width=0.45\textwidth]{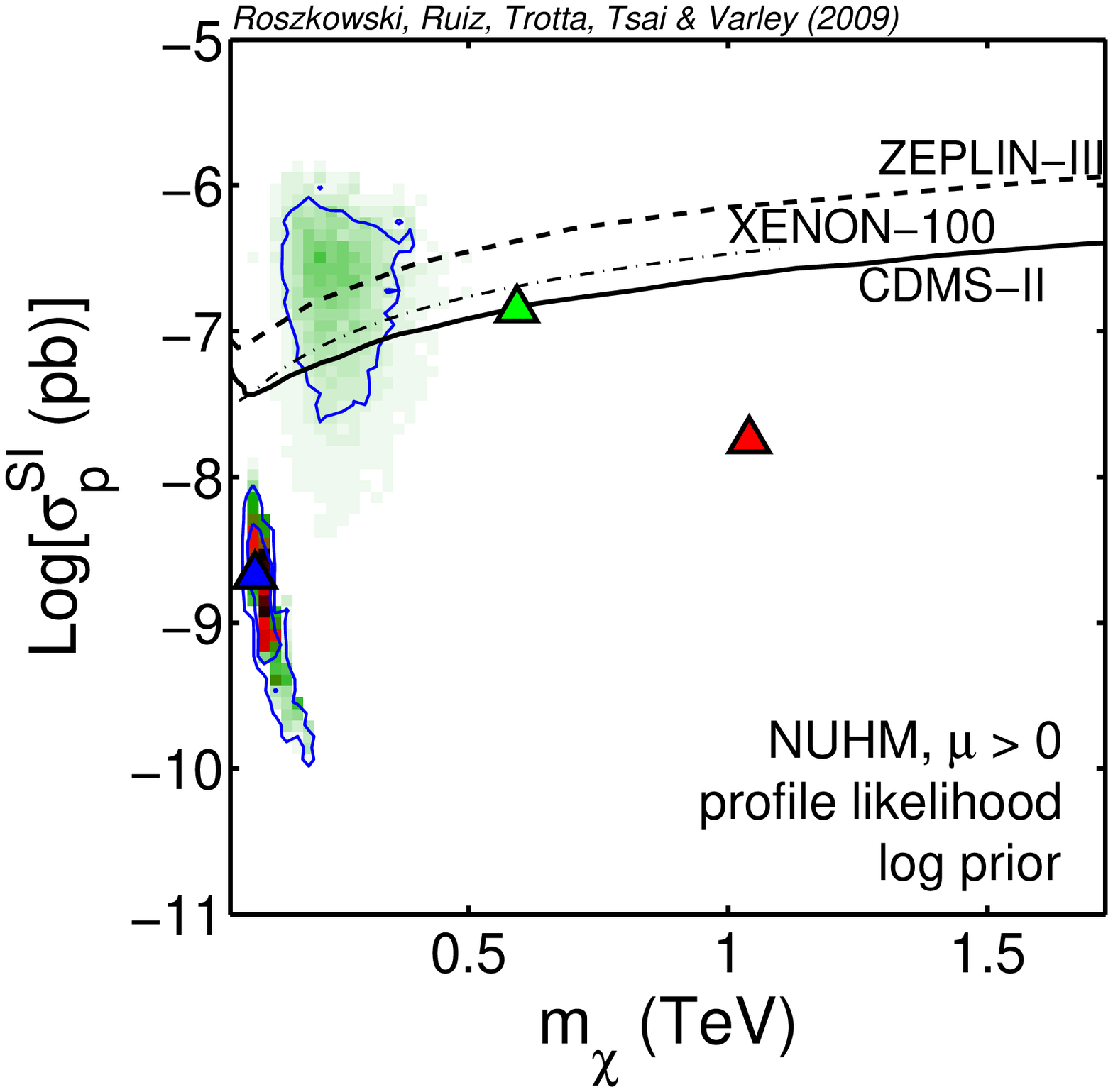}
\end{tabular}
\end{center}
\caption{\label{fig:sigsip_1d} Left panel: favoured regions (black
  contours, enclosing 68\% and 95\% probability) for the dark matter
  spin-independent cross section $\sigsip$ \vs\ the neutralino mass
  $\mchi$ in the NUHM model, together with some recent experimental
  upper limits (which have not been enforced in the
  likelihood). Samples from the posterior are plotted in different
  colours, highlighting regions of different DM composition: blue for
  gaugino-like, red for higgsino-like and green for the mixed
  type. The triangles denote the location of the best-fit points in
  each region. Right panel: profile likelihood for the same variables.
}
\end{figure}

Fig.~\ref{fig:sigsip_1d} shows some quantities of interest in the
plane spanned by the spin-independent cross section $\sigsip$ and the
neutralino mass $\mchi$.  The left panel shows 68\% and 95\% contours
from the 2D posterior pdf, with uniformly weighted samples from the
posterior colour-coded according to the neutralino composition. The
right panel shows the profile likelihood instead.  For comparison,
some of the most stringent 90\%~CL experimental upper limits are also
superimposed~\cite{zeplin3, xenon-100, cdms}. Due to the significant
uncertainties in comparing theoretical predictions with experimental
direct detection limits, 
we have chosen not to impose those limits in the likelihood at this
stage. However, current constraints are starting to impinge on the
favoured parameter space region, whose structure and extent is fairly
similar to what is found in the CMSSM~\cite{tfhrr1}. From the direct
detection point of view, therefore, there is little else than the
$\sim1\tev$ higgsino-like WIMP to tell the two models apart. We also
note that in terms of the posterior pdf all of the 68\% probability
region (inner contours) lies above $\sigsip\gsim10^{-10}\pb$ which
means that the favoured spin-independent cross section region in the
NUHM will be basically completely explored by future 1-tonne detectors
whose sensitivity reach is likely to be of that order or better. This
is encouraging in terms of being able to probe experimentally the
favoured parameter space of the model. As can be seen from the right
panel of Fig.~\ref{fig:sigsip_1d} , the profile likelihood result
essentially singles out a favoured region at small neutralino mass and
cross section between some $10^{-10}$ and $10^{-8}\pb$. The
higgsino-like region (red swarm of points in the left-hand panel)
appears ruled out in terms of the profile likelihood, but as noted
above this is mostly due to the $\gmtwo$  constraint.

%%

%lr* In the right-hand-side panel of fig.~\ref{fig:sigsip_1d} we further
%lr* highlight the different parameter space regions in terms of the
%lr* neutralino composition, as we have done above. The mixed region (green
%lr* samples) is the one that is going to be probed first, as it lies
%lr* completely above $\sigsip\gsim10^{-8}\pb$, a threshold within reach of
%lr* currently running 100-kg scale detector XENON-100. The higgsino region (red samples)
%lr* lies at a larger neutralino mass and above $\sim10^{-9}\pb$,
%lr* and is therefore essentially guaranteed to be explored by the next
%lr* generation of 1-tonne scale detectors.

\subsection{Indirect detection}
Next we discuss some indirect detection modes of much current
interest. We follow the formalism and procedure outlined in
ref.~\cite{rrst1} which we briefly summarize here for completeness. In
our numerical analysis we rely on DarkSusy~\cite{darksusy} to compute
the fluxes.  To start with, we compute the total $\gamma$-ray flux
$\fluxg$ produced by dark matter annihilations in the Galactic halo,
\beq
\fluxg(\Delta\Omega) =  \int^{\mchi}_{E_{\text{th}}}
d\egamma\, \dfluxgdetext(\egamma, \Delta\Omega),
\label{eq:totalgacflux}
\eeq
where the cone $\Delta\Omega$ is centered on the direction $\psi$
and the integration goes over the range of photon
energies from an energy threshold $E_{\text{th}}$ up to $\mchi$. The 
differential diffuse $\gamma$--ray flux arriving from a direction at
an angle $\psi$ from the Galactic center is given by
\beq
\dfluxgde (\egamma, \psi) = \sum_{i} \frac{\sigma_i
v}{8\pi\mchi^2}\, \frac{d N^i_\gamma}{d\egamma}
\int_{\text{l.o.s.}} dl\, \rhochi^2(r(l,\psi)),
\label{eq:diffgammaflux}
\eeq
where $\sigma_i v$ is a product of the WIMP pair-annihilation
cross section into a final state $i$ times the pair's relative
velocity and $d N^i_\gamma /d\egamma$ is the differential
$\gamma$--ray spectrum (including a branching ratio into photons)
following from the state $i$. Here we consider contributions from
the continuum (as opposed to photon lines coming from one loop
direct neutralino annihilation into $\gamma\gamma$ and $\gamma
Z$), resulting from cascade decays of all kinematically allowed final state
SM fermions and combinations of gauge and Higgs bosons. The integral
is taken along the line of sight (l.o.s.) from the detector. It is
convenient to separate factors depending on particle physics and
on halo properties by introducing the dimensionless quantity
$J(\psi)\equiv \left(1/8.5\kpc\right)\left(
0.3\gev/\cmeter^3\right)^2 \int_{\text{l.o.s.}} dl\,
\rhochi^2(r(l,\psi))$~\cite{bub97}. The flux is further averaged
over the solid angle $\Delta\Omega$ representing the acceptance
angle of the detector, and one defines the quantity ${\bar
J}(\Delta\Omega)=\left(1/\Delta\Omega\right) \int_{\Delta\Omega}
J(\psi) d\Omega$. Since we are interested in the Galactic center, we
set $\psi=0$.

The flux from the Galactic center critically depends on the dark matter halo
profile at small Galactic radius $r$ where dark matter density is
thought to be largest. In this analysis, as also previously in~\cite{rrst1},  we consider the NFW
model~\cite{nfwhalo95}, 
%%%%%%%%%%%%%%%%%%%%%%%%%%%%%%%%%%%%%%%%%%%%%%%%%%%%%%%%%%
\beq
\rhochi(r)=\frac{\rho_s}{ (r/r_s){\left[1+\left(r/r_{s}\right)\right]^2}},
\label{eq:halomodels}
\eeq
%%%%%%%%%%%%%%%%%%%%%%%%%%%%%%%%%%%%%%%%%%%%%%%%%%%%%%%%%%
where $\rho_s=0.285\gev/\cmeter^3$ and $r_s= 20.0\kpc$,
motivated by results from past numerical simulations~\cite{Navarro:2003ew,Graham:2005xx}.
In
addition also include the Einasto model which has recently become more
favored by numerical simulations of galactic
halos~\cite{einasto65,Navarro:2008kc},
\begin{equation}\label{einasto}
\rho(r)=\rho_s\exp\{\[\frac{2}{\alpha}
\frac{r_s^{\alpha}-r^{\alpha}}{r_{s}^{\alpha}}\]\}, 
\end{equation}
which has recently become favored by numerical simulations of the
Galactic halo. We adopt (thereafter called ``Einasto'') with
parameters $\rho_s=0.054 \gev/\cmeter^3$, the scale radius
$r_s=21.5\kpc$ and best-fit case for the slope
$\alpha=0.17$~\cite{Diemand:2008in}.  The inner radius density profile
for Einasto profile is not as steep as the NFW one with
$r^{-1}$. Close to solar radius both models become quite similar.

\begin{figure}[tbh!]
\begin{center}
\begin{tabular}{c c}
   \includegraphics[width=0.45\textwidth]{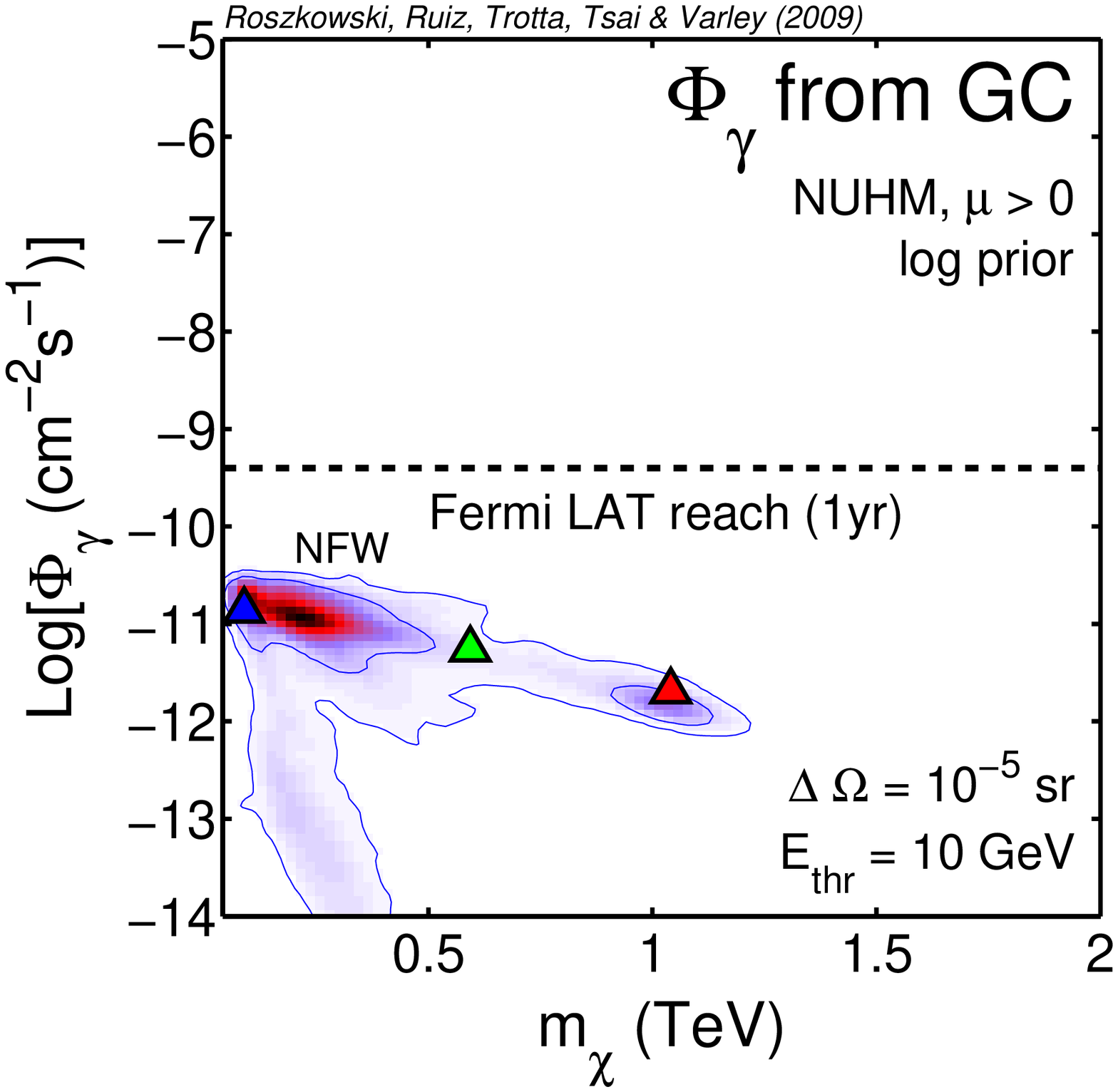}
   \includegraphics[width=0.45\textwidth]{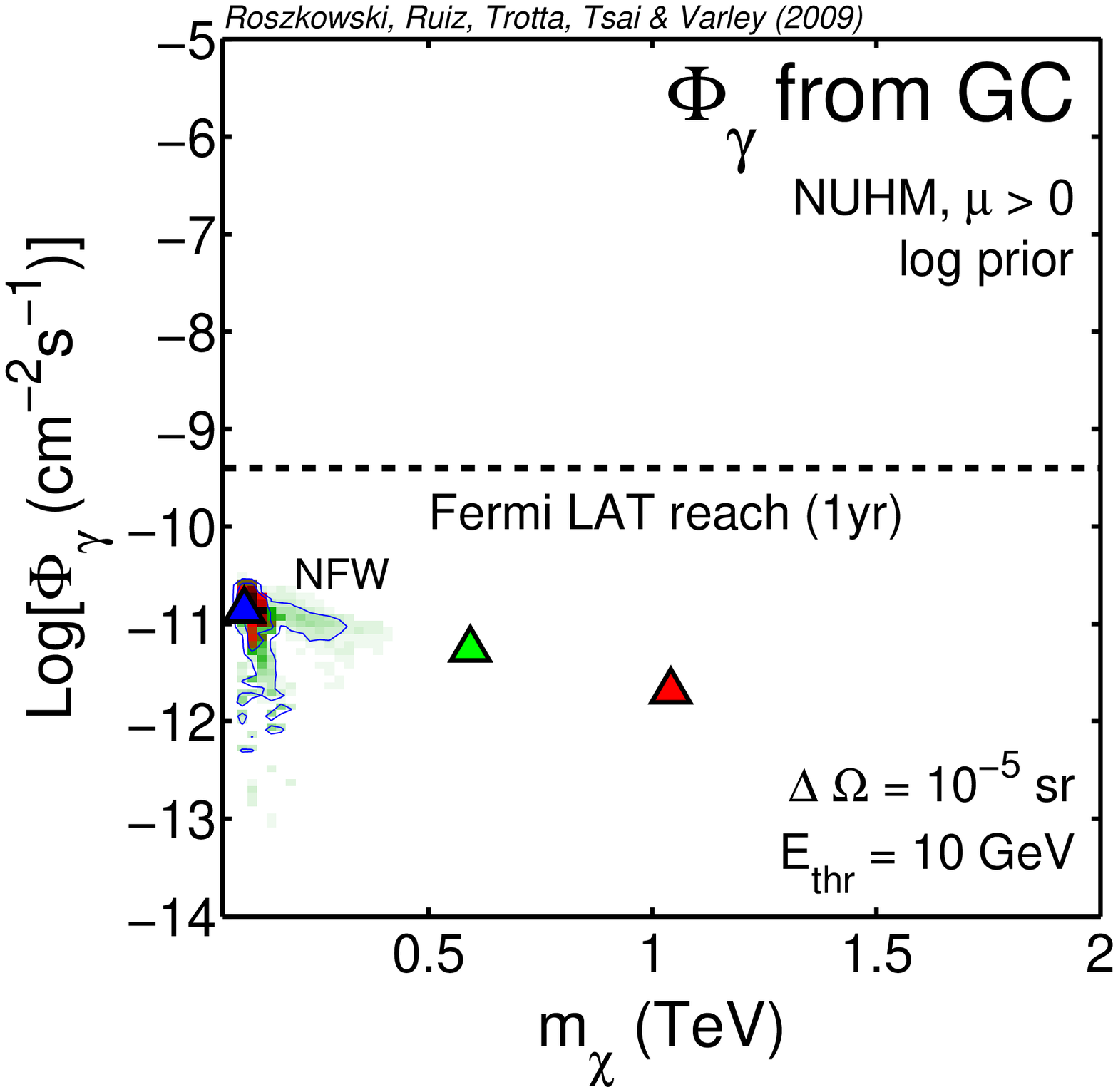}
\end{tabular}
\end{center}
\caption{\label{fig:nuhm_fluxg_1d} The diffuse $\gamma$-ray flux $\fluxg$ above $10\gev$
  produced by dark matter annihilation in the Galactic center \vs\ the
  neutralino mass $\mchi$ for some popular halo models. The left panel
  shows the posterior pdf (with 68\% and 95\% regions), while the
  right panel presents the profile likelihood. 
}
\end{figure}

In fig.~\ref{fig:nuhm_fluxg_1d} we present our predictions for the
diffuse $\gamma$-ray flux $\fluxg$ produced by dark matter
annihilation in the Galactic center \vs\ the neutralino mass $\mchi$
for the NFW halo model. We also assume a conservative energy threshold
$E_{\text{th}} = 10\gev$ and $\Delta\Omega = 10^{-5}\sr$ to match
Fermi's resolution.  Both the 68\% and the 95\% total probability
ranges (inner and outer contours, respectively) are shown. As
expected, the largest flux corresponds to the low-mass bino-like
neutralino region, but note the 68\% probability the higgsino-like
region at $\mchi \sim 1 \tev$. Blue, red and green trianges denote the
best-fit points for the bino, mixed and higgsino-like cases, as before.  Fermi's reach with one
year of data is also indicated (horizontal black/dashed
line~\cite{glast-reach-own}). Using the Einasto model instead gives
slightly lower fluxes, since the profile is less steep in the Galactic
center than the NFW one.

The emerging picture is fairly similar to the CMSSM~\cite{rrst1}.  In
particular, it is clear that the largest uncertainty in assessing
Fermi's prospects for detecting a signal in this model lies in the
cuspiness of the dark matter halo profile at small radii.  If the
steepness of the profile in the Galactic center is similar to that of
the NFW or Einasto model, then a signal in the NUHM or CMSSM is highly
unlikely.

\begin{figure}[tbh!]
\begin{center}
\begin{tabular}{c c}
  \includegraphics[width=0.45\textwidth]{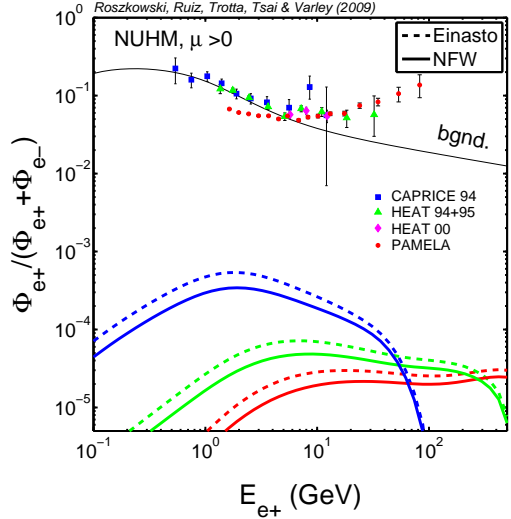}
%lr   \includegraphics[width=0.45\textwidth]{NUHMLog_Phiep-frac_vs_E_mchi_mup_BF1.ps}
\end{tabular}
\end{center}
\caption{\label{fig:nuhm_fluxep_2d_log} Positron flux fraction
  produced by dark matter annihilation in the Galactic halo \vs\ the
  positron energy $\epos$, for the three best-fit points corresponding
  to different DM composition: gaugino (blue), mixed (green) and
  higgsino (red). The solid lines assume the NFW profile, while the
  dashed lines are for the Einasto profile. Everywhere a boost factor
  BF=1 is assumed. }
\end{figure}

Finally we present the NUHM's predictions for positron flux from
neutralino dark matter annihilation in the local halo. Once produced,
positrons propagate through the Galactic medium and their spectrum is
distorted due to synchrotron radiation and inverse Compton scattering
at high energies, bremsstrahlung and ionization at lower energies.
The effects of positron propagation are computed following a standard
procedure described in~\cite{esu04,be98}, by solving numerically the
diffusion-loss equation for the number density of positrons per unit
energy $\dnposdepstext$. The diffusion coefficient is parameterized as
$K(\eps) = K_0(3^\alpha + \eps^{\alpha})$, with $K_0 = 2.1 \times
10^{28}~\cmeter^{2}\second^{-1}$, $\alpha = 0.6$ and $\eps=
\epos/1\gev$, mimicking re-acceleration effects. The energy loss rate
is given by $b(\eps) = \tau_E\eps^2$, with $\tau_E =
10^{-16}\second^{-1}$, and we describe the diffusion zone (\ie, the
Galaxy) as an infinite slab of height $L=4\kpc$, with free escape
boundary conditions. Changes in the above positron propagation model,
especially $K(\eps)$ (see, \eg,~\cite{ms98,be98}), can potentially
lead to variations by a factor of 5 to 10 in the spectral shape at low
positron energy, $\epos \lsim 20\gev$~\cite{hs04}.  Most high-energy
positrons, on the other hand, originate from the local neighborhood
the size of a few~$\kpc$s~\cite{be98,lpst06}, and their flux is less
dependent of the halo and propagation dynamics.  In order to reduce
the impact of solar winds and magnetosphere effects on the model's
predictions, it is useful to consider the positron fraction, defined
as $\fluxpos/(\fluxpos + \fluxelec)$, where $\fluxpos$ is the positron
differential flux from WIMP annihilation, while $\fluxelec$ is the
background electron flux. For background $e^-$ and $e^+$ fluxes we
follow the parametrization adopted in ref.~\cite{be98} from
ref.~\cite{ms98}.

In fig.~\ref{fig:nuhm_fluxep_2d_log} we present the positron flux
fraction produced by dark matter annihilation in the Galactic halo
\vs\ the positron energy $\epos$ for the bino (blue), mixed (green)
and higgsino-like (red) cases, and for the two halo models considered
above. Also included are the relevant experimental data~\cite{HEAT,
  pamelapositron08}, including the recent Pamela result. It is clear
that supersymmetric dark matter in the models like the NUHM (and also
the CMSSM~\cite{rrst1}) falls far short of reproducing the Pamela
result. This would remain true even if one would be prepared to
consider an unlikely existence of very dense local DM clumps for which
the boost factor would be unrealistically high,
$\sim10^3$. Although a more refined analysis fitting signal and
background simultaneously would be required to draw more quantitative
conclusions, it is clear that the spectrum predicted by the NUHM
appears to have a very different energy dependence from the flux
observed by Pamela. This is not necessarily a problem for the NUHM,
and other unified SUSY models like the CMSSM, as long as their signal
remains below the observed flux, since a more conventional
astrophysical explanation in terms of pulsar radiation may be entirely
sufficient to account for Pamela observations~\cite{pamelapulsars}.

%%%%%%%%%%%%%%%%%%%%%%%%%%%%%%%%%%%%%%%%%%%%%%%%%%%%%%%%%%%%%%%%%%%%%%%%%%%%%%
\section{Conclusions}\label{sec:summary}
%%%%%%%%%%%%%%%%%%%%%%%%%%%%%%%%%%%%%%%%%%%%%%%%%%%%%%%%%%%%%%%%%%%%%%%%%%%%%%

The MCMC analysis of the Non-Universal Higgs Model performed in this
paper reveals a remarkably rich and complex structure of its parameter
space. While the properties of the model are in some aspects fairly
similar to the CMSSM, we have found several interesting
differences. Perhaps the biggest one is the existence of higgsino-like
dark matter with a mass close to $1\tev$. The higgsino dark matter
results from having more parameters than in the CMSSM but also from
the focusing effect being less strong than in that model. A detailed
statistical analysis shows that this higgsino-like DM region is put
under pressure essentially only by the constraint from the anomalous
magnetic moment of the muon, while all other present-day constraints
cannot rule out this possibility very strongly.

In terms of observational consequences at colliders, while a more
detailed analysis would be required to make a more rigorous
quantitative statement, a simple comparison of the mass spectra of the
superpartners and the lightest Higgs leads us to believe that the NUHM
appears rather similar to the CMSSM, which will make it difficult to
experimentally distinguish the two models. Again, the best prospects
may be provided by finding in direct detection searches a $\sim1\tev$
dark matter WIMP, since such a case in the CMSSM is highly
unlikely~\cite{rrt3,tfhrr1}. Fermi's prospects for probing the model
strongly depend on the cuspiness of the dark matter profile in the
Galactic center, and with NFW-like profiles appear rather
unimpressive. Likewise, positrons produced in dark matter annihilation
 remain well below the Pamela result.

\medskip

\acknowledgments LR is partially supported by the EC 6th Framework
Programmes MRTN-CT-2004-503369 and MRTN-CT-2006-035505.  RRdA is
supported by the project PARSIFAL (FPA2007-60323) of the Ministerio de
Educaci\'{o}n y Ciencia of Spain.  TV is supported by STFC.
R.T. would like to thank the Galileo Galilei Institute for Theoretical
Physics for the hospitality and the INFN and the EU FP6 Marie Curie
Research and Training Network ``UniverseNet'' (MRTN-CT-2006-035863)
for partial support. The authors would like to thank the European
Network of Theoretical Astroparticle Physics ENTApP ILIAS/N6 under
contract number RII3-CT-2004-506222 for financial support. This
project benefited from the CERN-ENTApP joint visitor's programme on
dark matter, 2-6 February 2009. The use of the Iceberg computer
cluster at the University of Sheffield is gratefully acknowledged.

%%%%%%%%%%%%%%%%%%%%%%%%%%%%%%%%%%%%%%%%%%%%%%%%%%%%%%%%%%%%%%%%%%%%%%%%%%%%%%
\appendix

%%%%%%%%%%%%%%%%%%%%%%%%%%%%%%%%%%%%%%%%%%%%%%%%%%%%%%%%%%%%%%%%%%%%%%%%%%%

\section{Illustration of prior dependence}
\label{sec:priors}

We illustrate the dependence of our results on the choice of priors
for the NUHM parameters. As in any Bayesian analysis, the choice of
priors determines the metric with which the parameter space is
scanned, and therefore it can have an impact on the resulting
posterior probabilities. It is however expected that the prior choice
ought to become irrelevant once the constraining power of the data is
sufficient (see~\cite{Trotta:2008qt} for an illustration). Indeed, it
has been shown in a case study that future LHC
mass spectrum measurements are likely to lead to inferences on the
parameters of the CMSSM whose prior dependence will be strongly
reduced~\cite{rrt4}. 

We repeated our scans using a prior uniform in the NUHM masses
($\mhalf$, $\mzero$, $\mhu$ and $\mhd$), rather than uniform in their
log, as before. The prior on the remaining NUHM and SM parameters has
been left unchanged. We call this prior choice the {\em flat
  prior}. {\em A priori}, this choice expresses a state of
indifference with respect to the masses themselves, rather than with
respect to their order of magnitude, as with the log prior. However,
one has to bear in mind that a choice of a uniform metric on a linear
scale can lead to very strong ``volume effects'' even in a parameter
space of moderate dimensionality. This refers to the fact that on a
linear scale most of the volume of a $D$ dimensional cube is near its
edges. To estimate the magnitude of this volume effect, recall that,
for the flat prior, the volume encompassed by the mass range between
e.g. 1 and $4\tev$ is a factor $10^4$ larger than between 100 and
$400\gev$. For comparison, under the log prior the ratio of the
volumes of the two regions is unity. Therefore, in order to completely
override the prior volume via the likelihood, one would need at least
a $\sim 4.3\sigma$ preference for the low mass region. Clearly,
current data is not constraining enough to override the prior
preference for large mass in this case. Therefore, analogously to what
is observed in the CMSSM~\cite{tfhrr1,al05}, we expect that under a
flat prior choice our posterior probability mass will be shifted
towards larger masses.

\begin{figure}[tbh!]
\begin{center}
\begin{tabular}{c c c}
\includegraphics[width=0.31\textwidth]{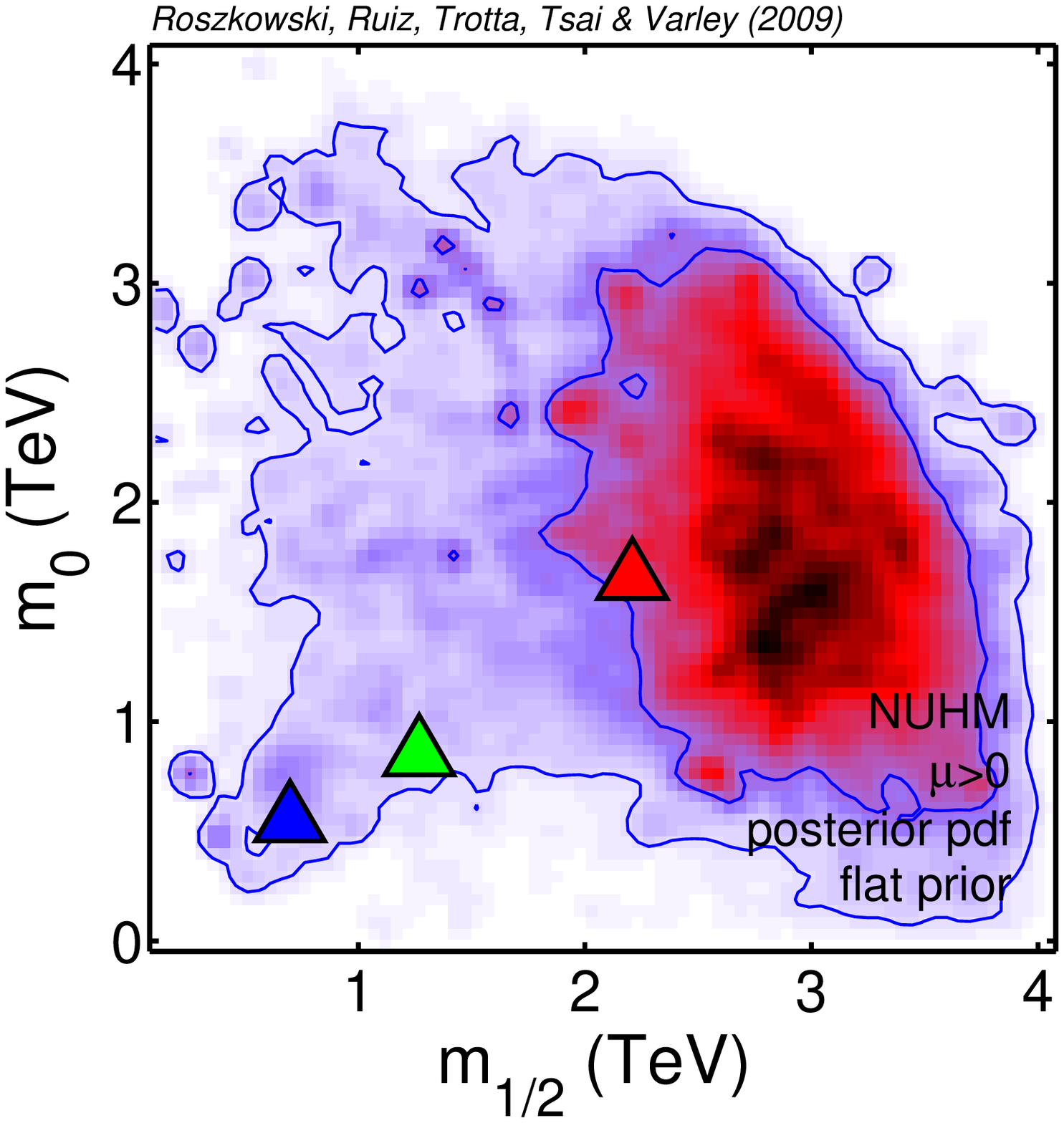}
 &\includegraphics[width=0.31\textwidth]{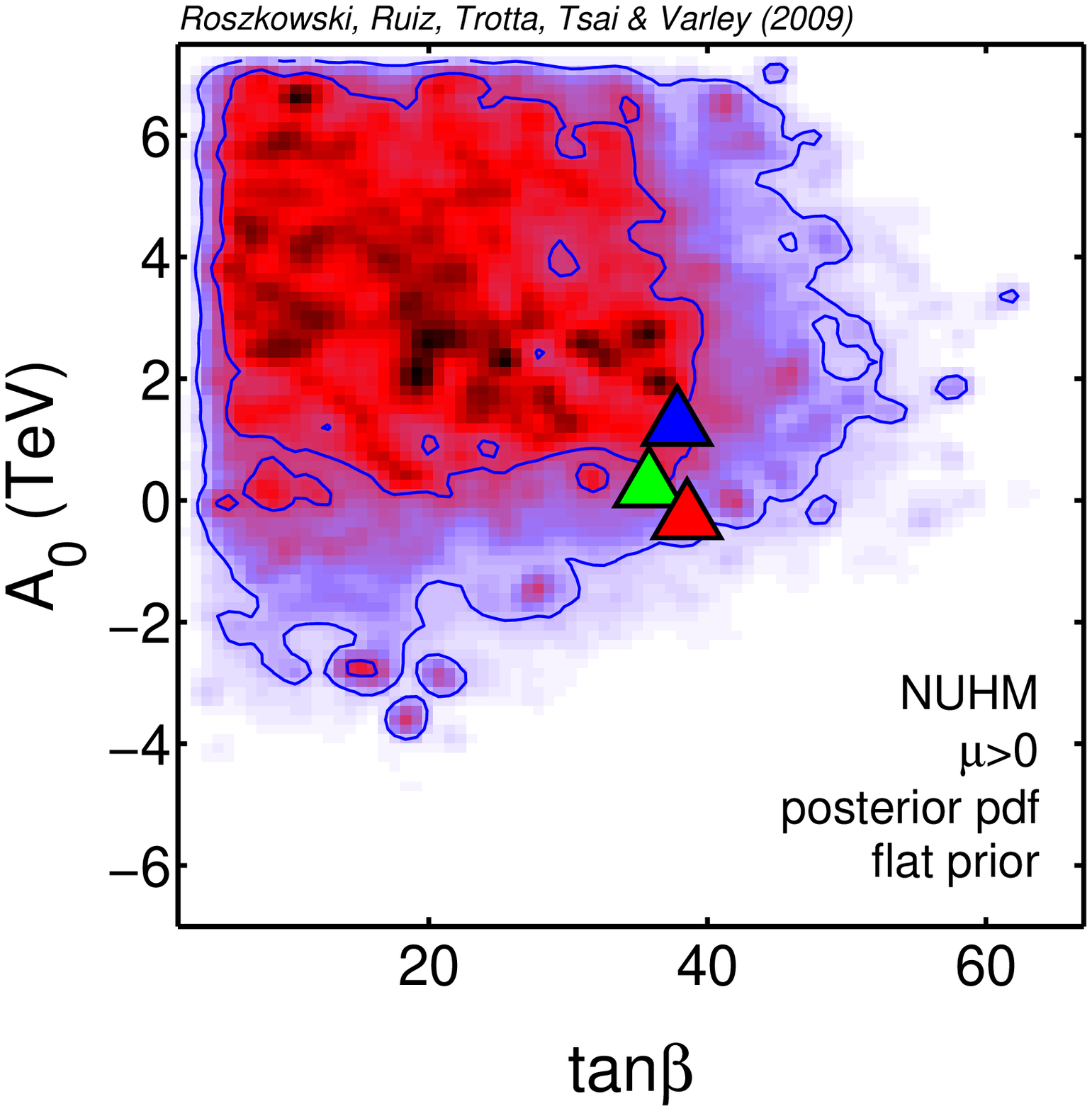}
 & \includegraphics[width=0.31\textwidth]{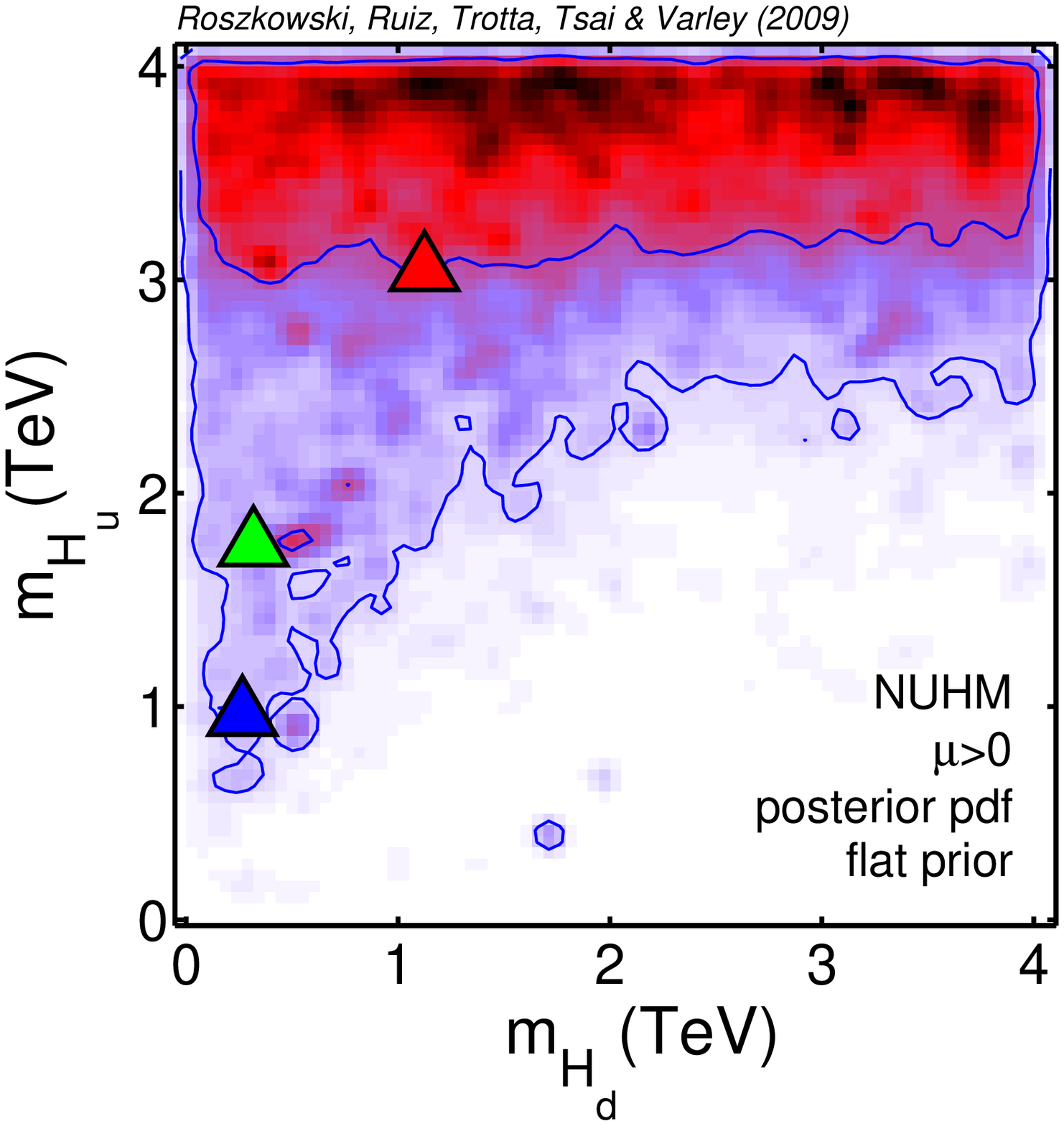}\\
\end{tabular}
%    %\\includegraphics[width=0.3\textwidth]{rrt3-colorbar.ps}
\includegraphics[width=0.3\textwidth]{figs/colorbar.ps}
\end{center}
\caption{ \label{fig:nuhmps2dpost_flat_mup} The same as in
  fig.~\protect\ref{fig:nuhmps2dpost_log_mup} but for the flat
  prior. Coloured triangles show the location of the best-fits for
  each of the DM composition (blue for gaugino, red for higssino and
  gree for mixed). The considerable shift in the probability density
  is largely due to a volume effect from the flat prior arising from
  the large number of samples at large mass for this choice of
  priors. } 
\end{figure}

This is indeed what is observed in
fig.~\ref{fig:nuhmps2dpost_flat_mup}, which is to be compared with
fig.~\protect\ref{fig:nuhmps2dpost_log_mup}. The overall effect of the
flat prior is to produce a large number of samples in the large mass
region, despite the fact that the best fit is still found in the low
mass region. A closer analysis
reveals that the average $\chisq$ of the 68\% probability cloud is
rather poor, ($\langle \chisq \rangle = 20.88$) hence pointing to a
strong volume effect. Thus the flat prior appears to override the
preference for the small mass region (where the average $\chisq$ is
generally better), in the sense that it imposes a measure on parameter
space that does not adequately explore the low mass region.

A detailed comparison of the results from the two prior choices shows
that generically the flat prior puts a stronger {\em a priori}
probability to the large mass region, therefore generally favouring
more strongly the higgsino-like DM scenario outlined before. Therefore
in a flat prior scan all of the distinctive higgsino-like features
highlighted above are strongly enhanced. The best-fit $\chisq$ found
under the flat prior is also quite a bit poorer than the one under the
log prior: $\chisq = 5.55$ under the flat prior vs $\chisq = 1.62$
under the log prior. This shows that the flat prior scan does not
sample accurately enough the low-mass region, which has a relatively
small extension given the choice of a uniform metric on a linear
scale. On the other hand, removing the $\gmtwo$ constraint brings the
best fit value for the flat prior ($\chisq = 2.25$) within less than
1$\sigma$ of the value obtained for the log prior without the $\gmtwo$
constraint ($\chisq = 1.60$), showing that it is indeed this latter
measurement that primarily drives the quality of fit in both scans. As mentioned above, this result shows that the best-fit values obtained from our MCMC scans can only be considered as representative of the quality of fit that can be achieved in each region. A dedicated investigation of the robustness of the best-fit values and of the ensuing profile likelihood is left to a future work.

The phenomenology ensuing from the flat prior choice remains
qualitatively similar to what has been presented above, although there
are fairly evident quantitative differences due to the volume effects
mentioned here. 

It must be emphasised that
the physical features of a model necessarily {\em do not} depend on
our choice of Bayesian prior -- in each case a given point in the
parameter space will give the same physical result.  What does depend
on the prior in the case of insufficiently constraining data is the
weighting that a given point contributes to producing a posterior,
which is a probabilistic and not physical result.  Therefore, while
the physical signatures for a given model are obviously
prior independent, the ensuing statistical conclusions may, and often
do, depend on the choice of priors and statistical approach used. Of
course in the limit of strongly constraining data, statistical
inferences are expected to become essentially independent of such
choices.  While we point out that current data are not sufficiently
constraining to eliminate prior dependence in the NUHM, we do consider
our results obtained under the log prior to be more robust and more
conservative.


\begin{thebibliography}{99}

%%%%%%%%%%%%%%%%%%%%%%%%%%%%%%%%%%%%%%%%%%%%%%%%%%%%%%%%%%%%%%%%%%%%%%%%%%
\bibitem{susy-reviews}
See, e.g., H.~E.~Haber and G.~L.~Kane,
{\it The Search for Supersymmetry: Probing Physics Beyond the Standard Model}
\prep{117}{1985}{75};\\
S.~P.~Martin, {\it A Supersymmetry Primer}, [\hepph{9709356}].

\bibitem{kkrw94}
G.~L.~Kane, C.~F.~Kolda, L.~Roszkowski and J.~D.~Wells,
{\it Study of constrained minimal supersymmetry},
\prd{49}{1994}{6173} [\hepph{9312272}].

%\bibitem{sugra-reviews}
%See, e.g., H.~P.~Nilles,
%{\it Supersymmetry, Supergravity and Particle Physics},
%\prep{110}{1984}{1};\\
%A.~Brignole, L.~E.~Iba\~{n}ez and C.~Mu\~{n}oz,
%{\it Soft supersymmetry breaking terms from supergravity
%and superstring models}, published in
%{\it Perspectives on Supersymmetry},  ed. G.~L.~Kane, 125 [\hepph{9707209]


\bibitem{raby-s010}
T.~Blazek, R.~Dermisek and S.~Raby,
{\it Predictions for the Higgs and supersymmetry spectra from
SO(10) Yukawa unification with $\mu$ greater than 0 },
\prl{88}{2002}{111804} [\hepph{0107097}];
and {\it Yukawa unification in SO(10)},
\prd{65}{2002}{115004} [\hepph{0201081}].

\bibitem{drrr1+2}
R.~Dermisek, S.~Raby, L.~Roszkowski and R.~Ruiz De Austri, {\it
Dark matter and $B(s) \to \mu^+ \mu^-$ SO(10) soft susy breaking},
\jhep{037}{2003}{0304} [\hepph{0304101}]; {\it Dark matter
and $B(s) \to \mu^+ \mu^-$ SO(10) soft susy breaking II},
\jhep{029}{2005}{0509} [\hepph{0507233}].



%\bibitem{ms06-bsg}
%M.~Misiak and M.~Steinhauser, {\it  NNLO QCD corrections to the $\bar{B}\to
%  X_s \gamma$ matrix elements using interpolation in $m_c$},
%\npb{764}{2007}{62} [\hepph{0609241].

%\bibitem{mm-prl06}
%M.~Misiak \etal, {\it  Estimate of $\bar{B}\to
%  X_s \gamma$ at ${\cal O}(\alpha_s^2)$}, \prl{98}{2007}{022002}, [\hepph{0609232].

%\bibitem{bn06}
%T.~Becher and M.~Neubert,
%{\it  Analysis of $Br(B\to X_s \gamma)$ at NNLO with a Cut on Photon Energy},
%\prl{98}{2007}{022003} [\hepph{0610067].
%\bibitem{mcmc}%See, e.g., B.~A.~Berg,%{\it Markov chain monte carlo simulations and their statistical analysis},%World Scientific, Singapore (2004).%\bibitem{bg04}%E.~A.~Baltz and P.~Gondolo,%{\it Markov chain monte carlo exploration of minimal supergravity
%with implications for dark matter},
%\jhep{0410}{2004}{052} [\hepph{0407039].

\bibitem{nuhmbasics}
 See, \eg,
 V.~Berezinsky, \etal
%lr A.~Bottino, J.~R.~Ellis, N.~Fornengo, G.~Mignola and S.~Scopel,
 {\it Neutralino dark matter in supersymmetric models with
   non-universal scalar mass terms}, Astropart.\ Phys.\ {\bf 5} (1996)
 1 [\hepph{9508249}];\\ P.~Nath and R.~L.~Arnowitt, {\it Non-universal
   soft SUSY breaking and dark matter}, Phys.\ Rev.\ D {\bf 56} (1997)
 2820 [\hepph{9701301}];\\ 
M.~Drees, \etal, 
% M.~M.~Nojiri, D.~P.~Roy and  Y.~Yamada, 
{\it Light Higgsino dark matter}, Phys.\ Rev.\ D {\bf 56}
 (1997) 276 [Erratum-ibid.\ D {\bf 64} (2001) 039901]
 [\hepph{9701219}].

\bibitem{Baer:2004fu}
  H.~Baer, \etal, 
% A.~Mustafayev, S.~Profumo, A.~Belyaev and X.~Tata,
  {\it Neutralino cold dark matter in a one parameter extension of the minismal
  supergravity model},
  Phys.\ Rev.\  D {\bf 71} (2005) 095008
  [\hepph{0412059}].

\bibitem{Baer:2005bu}
  H.~Baer, \etal, 
% A.~Mustafayev, S.~Profumo, A.~Belyaev and X.~Tata,
  {\it Direct, indirect and collider detection of neutralino dark matter in
  SUSY
  models with non-universal Higgs masses},
  \jhep{0507}{2005}{065}
  [\hepph{0504001}].

\bibitem{Ellis:2002wv}
  J.~R.~Ellis, K.~A.~Olive and Y.~Santoso,
  {\it The MSSM Parameter Space with Non-Universal Higgs Masses},
  Phys.\ Lett.\  B {\bf 539} (2002) 107
  [\hepph{0204192}].

\bibitem{Ellis:2002iu}
  J.~R.~Ellis, T.~Falk, K.~A.~Olive and Y.~Santoso,
  {\it Exploration of the MSSM with Non-Universal Higgs Masses},
  Nucl.\ Phys.\  B {\bf 652} (2003) 259
  [\hepph{0210205}].
  %%CITATION = NUPHA,B652,259;%%


\bibitem{Ellis:2007by}
  J.~R.~Ellis, S.~F.~King and J.~P.~Roberts,
  {\it The Fine-Tuning Price of Neutralino Dark Matter in Models with
  Non-Universal Higgs Masses},
  \jhep{0804}{2008}{099}
  [\hepph{0711.2741}].

  \bibitem{Baer:2008ih}
  H.~Baer, \etal, 
% A.~Mustafayev, E.~K.~Park and X.~Tata,
 {\it Collider signals and neutralino dark matter detection in
  relic-density-consistent models without universality},
  \jhep{0805}{2008}{058}
  [\hepph{0802.3384}].


 \bibitem{Ellis:lighthiggsino}
  J.~R.~Ellis, \etal, 
% T.~Falk, G.~Ganis, K.~A.~Olive and M.~Schmitt,
  {\it Charginos and Neutralinos in the Light of Radiative Corrections: Sealing
  the Fate of Higgsino Dark Matter},
  Phys.\ Rev.\  D {\bf 58} (1998) 095002
  [\hepph{9801445}];\\ 
J.~R.~Ellis, \etal, 
% T.~Falk, G.~Ganis and K.~A.~Olive,
  {\it Supersymmetric Dark Matter in the Light of LEP and the Tevatron Collider},
  Phys.\ Rev.\  D {\bf 62} (2000) 075010
  [\hepph{0004169}].



\bibitem{al05}
B.~C.~Allanach and C.~G.~Lester,
{\it Multi-dimensional MSUGRA likelihood maps},
\prd{73}{2006}{015013} [\hepph{0507283}];\\ B.~C.~Allanach,
{\it Naturalness priors and fits to the constrained minimal
supersymmetric standard model},
\plb{635}{2006}{123} [\hepph{0601089}];\\ B.~C.~Allanach, C.~G.~Lester and A
.~M.~Weber,
{\it The dark side of mSUGRA}, \jhep{0612}{2006}{065}
[\hepph{0609295}].

\bibitem{rtr1}
R.~Ruiz de Austri, R.~Trotta and L.~Roszkowski,
{\it A Markov Chain Monte Carlo analysis of the CMSSM},
\jhep{0605}{2006}{002} [\hepph{0602028}];
see also R.~Trotta, R.~Ruiz de Austri and L.~Roszkowski,
{\it Prospects for direct dark matter detection in the Constrained
  MSSM}, New Astron. Rev. {\bf 51} (2007) 316-320, [astro-ph/0609126]. %;
 %R. Trotta, R. Ruiz de Austri and C. P\'erez de los Heros, {\it
 %  Prospects for dark matter detection with IceCube in the context of
 %  the CMSSM }, [astro-ph.HE/0906.0366]. 
% Contribution to the proceedings of the Francesco Melchiorri Memorial
% conference, Rome, April 2006

\bibitem{buchmueller_chisq} O.~Buchmueller, \etal,
 {\it Predictions for Supersymmetric Particle Masses in the CMSSM using Indirect
 Experimental and Cosmological Constraints},
[\hepph{0808.4128}] and
 {\it Likelihood Functions for Supersymmetric Observables in
Frequentist Analyses of the CMSSM and NUHM1},
 O. Buchmueller,  \etal, 
% R. Cavanaugh, A. De Roeck, J.R. Ellis, H. Flächer,
%  S. Heinemeyer, G. Isidori, K.A. Olive, F.J. Ronga, G. Weiglein,
 Eur.Phys.J.C64:391-415,2009 [\hepph{0907.5568}].


\bibitem{superbayes}
See: \texttt{http://www.superbayes.org/}

\bibitem{rrt2}
L.~Roszkowski, R.~Ruiz de Austri and R.~Trotta, {\it  On the
  detectability of the CMSSM light Higgs boson at the Tevatron},
\jhep{0704}{2007}{084} [\hepph{0611173}].

\bibitem{BenModelComp}
F.~Feroz, \etal, 
% B.~C.~Allanach, M.~Hobson, S.~S.~AbdusSalam, R.~Trotta and A.~M.~Weber,
{\it Bayesian Selection of sign(mu) within mSUGRA in Global Fits
Including WMAP5 Results}, \jhep{10}{2008}{064}

\bibitem{tfhrr1}
  R.~Trotta, \etal, 
%F.~Feroz, M.~P.~Hobson, L.~Roszkowski and R.~Ruiz de Austri,
  {\it The impact of priors and observables on parameter inferences in the
  Constrained MSSM}, \jhep{0812}{2008}{024}
  [\hepph{0809.3792}].
  
\bibitem{rrt4}
L.~Roszkowski, R.~Ruiz de Austri and R. Trotta,
 {\it Efficient reconstruction of CMSSM parameters using LHC data - A Case Study}, [\hepph{0907.0594}]

\bibitem{GA}
  Y.~Akrami, \etal, 
% P.~Scott, J.~Edsjo, J.~Conrad and L.~Bergstrom,
  {\it A Profile Likelihood Analysis of the Constrained MSSM with Genetic
  Algorithms},
  JHEP {\bf 1004} (2010) 057
  [\hepph{0910.3950}].

\bibitem{coverage}
M.~Bridges, K.~Cranmer, F.~Feroz, M.~Hobson, R.~R.~de Austri and R.~Trotta,
  %``A Coverage Study of the CMSSM Based on ATLAS Sensitivity Using Fast Neural
  %Networks Techniques,''
  arXiv:1011.4306 [hep-ph]; 
  %%CITATION = ARXIV:1011.4306;%%
   Y.~Akrami, C.~Savage, P.~Scott, J.~Conrad and J.~Edsjo,
  %``Statistical coverage for supersymmetric parameter estimation: a case study
  %with direct detection of dark matter,''
  arXiv:1011.4297 [hep-ph].
  %%CITATION = ARXIV:1011.4297;%%

\bibitem{PL}
F.~Feroz \etal, in preparation.

 \bibitem{softsusy}
B.~C.~Allanach,
{\it SOFTSUSY: a C++ program for calculating supersymmetric spectra},
\cpc{143}{2002}{305} [\hepph{0104145}].

\bibitem{darksusy}
P.~Gondolo, \etal, 
% J.~Edsjo, P.~Ullio, L.~Bergstrom,  M.~Schelke and E.~A.~Baltz,
{\it DARKSUSY: computing supersymmetric dark matter
properties numerically}, \jcap{0407}{2004}{008} [astro-ph/0406204];
\texttt{http://www.physto.se/edsjo/darksusy/}.

\bibitem{rrt3}
  L.~Roszkowski, R.~Ruiz de Austri and R.~Trotta,
  {\it Implications for the Constrained MSSM from a new prediction for b to s
  gamma},
  \jhep{0707}{2007}{075}, [\hepph{0705.2012}].
  
 \bibitem{Feroz:2007kg}
F.~Feroz and  M.~P. Hobson
{\it  Multimodal nested sampling: an efficient and robust alternative to
MCMC methods for astronomical data analysis},
Mon. Not. Roy. Astron. Soc. \textbf{384} 449 (2008);\\
F.~Feroz, M.~P.~Hobson and M.~Bridges,
{\em MultiNest: an efficient and robust Bayesian inference tool for cosmology and particle physics} (2008), [astro-ph/0809.3437]


%\bibitem{focuspoint-fmm}
%J.~L.~Feng, K.~T.~Matchev and T.~Moroi,
%{\it Multi - TeV scalars are natural in minimal supergravity},
%\prl{84}{2000}{2322} [\hepph{9908309}]
%and
%{\it  Focus points and naturalness in supersymmetry},
%\prd{61}{2000}{075005} [\hepph{9909334}].

%\bibitem{cdf+dzero-mtop-06}
%The Tevatron Electroweak Working Group,
%{\it Combination of CDF and D0 results on the mass of the top quark},
%hep-ex/0608032.}

\bibitem{topmass:mar07}
Tevatron Electroweak Working Group (for the CDF and D0
Collaborations), {\it A Combination of CDF and D0 Results on the Mass
  of the Top Quark}, [hep-ex/0703034].

\bibitem{pdg06}
W.-M.~Yao \etal\ [Particle Data Group], {\em J. Phys.} {\bf G33} (2006) 1.

\bibitem{lepwwg}
See \texttt{http://lepewwg.web.cern.ch/LEPEWWG}.

\bibitem{gm2alt}
  K.~Hagiwara, A.~D.~Martin, D.~Nomura and T.~Teubner,
  {\it Improved predictions for g-2 of the muon and $\alpha_{\rm QED}(M_Z^2)$},
  Phys.\ Lett.\  B {\bf 649} (2007) 173
  [\hepph{0611102}].

\bibitem{bsgexp} Heavy Flavor Averaging Group (HFAG) (E. Barberio
\etal), {\it Averages of b-hadron properties at the end of 2005},
[hep-ex/0603003]; for a more recent update see Heavy Flavor Averaging
Group (HFAG) (E. Barberio \etal), {\it Averages of b-hadron properties
at the end of 2006}, [hep-ex/0704.3575].

\bibitem{cdf-deltambs}
The CDF Collaboration,
{\it Measurement of the  $B_s-\bar{B}_s$ oscillation frequency},
\prl{97}{2006}{062003} [hep-ex/0606027].
and
{\it Observation of $B_s-\bar{B}_s$ oscillations},
\prl{97}{2006}{242003} [hep-ex/0609040].

%\bibitem{wmap3yr}
%D.N.~Spergel \etal\ [The WMAP Collaboration],
%{\it Wilkinson Microwave Anisotropy Probe (WMAP) Three Year Results:
%  Implications for Cosmology}, astro-ph/0603449.


\bibitem{wmap5yr}
  E.~Komatsu, \etal  [WMAP Collaboration],
  {\it Five-Year Wilkinson Microwave Anisotropy Probe (WMAP)
  Observations:Cosmological Interpretation},
  Astrophys.\ J.\ Suppl.\  {\bf 180} (2009) 330
  [astro-ph/0803.0547].

\bibitem{cdf-bsmumu}
  T.~Aaltonen, etal [CDF Collaboration],
 {\it Search for $B_s \to \mu^+\mu^-$ and $B_d \to \mu^+\mu^-$ Decays with
  2fb$^{-1}$ of $p\bar{p}$ Collisions},
  Phys.\ Rev.\ Lett.\  {\bf 100} (2008) 101802
  [hep-ex/0712.1708].
%The CDF Collaboration,
%{\it Search for $B_s\to\mu^+\mu^-$ and
%$B_d\to\mu^+\mu^-$ decays in $p\bar{p}$ collisions with CDF-II}, CDF
%note 8176 (June 2006).

\bibitem{lhwg}
The LEP Higgs Working Group,
\texttt{http://lephiggs.web.cern.ch/LEPHIGGS};\\
G.~Abbiendi \etal\ [the ALEPH Collaboration, the DELPHI
  Collaboration, the L3 Collaboration and the OPAL Collaboration, The
  LEP Working Group for Higgs Boson Searches],
{\it Search for the standard model Higgs boson at LEP},
\plb{565}{2003}{61} [hep-ex/0306033].

\bibitem{aclw07} B.~C.~Allanach, K.~Cranmer, C.~G.~Lester and
  A.~M.~Weber, {\it Natural Priors, CMSSM Fits and LHC Weather
    Forecasts}, \jhep{0708}{2007}{023} [arXiv:0705.0487].
  
\bibitem{fhrrt1}
  F.~Feroz, M.~P.~Hobson, L.~Roszkowski, R.~Ruiz de Austri and R.~Trotta,
  {\it Are BR($b \rightarrow s \gamma$) and $(g-2)_\mu$ consistent
    within the Constrained MSSM? },  [\hepph{0903.2487}].

\bibitem{baer95} M.~Olechowski and S.~Pokorski, {\it Electroweak
    symmetry breaking with nonuniversal scalar soft terms and large
    tan beta solutions}, Phys.\ Lett.\ B {\bf 344} (1995) 201
  [\hepph{9407404}].

\bibitem{focuspoint-orig}
K.~L.~Chan, U.~Chattopadhyay and P.~Nath,
 {\it Naturalness, Weak Scale Supersymmetry and the Prospect for the
   Observation of Supersymmetry at the Tevatron and at the LHC},
\prd{58}{1998}{096004} [\hepph{9710473}];\\
J.~L.~Feng, K.~T.~Matchev and T.~Moroi,
{\it Multi - TeV scalars are natural in minimal supergravity},
\prl{84}{2000}{2322} [\hepph{9908309}];\\
{\it  Focus points and naturalness in supersymmetry},
\prd{61}{2000}{075005} [\hepph{9909334}];\\
J.~L.~Feng, K.~T.~Matchev and F.~Wilczek,
{\it Neutralino dark matter in focus point supersymmetry},
\plb{B482}{2000}{388} [\hepph{0004043}].

\bibitem{dn93scatt:ref}
M.~Drees and M.~Nojiri,
{\it Neutralino - nucleon scattering revisited},
\prd{48}{1993}{3483} [\hepph{9307208}].


\bibitem{susy-dm-reviews}
See, e.g., G.~Jungman, M.~Kamionkowski
and K.~Griest, {\it Supersymmetric dark matter}, \prep{267}{1996}{195};\\
C.~Mu\~{n}oz, {\it Dark Matter Detection in the Light of Recent
  Experimental Results}, \ijmpa{19}{2004}{3093} [\hepph{0309346}].
%   Journal-ref: Int.J.Mod.Phys. A19 (2004) 3093-3170


\bibitem{efo00}
J.~Ellis, A.~Ferstl, K.~A.~Olive, {\it Reevaluation of the elastic
  scattering of supersymmetric dark matter}, 
\plb{481}{2000}{304} [\hepph{0001005}].
%Phys. Lett. B {\bf 481}, 304 (2000), \hepph{0001005.
% reevaluation of the elastic scattering of supersymmetric dark matter.
\bibitem{knrr1}
Y.~G.~Kim, T.~Nihei, L.~Roszkowski and R.~Ruiz de Austri,
{\it Upper and lower limits on neutralino WIMP mass and spin-independent
scattering cross section, and impact of new (g-2)(mu) measurement},
\jhep{0212}{2002}{034}{2002} [\hepph{0208069}].

\bibitem{rrst1}
  L.~Roszkowski, R.~R.~de Austri, J.~Silk and R.~Trotta,
  {\it On prospects for dark matter indirect detection in the Constrained MSSM},
  Phys.\ Lett.\  B {\bf 671}, 10 (2009)
  [astro-ph/0707.0622].

\bibitem{Mambrini:2004ke}
  Y.~Mambrini and C.~Munoz,
  {\it Gamma-ray detection from neutralino annihilation in non-universal SUGRA
  scenarios},
  Astropart.\ Phys.\  {\bf 24} (2005) 208
  [\hepph{0407158}].

\bibitem{zeplin3}
  V.~N.~Lebedenko, \etal,
  {\it Result from the First Science Run of the ZEPLIN-III Dark Matter Search
  Experiment},
[astro-ph/0812.1150].

\bibitem{xenon-100}
  E.~Aprile, \etal  [XENON100 Collaboration],
 {\it First Dark Matter Results from the XENON100 Experiment},
  [astro-ph.CO/1005.0380].
 %\cite{Ahmed:2009zw}
\bibitem{cdms}
  Z.~Ahmed, \etal [The CDMS-II Collaboration],
  {\it Results from the Final Exposure of the CDMS II Experiment},
  [astro-ph.CO/0912.3592].

\bibitem{bub97}
L.~Bergstr\"{o}m, P.~Ullio and J.~Buckley,
{\it  Observability of Gamma Rays from Dark Matter Neutralino
Annihilations in the Milky Way Halo},
\app{9}{1998}{137} [astro-ph/9712318].
% Astropart.Phys. 9 (1998) 137-162

\bibitem{nfwhalo95}
J.~F.~Navarro, C.~S.~Frenk and S.~D.~M. White,
{\it The structure of cold dark matter halos},
% Astrophys.J.462:563-575,1996
\apj{462}{1996}{563} [astro-ph/9508025]
and
{\it A universal density profile from hierarchical clustering},
% Astrophys.J.490:493-508,1997
\apj{490}{1997}{493}.

%\cite{Navarro:2003ew}
\bibitem{Navarro:2003ew}
  J.~F.~Navarro, \etal,
  %``The Inner Structure of LambdaCDM Halos III: Universality and Asymptotic
  %Slopes,''
  Mon.\ Not.\ Roy.\ Astron.\ Soc.\  {\bf 349} (2004) 1039
  [arXiv:astro-ph/0311231].
  %%CITATION = MNRAA,349,1039;%%

%\cite{Graham:2005xx}
\bibitem{Graham:2005xx}
  A.~W.~Graham, \etal % D.~Merritt, B.~Moore, J.~Diemand and B.~Terzic,
  ``Empirical models for Dark Matter Halos. I. Nonparametric Construction of
  Density Profiles and Comparison with Parametric Models,''
  Astron.\ J.\  {\bf 132}, 2685 (2006)
  [arXiv:astro-ph/0509417].
  %%CITATION = ANJOA,132,2685;%%

\bibitem{einasto65}
J.~Einasto, Trudy Inst. Astroz. Alma-Ata, 51, 87, 1965.

\bibitem{Navarro:2008kc} %%% einasto model compare with
                         %%% arXiv:0904.3830 eg 27  and
                         %%% arXiv:0810.1522 
  J.~F.~Navarro, \etal,
  %``The Diversity and Similarity of Cold Dark Matter Halos,''
[astro-ph/0810.1522].

%\cite{Diemand:2008in}
\bibitem{Diemand:2008in}
  J.~Diemand, \etal,
% M.~Kuhlen, P.~Madau, M.~Zemp, B.~Moore, D.~Potter and J.~Stadel,
  ``Clumps and streams in the local dark matter distribution,''
  arXiv:0805.1244 [astro-ph].
  %%CITATION = ARXIV:0805.1244;%%

\bibitem{glast-reach-own}
See: \texttt{http://tinyurl.com/yp6g5w} (as of Dec 2007).
% http://www-glast.slac.stanford.edu/software/IS/ \newline
% glast\_lat\_performance.htm; http://www-glast.slac.stanford.edu/ \newline
% software/AnaGroup/burnett/performance.htm

\bibitem{esu04}
J.~Edsj\"{o}, M.~Schelke and P.~Ullio,
  {\it Direct versus indirect detection in mSUGRA with self-consistent halo
  models},
 \jcap{0409}{2004}{004} [astro-ph/0405414].
  %%CITATION = JCAPA,0409,004;%%
\bibitem{be98}
E.~A.~Baltz and J.~Edsj\"{o},
{\it Positron Propagation and Fluxes from Neutralino Annihilation in
the Halo},
\prd{59}{1999}{023511} [astro-ph/9808243].
% Phys.Rev. D59 (1999) 023511

\bibitem{ms98}
I.~V.~Moskalenko and A.~W.~Strong,
 {\it Production and propagation of cosmic ray positrons and electrons},
\apj{493}{1998}{694} [astro-ph/9710124];
% Astrophys.J.493:694-707,1998.
I.~V.~Moskalenko, \etal,
% A.~W.~Strong, S.~W.~Digel and T.~A.~Porter,
  {\it Developing the Galactic diffuse emission model for the GLAST Large Area
  Telescope},
[astro-ph/0704.1328].
  %%CITATION = ARXIV:0704.1328;%%

\bibitem{hs04}
D.~Hooper and J.~Silk,
 {\it Searching for Dark Matter with Future Cosmic Positron
 Experiments},
\prd{71}{2005}{083503} [\hepph{0409104}].

\bibitem{lpst06}
J.~Lavalle, J.~Pochon, P.~Salati and R.~Taillet,
{\it Clumpiness of Dark Matter and Positron Annihilation Signal:
 Computing the odds of the Galactic Lottery},
[astro-ph/0603796].
% accepted in A&A

\bibitem{HEAT}
  M.~A.~DuVernois, \etal,
  {\it Cosmic ray electrons and positrons from 1-GeV to 100-GeV: Measurements with
  HEAT and their interpretation},
  Astrophys.\ J.\  {\bf 559} (2001) 296.
\bibitem{pamelapositron08}
O.~Adriani, \etal, {\it Observation of an anomalous positron abundance
  in the cosmic radiation}, [astro-ph/0810.4995]. 

\bibitem{pamelapulsars}
D.~Hooper, P.~Blasi and P.~D.~Serpico,
  {\it Pulsars as the Sources of High Energy Cosmic Ray Positrons},
  JCAP {\bf 0901} (2009) 025
  [astro-ph/0810.1527];  S.~Profumo,
  {\it Dissecting Pamela (and ATIC) with Occam's Razor: existing, well-known
  Pulsars naturally account for the 'anomalous' Cosmic-Ray Electron and
  Positron Data},
  [astro-ph/0812.4457].
  
 \bibitem{Trotta:2008qt}
 R.~Trotta, {\em Bayes in the sky: Bayesian inference and model
   selection in cosmology}, {\em Contemporary Physics}, {\bf 49}, 2,
 71-104 (2008) [astro-ph/0803.4089] 
%%%%%%%%%%%%%%%%%%%%%%%%%%%%%%%%%%%%%%%%%%%%%%%%%%%%%%%%%%%%%%%%%%%%%%%%%%%%%%%%%


%lr* \bibitem{susy-reviews}
%lr* See, e.g., H.~E.~Haber and G.~L.~Kane,
%lr* {\it The Search for Supersymmetry: Probing Physics Beyond the Standard Model}
%lr* \prep{117}{1985}{75};\\
%lr* S.~P.~Martin, {\it A Supersymmetry Primer}, [\hepph{9709356}].
%lr* 
%lr* \bibitem{kkrw94}
%lr* G.~L.~Kane, C.~F.~Kolda, L.~Roszkowski and J.~D.~Wells,
%lr* {\it Study of constrained minimal supersymmetry},
%lr* \prd{49}{1994}{6173} [\hepph{9312272}].
%lr* 
%lr* %\bibitem{sugra-reviews}
%lr* %See, e.g., H.~P.~Nilles,
%lr* %{\it Supersymmetry, Supergravity and Particle Physics},
%lr* %\prep{110}{1984}{1};\\
%lr* %A.~Brignole, L.~E.~Iba\~{n}ez and C.~Mu\~{n}oz,
%lr* %{\it Soft supersymmetry breaking terms from supergravity
%lr* %and superstring models}, published in
%lr* %{\it Perspectives on Supersymmetry},  ed. G.~L.~Kane, 125 [\hepph{9707209]
%lr* 
%lr* 
%lr* \bibitem{raby-s010}
%lr* T.~Blazek, R.~Dermisek and S.~Raby,
%lr* {\it Predictions for the Higgs and supersymmetry spectra from
%lr* SO(10) Yukawa unification with $\mu$ greater than 0 },
%lr* \prl{88}{2002}{111804} [\hepph{0107097}];
%lr* {\it Yukawa unification in SO(10)},
%lr* \prd{65}{2002}{115004} [\hepph{0201081}].
%lr* 
%lr* \bibitem{drrr1+2}
%lr* R.~Dermisek, S.~Raby, L.~Roszkowski and R.~Ruiz De Austri, {\it
%lr* Dark matter and $B(s) \to \mu^+ \mu^-$ SO(10) soft susy breaking},
%lr* \jhep{037}{2003}{0304} [\hepph{0304101}]; {\it Dark matter
%lr* and $B(s) \to \mu^+ \mu^-$ SO(10) soft susy breaking II},
%lr* \jhep{029}{2005}{0509} [\hepph{0507233}].
%lr* 
%lr* 
%lr* 
%lr* %\bibitem{ms06-bsg}
%lr* %M.~Misiak and M.~Steinhauser, {\it  NNLO QCD corrections to the $\bar{B}\to
%lr* %  X_s \gamma$ matrix elements using interpolation in $m_c$},
%lr* %\npb{764}{2007}{62} [\hepph{0609241].
%lr* 
%lr* %\bibitem{mm-prl06}
%lr* %M.~Misiak \etal, {\it  Estimate of $\bar{B}\to
%lr* %  X_s \gamma$ at ${\cal O}(\alpha_s^2)$}, \prl{98}{2007}{022002}, [\hepph{0609232].
%lr* 
%lr* %\bibitem{bn06}
%lr* %T.~Becher and M.~Neubert,
%lr* %{\it  Analysis of $Br(B\to X_s \gamma)$ at NNLO with a Cut on Photon Energy},
%lr* %\prl{98}{2007}{022003} [\hepph{0610067].
%lr* %\bibitem{mcmc}%See, e.g., B.~A.~Berg,%{\it Markov chain monte carlo simulations and their statistical analysis},%World Scientific, Singapore (2004).%\bibitem{bg04}%E.~A.~Baltz and P.~Gondolo,%{\it Markov chain monte carlo exploration of minimal supergravity
%lr* %with implications for dark matter},
%lr* %\jhep{0410}{2004}{052} [\hepph{0407039].
%lr* 
%lr* \bibitem{nuhmbasics}
%lr*  See, \eg,
%lr*  V.~Berezinsky, \etal
%lr* %lr A.~Bottino, J.~R.~Ellis, N.~Fornengo, G.~Mignola and S.~Scopel,
%lr*  {\it Neutralino dark matter in supersymmetric models with
%lr*    non-universal scalar mass terms}, Astropart.\ Phys.\ {\bf 5} (1996)
%lr*  1 [\hepph{9508249}];\\ P.~Nath and R.~L.~Arnowitt, {\it Non-universal
%lr*    soft SUSY breaking and dark matter}, Phys.\ Rev.\ D {\bf 56} (1997)
%lr*  2820 [\hepph{9701301}];\\ M.~Drees, M.~M.~Nojiri, D.~P.~Roy and
%lr*  Y.~Yamada, {\it Light Higgsino dark matter}, Phys.\ Rev.\ D {\bf 56}
%lr*  (1997) 276 [Erratum-ibid.\ D {\bf 64} (2001) 039901]
%lr*  [\hepph{9701219}].
%lr* \bibitem{Baer:2004fu}
%lr*   H.~Baer, A.~Mustafayev, S.~Profumo, A.~Belyaev and X.~Tata,
%lr*   {\it Neutralino cold dark matter in a one parameter extension of the minismal
%lr*   supergravity model},
%lr*   Phys.\ Rev.\  D {\bf 71} (2005) 095008
%lr*   [\hepph{0412059}].
%lr* \bibitem{Baer:2005bu}
%lr*   H.~Baer, A.~Mustafayev, S.~Profumo, A.~Belyaev and X.~Tata,
%lr*   {\it Direct, indirect and collider detection of neutralino dark matter in
%lr*   SUSY
%lr*   models with non-universal Higgs masses},
%lr*   \jhep{0507}{2005}{065}
%lr*   [\hepph{0504001}].
%lr* \bibitem{Ellis:2002wv}
%lr*   J.~R.~Ellis, K.~A.~Olive and Y.~Santoso,
%lr*   {\it The MSSM Parameter Space with Non-Universal Higgs Masses},
%lr*   Phys.\ Lett.\  B {\bf 539} (2002) 107
%lr*   [\hepph{0204192}].
%lr* 
%lr* \bibitem{Ellis:2002iu}
%lr*   J.~R.~Ellis, T.~Falk, K.~A.~Olive and Y.~Santoso,
%lr*   {\it Exploration of the MSSM with Non-Universal Higgs Masses},
%lr*   Nucl.\ Phys.\  B {\bf 652} (2003) 259
%lr*   [\hepph{0210205}].
%lr*   %%CITATION = NUPHA,B652,259;%%
%lr*   
%lr*   
%lr* \bibitem{Ellis:2007by}
%lr*   J.~R.~Ellis, S.~F.~King and J.~P.~Roberts,
%lr*   {\it The Fine-Tuning Price of Neutralino Dark Matter in Models with
%lr*   Non-Universal Higgs Masses},
%lr*   \jhep{0804}{2008}{099}
%lr*   [\hepph{0711.2741}].
%lr*   \bibitem{Baer:2008ih}
%lr*   H.~Baer, A.~Mustafayev, E.~K.~Park and X.~Tata,
%lr*  {\it Collider signals and neutralino dark matter detection in
%lr*   relic-density-consistent models without universality},
%lr*   \jhep{0805}{2008}{058}
%lr*   [\hepph{0802.3384}].
%lr* 
%lr*   
%lr*  \bibitem{Ellis:lighthiggsino} 
%lr*   J.~R.~Ellis, T.~Falk, G.~Ganis, K.~A.~Olive and M.~Schmitt,
%lr*   {\it Charginos and Neutralinos in the Light of Radiative Corrections: Sealing
%lr*   the Fate of Higgsino Dark Matter},
%lr*   Phys.\ Rev.\  D {\bf 58} (1998) 095002
%lr*   [\hepph{9801445}];\\ J.~R.~Ellis, T.~Falk, G.~Ganis and K.~A.~Olive,
%lr*   {\it Supersymmetric Dark Matter in the Light of LEP and the Tevatron Collider},
%lr*   Phys.\ Rev.\  D {\bf 62} (2000) 075010
%lr*   [\hepph{0004169}].
%lr* 
%lr* \bibitem{superbayes} 
%lr* See: \texttt{http://www.superbayes.org/} 
%lr*   
%lr* 
%lr* \bibitem{al05}
%lr* B.~C.~Allanach and C.~G.~Lester,
%lr* {\it Multi-dimensional MSUGRA likelihood maps},
%lr* \prd{73}{2006}{015013} [\hepph{0507283}];\\ B.~C.~Allanach,
%lr* {\it Naturalness priors and fits to the constrained minimal
%lr* supersymmetric standard model},
%lr* \plb{635}{2006}{123} [\hepph{0601089}];\\ B.~C.~Allanach, C.~G.~Lester and A
%lr* .~M.~Weber,
%lr* {\it The dark side of mSUGRA}, \jhep{0612}{2006}{065}  [\hepph{0609295}].
%lr* 
%lr* \bibitem{aclw07} B.~C.~Allanach, K.~Cranmer, C.~G.~Lester and
%lr*   A.~M.~Weber, {\it Natural Priors, CMSSM Fits and LHC Weather
%lr*     Forecasts}, \jhep{0708}{2007}{023} [arXiv:0705.0487].
%lr* 
%lr* \bibitem{rtr1}
%lr* R.~Ruiz de Austri, R.~Trotta and L.~Roszkowski,
%lr* {\it A Markov Chain Monte Carlo analysis of the CMSSM},
%lr* \jhep{0605}{2006}{002} [\hepph{0602028}];
%lr* see also R.~Trotta, R.~Ruiz de Austri and L.~Roszkowski,
%lr* {\it Prospects for direct dark matter detection in the Constrained
%lr*   MSSM}, New Astron. Rev. {\bf 51} (2007) 316-320, [astro-ph/0609126]; R. Trotta, R. Ruiz de Austri and C. P\'erez de los Heros, {\it Prospects for dark matter detection with IceCube in the context of the CMSSM }, [astro-ph.HE/0906.0366].
%lr* % Contribution to the proceedings of the Francesco Melchiorri Memorial
%lr* % conference, Rome, April 2006
%lr* 
%lr* \bibitem{buchmueller_chisq} O.~Buchmueller {\it et al.}, 
%lr* %l {\it Predictions for Supersymmetric Particle Masses in the CMSSM using Indirect
%lr* %l Experimental and Cosmological Constraints}, 
%lr* arXiv:0808.4128 [hep-ph] and
%lr* %l {\it Likelihood Functions for Supersymmetric Observables in
%lr* %Frequentist Analyses of the CMSSM and NUHM1}, 
%lr* % O. Buchmueller, R. Cavanaugh, A. De Roeck, J.R. Ellis, H. Flächer,
%lr* % S. Heinemeyer, G. Isidori, K.A. Olive, F.J. Ronga, G. Weiglein 
%lr* % Eur.Phys.J.C64:391-415,2009 
%lr* arXiv:0907.5568 [hep-ph].
%lr* 
%lr* \bibitem{rrt2}
%lr* L.~Roszkowski, R.~Ruiz de Austri and R.~Trotta, {\it  On the
%lr*   detectability of the CMSSM light Higgs boson at the Tevatron},
%lr* \jhep{0704}{2007}{084} [\hepph{0611173}].
%lr* 
%lr* \bibitem{BenModelComp}
%lr* F.~Feroz, B.~C.~Allanach, M.~Hobson, S.~S.~AbdusSalam, R.~Trotta and 
%lr* A.~M.~Weber, 
%lr* {\it Bayesian Selection of sign(mu) within mSUGRA in Global Fits 
%lr* Including WMAP5 Results}, \jhep{10}{2008}{064}
%lr* 
%lr* 
%lr* \bibitem{rrt3}
%lr*   L.~Roszkowski, R.~Ruiz de Austri and R.~Trotta,
%lr*   {\it Implications for the Constrained MSSM from a new prediction for b to s
%lr*   gamma},
%lr*   \jhep{0707}{2007}{075}
%lr*   [\hepph{0705.2012}].
%lr* 
%lr* 
%lr* \bibitem{tfhrr1}
%lr*   R.~Trotta, F.~Feroz, M.~P.~Hobson, L.~Roszkowski and R.~Ruiz de Austri,
%lr*   {\it The impact of priors and observables on parameter inferences in the
%lr*   Constrained MSSM}, \jhep{0812}{2008}{024}
%lr*   [\hepph{0809.3792}]. 
%lr*   
%lr* 
%lr* \bibitem{GA}
%lr*   Y.~Akrami, P.~Scott, J.~Edsjo, J.~Conrad and L.~Bergstrom,
%lr*   {\it A Profile Likelihood Analysis of the Constrained MSSM with Genetic
%lr*   Algorithms},
%lr*   JHEP {\bf 1004} (2010) 057
%lr*   [\hepph{0910.3950}].
%lr* 
%lr*  \bibitem{rrt4}   L.~Roszkowski, R.~Ruiz de Austri and R. Trotta, 
%lr*  {\it Efficient reconstruction of CMSSM parameters using LHC data - A Case Study}, [\hepph{0907.0594}]
%lr*  
%lr* %lr  \bibitem{st09} L. Strigari and R. Trotta, {\it Reconstructing WIMP
%lr* %lr      Properties in Direct Detection Experiments Including Galactic
%lr* %lr      Dark Matter Distribution Uncertainties}, [astro-ph.HE/0906.5361] 
%lr*  
%lr* % lr \bibitem{uc09} P. Ullio and R. Catena, {\it A novel determination
%lr* %   of the local dark matter density}, [astro-ph.CO)/0907.0018]
%lr*  
%lr* \bibitem{fhrrt1}
%lr*   F.~Feroz, M.~P.~Hobson, L.~Roszkowski, R.~Ruiz de Austri and R.~Trotta,
%lr*   {\it Are BR($b \rightarrow s \gamma$) and $(g-2)_\mu$ consistent within the Constrained MSSM? },
%lr*   [\hepph{0903.2487}]. 
%lr*  
%lr*  %**** JHEP0812:024,2008 ***
%lr*  
%lr* %\bibitem{focuspoint-fmm}
%lr* %J.~L.~Feng, K.~T.~Matchev and T.~Moroi,
%lr* %{\it Multi - TeV scalars are natural in minimal supergravity},
%lr* %\prl{84}{2000}{2322} [\hepph{9908309}]
%lr* %and
%lr* %{\it  Focus points and naturalness in supersymmetry},
%lr* %\prd{61}{2000}{075005} [\hepph{9909334}].
%lr* 
%lr* %\bibitem{cdf+dzero-mtop-06}
%lr* %The Tevatron Electroweak Working Group,
%lr* %{\it Combination of CDF and D0 results on the mass of the top quark},
%lr* %hep-ex/0608032.}
%lr* 
%lr* \bibitem{softsusy}
%lr* B.~C.~Allanach,
%lr* {\it SOFTSUSY: a C++ program for calculating supersymmetric spectra},
%lr* \cpc{143}{2002}{305} [\hepph{0104145}].
%lr* 
%lr* \bibitem{darksusy}
%lr* P.~Gondolo, J.~Edsjo, P.~Ullio, L.~Bergstrom,  M.~Schelke and
%lr* E.~A.~Baltz,
%lr* {\it DARKSUSY: computing supersymmetric dark matter
%lr* properties numerically}, \jcap{0407}{2004}{008} [astro-ph/0406204];
%lr* \texttt{http://www.physto.se/edsjo/darksusy/}.
%lr* 
%lr* \bibitem{Feroz:2007kg}
%lr* F.~Feroz and  M.~P. Hobson  
%lr* {\it  Multimodal nested sampling: an efficient and robust alternative to 
%lr* MCMC methods for astronomical data analysis},
%lr* Mon. Not. Roy. Astron. Soc. \textbf{384} 449 (2008);\\
%lr* F.~Feroz, M.~P.~Hobson and M.~Bridges, 
%lr* {\em MultiNest: an efficient and robust Bayesian inference tool for cosmology and particle physics} (2008), [astro-ph/0809.3437]
%lr* 
%lr* 
%lr* \bibitem{topmass:mar07}
%lr* Tevatron Electroweak Working Group (for the CDF and D0
%lr* Collaborations), {\it A Combination of CDF and D0 Results on the Mass
%lr*   of the Top Quark}, [hep-ex/0703034].
%lr*  
%lr* \bibitem{pdg06}
%lr* W.-M.~Yao \etal\ [Particle Data Group], {\em J. Phys.} {\bf G33} (2006) 1.
%lr*    
%lr* \bibitem{lepwwg}
%lr* See \texttt{http://lepewwg.web.cern.ch/LEPEWWG}.
%lr* 
%lr* \bibitem{gm2alt}
%lr*   K.~Hagiwara, A.~D.~Martin, D.~Nomura and T.~Teubner,
%lr*   {\it Improved predictions for g-2 of the muon and $\alpha_{\rm QED}(M_Z^2)$},
%lr*   Phys.\ Lett.\  B {\bf 649} (2007) 173
%lr*   [\hepph{0611102}].
%lr* 
%lr* \bibitem{bsgexp} Heavy Flavor Averaging Group (HFAG) (E. Barberio
%lr* \etal), {\it Averages of b-hadron properties at the end of 2005},
%lr* [hep-ex/0603003]; for a more recent update see Heavy Flavor Averaging
%lr* Group (HFAG) (E. Barberio \etal), {\it Averages of b-hadron properties
%lr* at the end of 2006}, [hep-ex/0704.3575].
%lr* 
%lr* \bibitem{cdf-deltambs}
%lr* The CDF Collaboration,
%lr* {\it Measurement of the  $B_s-\bar{B}_s$ oscillation frequency},
%lr* \prl{97}{2006}{062003} [hep-ex/0606027].
%lr* and
%lr* {\it Observation of $B_s-\bar{B}_s$ oscillations},
%lr* \prl{97}{2006}{242003} [hep-ex/0609040].
%lr* 
%lr* %\bibitem{wmap3yr}
%lr* %D.N.~Spergel \etal\ [The WMAP Collaboration],
%lr* %{\it Wilkinson Microwave Anisotropy Probe (WMAP) Three Year Results:
%lr* %  Implications for Cosmology}, astro-ph/0603449.
%lr*   
%lr*   
%lr* \bibitem{wmap5yr}
%lr*   E.~Komatsu {\it et al.}  [WMAP Collaboration],
%lr*   {\it Five-Year Wilkinson Microwave Anisotropy Probe (WMAP)
%lr*   Observations:Cosmological Interpretation},
%lr*   Astrophys.\ J.\ Suppl.\  {\bf 180} (2009) 330
%lr*   [astro-ph/0803.0547].
%lr* 
%lr* \bibitem{cdf-bsmumu}
%lr*   T.~Aaltonen {\it et al.}  [CDF Collaboration],
%lr*  {\it Search for $B_s \to \mu^+\mu^-$ and $B_d \to \mu^+\mu^-$ Decays with
%lr*   2fb$^{-1}$ of $p\bar{p}$ Collisions},
%lr*   Phys.\ Rev.\ Lett.\  {\bf 100} (2008) 101802
%lr*   [hep-ex/0712.1708].
%lr* %The CDF Collaboration,
%lr* %{\it Search for $B_s\to\mu^+\mu^-$ and
%lr* %$B_d\to\mu^+\mu^-$ decays in $p\bar{p}$ collisions with CDF-II}, CDF
%lr* %note 8176 (June 2006).
%lr* 
%lr* \bibitem{lhwg}
%lr* The LEP Higgs Working Group,
%lr* \texttt{http://lephiggs.web.cern.ch/LEPHIGGS};\\
%lr* G.~Abbiendi \etal\ [the ALEPH Collaboration, the DELPHI
%lr*   Collaboration, the L3 Collaboration and the OPAL Collaboration, The
%lr*   LEP Working Group for Higgs Boson Searches],
%lr* {\it Search for the standard model Higgs boson at LEP},
%lr* \plb{565}{2003}{61} [hep-ex/0306033].
%lr* 
%lr* \bibitem{focuspoint-orig}
%lr* K.~L.~Chan, U.~Chattopadhyay and P.~Nath, 
%lr*  {\it Naturalness, Weak Scale Supersymmetry and the Prospect for the
%lr*    Observation of Supersymmetry at the Tevatron and at the LHC},
%lr* \prd{58}{1998}{096004} [\hepph{9710473}];\\ 
%lr* J.~L.~Feng, K.~T.~Matchev and T.~Moroi,
%lr* {\it Multi - TeV scalars are natural in minimal supergravity},
%lr* \prl{84}{2000}{2322} [\hepph{9908309}];\\
%lr* {\it  Focus points and naturalness in supersymmetry},
%lr* \prd{61}{2000}{075005} [\hepph{9909334}];\\
%lr* J.~L.~Feng, K.~T.~Matchev and F.~Wilczek,
%lr* {\it Neutralino dark matter in focus point supersymmetry},
%lr* \plb{B482}{2000}{388} [\hepph{0004043}].
%lr*   
%lr*  \bibitem{baer95}
%lr*   M.~Olechowski and S.~Pokorski,
%lr*   {\it Electroweak symmetry breaking with nonuniversal scalar soft terms and large tan beta solutions},
%lr*   Phys.\ Lett.\  B {\bf 344} (1995) 201
%lr*   [\hepph{9407404}].
%lr*   
%lr*    \bibitem{Trotta:2008qt}
%lr*  R.~Trotta, {\em Bayes in the sky: Bayesian inference and model selection in cosmology}, {\em Contemporary Physics}, {\bf 49}, 2, 71-104 (2008) [astro-ph/0803.4089]
%lr* 
%lr* 
%lr* %\bibitem{awramik-acfw04}
%lr* %M.~Awramik, M.~Czakon, A.~Freitas and G.~Weiglein,
%lr* %{\it Precise prediction for the W boson mass in the standard model},
%lr* %\prd{69}{2004}{053006} [\hepph{0311148]; and
%lr* %{\it Complete two-loop electroweak fermionic corrections to
%lr* %$\sineff$ and indirect determination of the
%lr* %Higgs boson mass}, \prl{93}{2004}{201805}  [\hepph{0407317].
%lr* 
%lr* %\bibitem{dghhjw97}
%lr* %A.~Djouadi, P.~Gambino, S.~Heinemeyer, W.~Hollik,
%lr* %C.~Junger and G.~Weiglein, {\it Leading QCD corrections to scalar
%lr* %quark contributions to electroweak precision observables},
%lr* %\prd{57}{1998}{4179} [\hepph{9710438].
%lr* %\bibitem{hmnt06}
%lr* %K.~Hagiwara, A.~D.~Martin, D.~Nomura, T.~Teubner,
%lr* %{\it Improved predictions for g-2 of the muon and $\alpha_{\rm
%lr* %    QED}(M_Z^2)$},
%lr* %\hepph{0611102v3.
%lr* 
%lr* %\bibitem{dgg00}
%lr* %C.~Degrassi, P.~Gambino and G.~F.~Giudice,
%lr* %{\it $B \to X_s \gamma$ in supersymmetry:
%lr* %large contributions beyond the leading order},
%lr* %\jhep{0012}{2000}{009} [\hepph{0009337].
%lr* 
%lr* %\bibitem{gm01}
%lr* %P.~Gambino and M.~Misiak,
%lr* %{\it Quark mass effects in $\bar{B} \to X_s \gamma$},
%lr* %\npb{611}{2001}{338} [\hepph{0104034].
%lr* 
%lr* %\bibitem{or1+2}
%lr* %K.~Okumura and L.~Roszkowski,
%lr* %{\it Deconstraining supersymmetry from $b \to s \gamma$},
%lr* %\prl{92}{2004}{161801} [\hepph{0208101];
%lr* %{\it Large beyond leading order effects in $b \to s \gamma$
%lr* %in supersymmetry with general flavour mixing},
%lr* %\jhep{0310}{2003}{024} [\hepph{0308102].
%lr* 
%lr* %\bibitem{for1+2}
%lr* %J.~Foster, K.~Okumura and L.~Roszkowski, {\it New Higgs effects in
%lr* %B--physics in supersymmetry with general flavour mixing},
%lr* %\plb{609}{2005}{102} [\hepph{0410323] % .
%lr* %and
%lr* %{\it Probing the
%lr* %flavour structure of supersymmetry breaking with rare
%lr* %B-processes: a beyond leading order analysis},
%lr* %\jhep{0508}{2005}{094} [\hepph{0506146].
%lr* 
%lr* %\bibitem{for3+4}
%lr* %J.~Foster, K.~Okumura and L.~Roszkowski,
%lr* %{\it Current and future limits on general flavor violation
%lr* %in $b \to s$ transitions in minimal supersymmetry},
%lr* % JHEP 0603:044,2006
%lr* %\jhep{0603}{2006}{044} [\hepph{0510422] %.
%lr* %and
%lr* %{\it New Constraints
%lr* %on SUSY Flavour Mixing in Light of Recent Measurements at the
%lr* %Tevatron}, \plb{641}{2006}{452} [\hepph{0604121].
%lr* 
%lr* 
%lr* %\bibitem{ehow}
%lr* % \bibitem{ehow04}
%lr* %J.~R.~Ellis, S.~Heinemeyer, K.~A.~Olive and G.~Weiglein,
%lr* %{\it Indirect sensitivities to the scale of supersymmetry},
%lr* %\jhep{0502}{2005}{013} [\hepph{0411216] % .
%lr* %and
%lr* % \bibitem{ehow06}
%lr* % J.~Ellis, S.~Heinemeyer, K.~A.~Olive and G.~Weiglein,
%lr* %{\it Phenomenological indications of the scale of supersymmetry},
%lr* %\jhep{0605}{2006}{005} [\hepph{0602220].
%lr* 
%lr* %\bibitem{trotta07}
%lr* %R.~Trotta,
%lr* %{\it Applications of Bayesian model selection to cosmological
%lr* %  parameters},
%lr* %\mnras{378}{2007}{72} [astro-ph/0504022v3].
%lr* % Mon. Not. R. Astron. Soc., 378, 72-82 (2007)
%lr* %\bibitem{gt07}
%lr* %C.~Gordon and R.~Trotta,
%lr* %{\it Bayesian Calibrated Significance Levels Applied to the Spectral Tilt
%lr* %and Hemispherical Asymmetry}, arxiv:0706.3014.
%lr* 
%lr* \bibitem{dwarfs}
%lr*  G.~D.~Martinez, J.~S.~Bullock, M.~Kaplinghat, L.~E.~Strigari and R.~Trotta,
%lr*   {\it Indirect Dark Matter Detection from Dwarf Satellites: Joint Expectations
%lr*   from Astrophysics and Supersymmetry}
%lr*   [[astro-ph.HE/0902.4715]
%lr*   %%CITATION = ARXIV:0902.4715;%%
%lr*   
%lr* \bibitem{susy-dm-reviews}
%lr* See, e.g., G.~Jungman, M.~Kamionkowski
%lr* and K.~Griest, {\it Supersymmetric dark matter}, \prep{267}{1996}{195};\\
%lr* C.~Mu\~{n}oz, {\it Dark Matter Detection in the Light of Recent
%lr*   Experimental Results}, \ijmpa{19}{2004}{3093} [\hepph{0309346}].
%lr* %   Journal-ref: Int.J.Mod.Phys. A19 (2004) 3093-3170
%lr* 
%lr* 
%lr* 
%lr* \bibitem{dn93scatt:ref}
%lr* M.~Drees and M.~Nojiri,
%lr* {\it Neutralino - nucleon scattering revisited},
%lr* \prd{48}{1993}{3483} [\hepph{9307208}].
%lr* 
%lr* \bibitem{efo00}
%lr* J.~Ellis, A.~Ferstl, K.~A.~Olive, {\it Reevaluation of the elastic scattering of supersymmetric dark matter}, 
%lr* \plb{481}{2000}{304} [\hepph{0001005}].
%lr* %Phys. Lett. B {\bf 481}, 304 (2000), \hepph{0001005.
%lr* % reevaluation of the elastic scattering of supersymmetric dark matter.
%lr* \bibitem{knrr1}
%lr* Y.~G.~Kim, T.~Nihei, L.~Roszkowski and R.~Ruiz de Austri,
%lr* {\it Upper and lower limits on neutralino WIMP mass and spin-independent
%lr* scattering cross section, and impact of new (g-2)(mu) measurement},
%lr* \jhep{0212}{2002}{034}{2002} [\hepph{0208069}].
%lr* 
%lr* \bibitem{rrst1}
%lr*   L.~Roszkowski, R.~R.~de Austri, J.~Silk and R.~Trotta,
%lr*   {\it On prospects for dark matter indirect detection in the Constrained MSSM},
%lr*   Phys.\ Lett.\  B {\bf 671}, 10 (2009)
%lr*   [astro-ph/0707.0622].
%lr* \bibitem{Mambrini:2004ke}
%lr*   Y.~Mambrini and C.~Munoz,
%lr*   {\it Gamma-ray detection from neutralino annihilation in non-universal SUGRA
%lr*   scenarios},
%lr*   Astropart.\ Phys.\  {\bf 24} (2005) 208
%lr*   [\hepph{0407158}].
%lr* %\bibitem{lewin+smith96}
%lr* %J.~D.~Lewin and P.~F.~Smith,
%lr* %{\it  Review of mathematics, numerical factors, and corrections for
%lr* %  dark matter experiments based on elastic nuclear recoil},
%lr* %\app{6}{87}{1996}.
%lr* % Astropart. Phys. 6, 87 (1996).
%lr* %\bibitem{tggrr00}
%lr* %D.~R.~Tovey, R.~J.~Gaitskell, P.~Gondolo, Y.~Ramachers, L.~Roszkowski,
%lr* %{\it A New Model-Independent Method for Extracting Spin-Dependent
%lr* %  Cross Section Limits from Dark Matter Searches},
%lr* %\plb{488}{2000}{17} [\hepph{0005041].
%lr* 
%lr* %  Published in Phys.Lett.B482:388-399,2000.
%lr* 
%lr* % \bibitem{cdms-sep05}
%lr* % The CDMS Collaboration,
%lr* % {\it Limits on spin-independent WIMP-nucleon interactions from the
%lr* % two-tower run of the Cryogenic Dark Matter Search}, 
%lr* % \prl{96}{2006}{011302} [astro-ph/0509259].
%lr* % % Phys.Rev.Lett. 96 (2006) 011302
%lr* % \bibitem{edelweiss-one-final}
%lr* % V.~Sanglard \etal [EDELWEISS Collaboration], {\it Final results of
%lr* % the EDELWEISS-I dark matter search with cryogenic
%lr* % heat-and-ionization Ge detectors}, \prd{71}{2005}{122002}
%lr* % [astro-ph/0503265].
%lr* 
%lr* % \bibitem{zeplin-one-final}
%lr* % G.~J.~Alner \etal [UK Dark Matter Collaboration],
%lr* % {\it First limits on nuclear recoil events from
%lr* % the ZEPLIN-I galactic dark matter detector},
%lr* % \app{23}{2005}{444}.
%lr* % \bibitem{zeplin2}
%lr* %   G.~J.~Alner {\it et al.},
%lr* %   {\it First limits on WIMP nuclear recoil signals in ZEPLIN-II: A two phas
%lr* % e xenon
%lr* %   detector for dark matter detection},
%lr* %   Astropart.\ Phys.\  {\bf 28} (2007) 287
%lr* %   [astro-ph/0701858].
%lr* %   
%lr* \bibitem{zeplin3}
%lr*   V.~N.~Lebedenko {\it et al.},
%lr*   {\it Result from the First Science Run of the ZEPLIN-III Dark Matter Search
%lr*   Experiment},
%lr* [astro-ph/0812.1150].
%lr* 
%lr* \bibitem{xenon-100}
%lr*  \bibitem{Aprile:2010um}
%lr*   E.~Aprile {\it et al.}  [XENON100 Collaboration],
%lr*  {\it First Dark Matter Results from the XENON100 Experiment},
%lr*   [astro-ph.CO/1005.0380].
%lr*  %\cite{Ahmed:2009zw}
%lr* \bibitem{cdms}
%lr*   Z.~Ahmed {\it et al.}  [The CDMS-II Collaboration],
%lr*   {\it Results from the Final Exposure of the CDMS II Experiment},
%lr*   [astro-ph.CO/0912.3592].
%lr* 
%lr* %\bibitem{picasso-sd-05}
%lr* %M.~Barnabe-Heider et al. [PICASSO Collaboration],
%lr* %{\it Improved spin dependent limits from the PICASSO dark matter search experiment},
%lr* %\plb{624}{2005}{186} [hep-ex/0502028].
%lr* % Phys. Lett. B624 (2005), 186-194.
%lr* 
%lr* %\bibitem{simple-sd-05}
%lr* %T.~A.~Girard \etal\ [SIMPLE Collaboration],
%lr* %{\it   Simple dark matter search results}
%lr* %\plb{621}{2005}{233} [hep-ex/0505053].
%lr* % Phys. Lett. B621 (2005), 233-238
%lr* %\bibitem{naiad-sdlimit-05}
%lr* %G.~J.~Alner \etal\ [UKDM Collaboration],
%lr* %{\it Limits on WIMP cross-sections from the NAIAD experiment at the
%lr* %  Boulby Underground Laboratory},
%lr* %\plb{616}{2005}{17} [hep-ex/0504031].
%lr* %Phys. Lett. B 616 (2005), 17.
%lr* %\bibitem{kims-sdlimit-07}
%lr* %H.~S Lee \etal\ [KIMS Collaboration],
%lr* %{\it Limits on WIMP-nucleon
%lr* %cross section with CsI(Tl) crystal detectors}, arXiv:0704.0423.
%lr* %\bibitem{superk-sdlimit-04}
%lr* %S.~Desai \etal\ [Super-Kamiokande Collaboration],
%lr* %{\it Search for dark matter WIMPs using upward through-going muons in
%lr* % Super-Kamiokande},
%lr* %\prd{70}{2004}{083523} [hep-ex/0404025].
%lr* % Phys. Rev. D 70 (2004), 083523.
%lr* 
%lr* \bibitem{bub97}
%lr* L.~Bergstr\"{o}m, et al,
%lr* % P.~Ullio and J.~Buckley,
%lr* {\it  Observability of Gamma Rays from Dark Matter Neutralino
%lr* Annihilations in the Milky Way Halo},
%lr* \app{9}{1998}{137} [astro-ph/9712318].
%lr* % Astropart.Phys. 9 (1998) 137-162
%lr* 
%lr* \bibitem{nfwhalo95}
%lr* J.~F.~Navarro, C.~S.~Frenk and S.~D.~M. White,
%lr* {\it The structure of cold dark matter halos},
%lr* % Astrophys.J.462:563-575,1996
%lr* \apj{462}{1996}{563} [astro-ph/9508025]
%lr* and
%lr* {\it A universal density profile from hierarchical clustering},
%lr* % Astrophys.J.490:493-508,1997
%lr* \apj{490}{1997}{493}.
%lr* 
%lr* \bibitem{einasto65} 
%lr* J.~Einasto, Trudy Inst. Astroz. Alma-Ata, 51, 87, 1965. 
%lr* 
%lr* \bibitem{Navarro:2008kc} %%% einasto model compare with arXiv:0904.3830 eg 27  and arXiv:0810.1522
%lr*   J.~F.~Navarro, {\it et al.},
%lr*   %``The Diversity and Similarity of Cold Dark Matter Halos,''
%lr* [astro-ph/0810.1522].
%lr* 
%lr* %\cite{Diemand:2008in}
%lr* \bibitem{Diemand:2008in}
%lr*   J.~Diemand, \etal,
%lr* % M.~Kuhlen, P.~Madau, M.~Zemp, B.~Moore, D.~Potter and J.~Stadel,
%lr*   ``Clumps and streams in the local dark matter distribution,''
%lr*   arXiv:0805.1244 [astro-ph].
%lr*   %%CITATION = ARXIV:0805.1244;%%
%lr* 
%lr* %lr *** remove *** \bibitem{klypinmodel} 
%lr* %lr A.~Klypin (private communication) ??and F.~Prada, A.~Klypin, J.~Flix, M
%lr* %lr .~Martinez and E.~Simonneau,
%lr* %lr   {\it Astrophysical inputs on the SUSY dark matter annihilation  detectabi
%lr* %lr lity},
%lr* %lr   Phys.\ Rev.\ Lett.\  {\bf 93} (2004) 241301
%lr* %lr   [astro-ph/0401512].
%lr* 
%lr* 
%lr* \bibitem{glast-reach-own}
%lr* See: \texttt{http://tinyurl.com/yp6g5w} (as of Dec 2007).
%lr* % http://www-glast.slac.stanford.edu/software/IS/ \newline
%lr* % glast\_lat\_performance.htm; http://www-glast.slac.stanford.edu/ \newline
%lr* % software/AnaGroup/burnett/performance.htm
%lr* 
%lr* \bibitem{esu04}
%lr* J.~Edsj\"{o}, M.~Schelke and P.~Ullio,
%lr*   {\it Direct versus indirect detection in mSUGRA with self-consistent halo
%lr*   models},
%lr*  \jcap{0409}{2004}{004} [astro-ph/0405414].
%lr*   %%CITATION = JCAPA,0409,004;%%
%lr* \bibitem{be98}
%lr* E.~A.~Baltz and J.~Edsj\"{o},
%lr* {\it Positron Propagation and Fluxes from Neutralino Annihilation in
%lr* the Halo},
%lr* \prd{59}{1999}{023511} [astro-ph/9808243].
%lr* % Phys.Rev. D59 (1999) 023511
%lr* 
%lr* \bibitem{ms98}
%lr* I.~V.~Moskalenko and A.~W.~Strong,
%lr*  {\it Production and propagation of cosmic ray positrons and electrons},
%lr* \apj{493}{1998}{694} [astro-ph/9710124];
%lr* % Astrophys.J.493:694-707,1998.
%lr* I.~V.~Moskalenko, et al,
%lr* % A.~W.~Strong, S.~W.~Digel and T.~A.~Porter,
%lr*   {\it Developing the Galactic diffuse emission model for the GLAST Large Area
%lr*   Telescope},
%lr* [astro-ph/0704.1328].
%lr*   %%CITATION = ARXIV:0704.1328;%%
%lr* \bibitem{hs04}
%lr* D.~Hooper and J.~Silk,
%lr*  {\it Searching for Dark Matter with Future Cosmic Positron
%lr*  Experiments},
%lr* \prd{71}{2005}{083503} [\hepph{0409104}].
%lr* 
%lr* \bibitem{lpst06}
%lr* J.~Lavalle, J.~Pochon, P.~Salati and R.~Taillet,
%lr* {\it Clumpiness of Dark Matter and Positron Annihilation Signal:
%lr*  Computing the odds of the Galactic Lottery},
%lr* [astro-ph/0603796].
%lr* % accepted in A&A
%lr* 
%lr* \bibitem{HEAT}
%lr*   M.~A.~DuVernois {\it et al.},
%lr*   {\it Cosmic ray electrons and positrons from 1-GeV to 100-GeV: Measurements with
%lr*   HEAT and their interpretation},
%lr*   Astrophys.\ J.\  {\bf 559} (2001) 296.
%lr* \bibitem{pamelapositron08} 
%lr* O.~Adriani et al., {\it Observation of an anomalous positron abundance in the cosmic radiation}, [astro-ph/0810.4995].
%lr* \bibitem{pamelapulsars} 
%lr* D.~Hooper, P.~Blasi and P.~D.~Serpico,
%lr*   {\it Pulsars as the Sources of High Energy Cosmic Ray Positrons},
%lr*   JCAP {\bf 0901} (2009) 025
%lr*   [astro-ph/0810.1527];  S.~Profumo,
%lr*   {\it Dissecting Pamela (and ATIC) with Occam's Razor: existing, well-known
%lr*   Pulsars naturally account for the 'anomalous' Cosmic-Ray Electron and
%lr*   Positron Data},
%lr*   [astro-ph/0812.4457].

\end{thebibliography}
\end{document}